\documentstyle[epsf]{l-aa}

\def\nix{\discretionary{}{}{}}
\def\ppmm{$\!\!\!\!\pm\!\!\!\!$}

\begin{document}

\thesaurus{06(02.01.2, 02.08.1, 02.09.1, 02.19.1, 03.13.4, 08.02.1)}

\title{Non-axisymmetric wind-accretion simulations}
\subtitle{I. Velocity gradients of 3\% and 20\% over one accretion radius}

\author{M. Ruffert\thanks{e-mail: {\tt mruffert@mpa-garching.mpg.de}}}
\institute{Max-Planck-Institut f\"ur Astrophysik, Postfach~1523,
D - 85740 Garching, Germany}

%\date{Received; accepted}

\maketitle

\begin{abstract}
We investigate the hydrodynamics of a variant of classical
Bondi-Hoyle-Lyttleton accretion: a totally absorbing sphere
moves at various Mach numbers (3 and 10) relative to a
medium, which is taken to be an ideal gas having a velocity
gradient (of 3\% or 20\% over one accretion radius)
perpendicular to the relative motion.
We examine the influence of the Mach number of the flow and the
strength of the gradient upon the physical behaviour of the flow
and the accretion rates of the angular momentum in particular.
The hydrodynamics is modeled by the ``Piecewise Parabolic Method'' (PPM).
The resolution in the vicinity of the accretor is increased by
multiply nesting several grids around the sphere.

Similarly to the 3D models without gradients published previously,
models exhibit non-stationary flow patterns,
although the Mach cone remains fairly stable.
The accretion rates of mass, linear and angular momenta
do not fluctuate as strongly as published previously for 2D
models, but similarly to the 2D models, transient disks form around
the accretor that alternate their direction of rotation with time. 
The average specific angular momentum accreted is roughly between 7\%
and 70\% of the total angular momentum available in the accretion
cylinder and is always smaller than the value of a vortex with Kepler
velocity around the surface of the accretor. 
The fluctuations of the mass accretion rate in the models with small
gradients (2\%) are similar to the values of the models without
gradients, while the models with large gradients (20\%) exhibit larger
fluctuations.
The mass accretion rate is maximal when the specific angular momentum
is zero, while the specific entropy tends to be smaller when the disks
are prograde.

\end{abstract}

\section{Introduction\label{sec:intro}}

The simplicity of the classic Bondi-Hoyle-Lyttleton (BHL) accretion model
makes its use attractive in order to roughly estimate accretion rates
and drag forces in many different astrophysical contexts, ranging from
wind-fed X-ray binaries (e.g.~Anzer \& B\"orner~1996), over
supernovae (e.g.~Chevalier~1996), and galaxies moving through
intracluster gas in a cluster of galaxies (Balsara at al.~(1994), 
to the black hole believed to be at the center of our Galaxy
(Ruffert \& Melia~1994; Mirabel et al.~1991).
In the BHL scenario a totally absorbing sphere of mass $M$ 
moves with velocity $v_\infty$ relative to a surrounding homogeneous
medium of density $\rho_\infty$ and sound speed  $c_\infty$.
It has been investigated numerically by many workers
(e.g.~Ruffert 1994 and 1995, and references therein).
Usually, the accretion rates of various quantities, like mass, angular 
momentum, etc., including drag forces are of interest as well as the
properties of the flow, (e.g.~distribution of matter and velocity,
stability, etc.). 
All results pertaining to total accretion rates are in qualitative
agreement (to within factors of two, ignoring the instablitites
of the flow) with the original calculations of Bondi, Hoyle and
Lyttleton (e.g.~Ruffert \& Arnett~1994).

The BHL recipe for accretion in the axisymmetric case for pressureless
matter is the following.
A ring of material with radius $b$ (which is identical to the impact
parameter) far upstream from the accretor and thickness d$b$ will be
focussed gravitationally to a point along the radial accretion line
downstream of the accretor. 
At this point the linear momentum perpendicular to the radial
direction is assumed to be cancelled.
Then, if the remaining energy of the matter at this point is not
sufficient for escape from the potential, this material is assumed to
be accreted.
The largest radius $b$ from which matter is still accreted by this
procedure turns out to be the so-called Hoyle-Lyttleton accretion
radius (Hoyle \& Lyttleton~1939, 1940a, 1940b, 1940c; 
Bondi \& Hoyle~1944)
\begin{equation}
   R_{\rm A} = \frac{2GM}{v_\infty^2}  \quad,
\label{eq:accrad1}
\end{equation}
where $G$ is the gravitational constant.
The mass accretion rate follows to be
\begin{equation}
   \dot{M}_{\rm HL} = \pi R^2_{\rm A} \rho_\infty v_\infty  \quad.
  \label{eq:accmass1}
\end{equation}
I will refer to the volume upstream of the accretor from which matter
is accreted as accretion cylinder.

However, if the assumption of homogeneity of the surrounding medium
is dropped, e.g.~by assuming some constant gradient in the density
or the velocity distribution, the consequences on the accretion flow
remain very unclear.
Using the same conceptual procedures, one can calculate 
(Dodd \& McCrea, 1952; Illarionov \& Sunyaev, 1975; 
Shapiro \& Lightman, 1976; Wang, 1981)
how much angular momentum is present in the accretion
cylinder for a non-axisymmetric flow which has a gradient in its
density or velocity perpendicular to the mean velocity direction.
Then, assuming that the angular momentum will be accreted together
with the mass, it is only a small step to conclude that the amount of
angular momentum accreted is equal to (or at least is a large fraction
of) the angular momentum present in the accretion cylinder.
Note, that if the velocity is a function of position, then by virtue
of Eq.~(\ref{eq:accrad1}) also the accretion radius varies in space.
Thus the cross section of the accretion cylinder (perpendicular to the
axis) is not circular.

However, the reasoning of BHL calls for a cancelling of linear
momentum perpendicular to the radial accretion line before matter is
accreted.
Together with this linear momentum also angular momentum is cancelled
and so the matter accreted has zero angular momentum by construction!
This point was first discussed by Davies \& Pringle~(1980), who were
able to construct two-dimensional flows with small non-vanishing
gradients for which the accreted angular momentum was exactly zero, by 
placing the accretion line appropriately.
Thus, following these analytic investigations two opposing views are
voiced about how much angular momentum can be accreted: either a large
or a very small fraction of what is present in the accretion cylinder.
Numerical simulations thus are called for to help solve the problem.

In this paper I would like to compare the accretion rates of several
quantities (especially angular momentum) of numerically modeled
accretion flows with gradients to the previous results of accretion
without gradients (e.g.~Ruffert~1994).
One has to change some of the parameters of the flow (Mach number,
size of the accretor) in order to get a good overview of which
features are generic and which specific to that combination
of parameters.
Although several investigations of {\it two}-dimensional flows with
velocity gradients exist (Anzer et al.~1987; Fryxell \& Taam~1988; 
Taam \& Fryxell~1989; Ho et al.~1989),
{\it three}-dimensional simulations are scarse
due to their inherently high computational load.
Livio et al.~(1986) first attempted a three-dimensional model
including gradients, but due to their 
low numerical resolution the results were only tentative.
Also in the models of Ishii et al.~(1993) was the accretor only
coarsly resolved, while the results of Boffin (1991) and 
Sawada et al.~(1989) are only indicative, because due to the numerical
procedure the flows remained stable (too few SPH particles in
Boffin~1991 and local time stepping in Sawada et al.~1989 which is
appropriate only for stationary flows).
A simulation that was numerically better resolved was performed later
by Ruffert \& Anzer (1995), but since only one model was presented,
the results cannot be taken as conclusive either.
I intend to remedy these shortcomings in the present paper.

In section~\ref{sec:numer} I give only a short summary of the
numerical procedure used. 
Sections~\ref{sec:descr1} to~\ref{sec:descr3} present the results,
which I analyze and interpret in Sect.~\ref{sec:analy}.
Section~\ref{sec:conc} summarizes the implications of this work.

\begin{table*}
\caption[] {
Parameters and some computed quantities for all models.
${\cal  M}_\infty$ is the Mach number of the unperturbed flow,
$\varepsilon_{\rm v}$ the parameter specifying the strength of the
gradient,
$\gamma$ the ratio of specific heats,
$R_\star$ the radius of the accretor,
$g$ the number of grid nesting depth levels,
$\delta$ the size of one zone on the finest grid,
$\epsilon$ the softening parameter (zones) for the potential of the
accretor (see Ruffert, 1994),
$t_{\rm f}$ the total time of the run (units: $R_{\rm  A}/c_{\infty}$),
$\overline{\dot{M}}$ the integral average of the mass accretion rate,
$S$ one standard deviation around the mean $\overline{\dot{M}}$
of the mass accretion rate fluctuations,
$\widehat{\dot{M}}$ the maximum mass accretion rate,
$\dot{M}_{\rm BH}$ is defined in Eq.~(3) of Ruffert \& Arnett (1994),
$l_{\rm x}$, $l_{\rm y}$, $l_{\rm z}$,
are the averages of specific angular momentum components
together with their respective standard deviations
$\sigma_{\rm x}$, $\sigma_{\rm y}$, $\sigma_{\rm z}$,
$s$ is the entropy (Eq.~(4) in Ruffert \& Arnett~1994),
the number $N$ of zones per grid dimension is 32,
and the size of the largest grid is $L=32R_{\rm A}$ (except for
model~RL for which it is $L=128R_{\rm A}$).
}
\label{tab:models}
\tabcolsep = 1.6mm
\begin{flushleft}
\begin{tabular}{lccccccccrclcrclrclrclc}
\hline\\[-3mm]
   Model & ${\cal M}_\infty$ & $\varepsilon_{\rm v}$ 
   & $\gamma$ & $R_\star$ & $g$ & $\delta$ & $\epsilon$
   & $t_{\rm f}$ & $\overline{\dot{M}}$&\ppmm&$S$ & $\widehat{\dot{M}}$
   & $l_{\rm x}$&\ppmm&$\sigma_{\rm x}$ 
   & $l_{\rm y}$&\ppmm&$\sigma_{\rm y}$
   & $l_{\rm z}$&\ppmm&$\sigma_{\rm z}$ & $s$ \\
    & & & & ($R_{\rm A}$) & & ($R_{\rm A}$) & & 
      & \multicolumn{3}{c}{($\dot{M}_{\rm BH}$)}  & ($\dot{M}_{\rm BH}$)
%      & \multicolumn{3}{c}{~~($R_{\rm A}c_\infty$)}
%      & \multicolumn{3}{c}{~~($R_{\rm A}c_\infty$)}
%      & \multicolumn{3}{c}{~~($R_{\rm A}c_\infty$)}
      & \multicolumn{3}{c}{~($1.5\,\varepsilon_{\rm v}R_{\rm A}v_0$)}
      & \multicolumn{3}{c}{~($1.5\,\varepsilon_{\rm v}R_{\rm A}v_0$)}
      & \multicolumn{3}{c}{~($1.5\,\varepsilon_{\rm v}R_{\rm A}v_0$)}
      & (${\cal R}$)
     \\[0.5mm] \hline\\[-3mm]
IT & 3 &-0.03&5/3& 0.02&10 & $1/512$& 8 & 4.82 & 0.72&\ppmm&0.04 & 0.78
      &   0.00&\ppmm&0.01 &   0.00&\ppmm&0.02 &  +0.20&\ppmm&0.04
      & 2.1  \\  % RAIT
IS & 3 &-0.03&5/3& 0.02& 9 & $1/256$& 3 & 13.9 & 0.53&\ppmm&0.09 & 0.80
      &   0.00&\ppmm&0.13 &   0.05&\ppmm&0.17 &  +0.12&\ppmm&0.24
      & 2.2  \\ % RAIS
IM & 3 &-0.03&5/3& 0.10& 7 & $1/64$ & 4 & 26.6 & 0.79&\ppmm&0.06 & 0.90
      &  -0.01&\ppmm&0.11 &   0.00&\ppmm&0.13 &  +0.68&\ppmm&0.33
      & 2.2  \\  % RAIM
IM*& 3 &-0.03&5/3& 0.10& 7 & $1/64$ & 4 & 8.37 & 0.82&\ppmm&0.08 & 0.91
      &  +0.02&\ppmm&0.06 &  -0.08&\ppmm&0.14 &  +0.69&\ppmm&0.45
      & 2.2  \\[1mm]  % RAIN
JS & 10&-0.03&5/3& 0.02& 9 & $1/256$& 3 & 2.93 & 0.45&\ppmm&0.09 & 0.68
      &  -0.04&\ppmm&0.17 &  -0.08&\ppmm&0.35 &  +0.18&\ppmm&0.28
      & 5.3  \\  % RAJS
JM & 10&-0.03&5/3& 0.10& 7 & $1/64$ & 4 & 10.3 & 0.72&\ppmm&0.05 & 0.79
      &  +0.01&\ppmm&0.06 &  -0.07&\ppmm&0.33 &  +0.26&\ppmm&0.35
      & 5.2  \\[1mm]  % RAJM
KS & 3 &-0.20&5/3& 0.02& 9 & $1/256$& 3 & 6.89 & 0.54&\ppmm&0.09 & 0.78
      &   0.00&\ppmm&0.02 &  -0.01&\ppmm&0.03 &  +0.07&\ppmm&0.02
      & 1.8  \\  % RAKS
KM & 3 &-0.20&5/3& 0.10& 7 & $1/64$ & 4 & 20.3 & 0.95&\ppmm&0.19 & 1.30
      &   0.00&\ppmm&0.02 &   0.00&\ppmm&0.02 &  +0.26&\ppmm&0.06
      & 1.6  \\[1mm] % RAKM
LS & 10&-0.20&5/3& 0.02& 9 & $1/256$& 3 & 1.94 & 0.35&\ppmm&0.08 & 0.53
      &  +0.02&\ppmm&0.03 &  +0.01&\ppmm&0.05 &  +0.09&\ppmm&0.03
      & 5.2  \\  % RALS
%LM & 10&-0.20&5/3& 0.10& 7 & $1/64$ & 4 & 6.95 & 2.18&\ppmm&0.32 & 3.04
%      &  +0.03&\ppmm&0.06 &  +0.01&\ppmm&0.09 &  +0.34&\ppmm&0.05
%      & 2.5  \\  % RALM
LM & 10&-0.20&5/3& 0.10& 7 & $1/64$ & 4 & 8.54 & 0.72&\ppmm&0.17 & 1.11
      &   0.00&\ppmm&0.04 &   0.00&\ppmm&0.05 &  +0.25&\ppmm&0.09
      & 4.7 \\[1mm] % RALN
RL &0.6&-0.20& 5/3& 1.00& 6 & $1/8$  & 5 & 63.1 & 36.3&\ppmm&0.14 & 36.4
      &   0.00&\ppmm&0.00 &   0.00&\ppmm&0.00 &  -0.49&\ppmm&0.04
      & 0.18\\[1mm] % RARL
ST & 3 &-0.03&4/3& 0.02& 10 & $1/512$& 8 & 4.60 & 1.01&\ppmm&0.09 & 1.29
      &   0.00&\ppmm&0.05 &  +0.01&\ppmm&0.07 &  +0.51&\ppmm&0.23
      & 4.4  \\  % RAST
SS & 3 &-0.03&4/3& 0.02& 9 & $1/256$& 3 & 9.24 & 1.01&\ppmm&0.12 & 1.46
      &  +0.02&\ppmm&0.15 &  +0.03&\ppmm&0.25 &  +0.36&\ppmm&0.40
      & 5.2  \\  % RASS
\hline
\end{tabular}
\end{flushleft}
\end{table*}

\section{Numerical Procedure and Initial Conditions
     \label{sec:numer}}

Since the numerical procedures and initial conditions are mostly
identical to what has already been described and used in previous
papers (e.g.~Ruffert, 1996; Ruffert \& Anzer, 1994) I will refrain
from repeating every detail, but only give a brief summary.

\subsection{Numerical Procedure\label{sec:numproc}}

The distribution of matter is discretised on multiply nested
equidistant Cartesian grids (e.g.~Berger \& Colella, 1989) 
with zone size $\delta$ and is evolved using the ``Piecewise
Parabolic Method'' (PPM) of Colella \& Woodward (1984).
The equation of state is that of a perfect gas with a specific
heat ratio of $\gamma=5/3$ or $\gamma=4/3$ (see Table~\ref{tab:models}).
The model of the maximally accreting, vacuum sphere in a softened
gravitational potential is summarized in Ruffert \& Arnett (1994) and
Ruffert \& Anzer (1995).

A gravitating, totally absorbing ``sphere'' moves relative to a medium
that far upstream has a distribution of density and
velocity given by
\begin{equation}
 \rho_\infty = \rho_{\rm 0} 
        \left( 1 + \varepsilon_\rho \frac{y}{R_{\rm A}} \right)   \quad, 
  \label{eq:rhograd}
\end{equation}
\begin{equation}
 v_{{\rm x}\infty} 
     = v_{\rm 0} \left( 1 + \frac{1}{2} 
       \tanh \left[ 2 \varepsilon_{\rm v} \frac{y}{R_{\rm A}} \right] \right),
  \,\, v_{{\rm y}\infty} = 0, \,\, v_{{\rm z}\infty} = 0 \,\,,
  \label{eq:vgrad}
\end{equation}
with the redefined accretion radius
\begin{equation}
    R_{\rm A}= \frac{2GM}{v_0^2} \quad.
  \label{eq:accrad}
\end{equation}

In this paper I only investigate models with gradients of the
velocity distribution; the values of $\varepsilon_{\rm v}$ can be
found in Table~\ref{tab:models}.
Thus for all models I set $\varepsilon_\rho\equiv0$.
Additionally, if only a density gradient is introduced without varying
some other thermodynamic variable (e.g.~temperature, entropy, etc.) at
the same rate, pressure will not be in equilibrium 
(cf.~e.g.~Ho et al.~1989), so an additional thermodynamic variable
should be varied, which complicates matters.

The function ``$\tanh$'' is introduced in Eq.~(\ref{eq:vgrad}) to serve
as a cutoff at large distances $y$ for large gradients
$\varepsilon_{\rm v}$: In some models I imposed a gradient of
$\varepsilon_{\rm v}=0.2$ which at distances beyond 5$R_{\rm A}$ would
produce negative velocities if a linear distribution were used.
In the limit of small $y\ll R_{\rm A}$, Eq.~(\ref{eq:vgrad}) transforms
to a shape 
very similar to Eq.~(\ref{eq:rhograd}) with $\rho$ replaced by $v$ 
(as in Ruffert \& Anzer~1995, Eq.~(2)).
The relative velocity $v_0$ is varied in different models:
I perform simulations with Mach numbers 
${\cal M}_\infty\equiv v_0/c_\infty$ of 3.0 and~10.
In the reference frame of the accretor the surrounding
matter flows in +x-direction.
Our units are (1) the sound speed $c_{\infty}$ as velocity unit;
(2) the accretion radius (Eq.~(\ref{eq:accrad}))
as unit of length, and
(3) $\rho_\infty=\rho_0$ as density unit.
Thus the unit of time is $R_{\rm A}/c_{\infty}$.

When using the density and velocity distributions
(Eqs.~(\ref{eq:rhograd}) and~(\ref{eq:vgrad})) to calculate the mass
accretion rate, assuming that all mass within the accretion cylinder is
accreted, one obtains to lowest order in $\varepsilon_{\rm v}$ 
\begin{equation}
   \dot{M} = \pi R^2_{\rm A} \rho_0 v_0  \quad,
  \label{eq:accmass}
\end{equation}
an equation very similar to Eq.~(\ref{eq:accmass1}).
Further assuming that all angular momentum within the deformed
accretion cylinder is accreted too, the specific angular momentum of
the accreted matter follows to be
(Ruffert \& Anzer 1994; Shapiro \& Lightman 1976; again to lowest
order in $\varepsilon_{\rm v}$)
\begin{equation}
  j_{\rm z} = \frac{1}{4}
     \left( 6 \varepsilon_{\rm v} - \varepsilon_\rho \right) v_0 R_{\rm A}  
  \label{eq:specmomang}\quad.
\end{equation}
Note the different signs with which the two $\varepsilon$ enter this
equation.
The density gradient $\varepsilon_\rho$ acts in the ``expected''
direction: if the density is higher on the positive side of the
$y$-axis, then the vortex formed around the accretor is in the
counter-clockwise direction, i.e.~the angular momentum is negative.
Contrary to this, if the velocity is larger on the positive $y$-side,
then the shortened accretion radius on this side reduces the
cross-section for the higher specific angular momentum to such an
extent that the rotational direction of the vortex is reversed: the
angular momentum is positive.

\begin{figure}
\epsfxsize=8.8cm \epsfclipon \epsffile{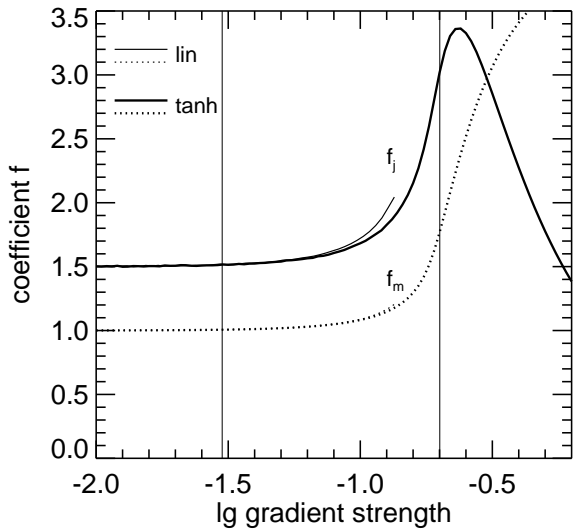}
%\picplace{6.3cm}
\caption[]{The coefficient for the mass accretion rate $f_{\rm m}$
(dotted), defined by Eq.~(\ref{eq:coeffmass}), and for the specific
angular momentum $f_{\rm j}$ (solid), defined by
Eq.~(\ref{eq:coeffspec}), as a function of $\varepsilon_{\rm v}$.
The two thin curves that end at around -0.8 show the values of $f$ for
a simple linear relation between $v_{{\rm x}\infty}$ 
and $\varepsilon_{\rm v}$, while the bold curves apply to a relation
including the ``tanh''-term as given in Eq.~(\ref{eq:vgrad}).
The two vertical straight lines indicate the gradients that were used
in the numerical models (cf.~Table~\ref{tab:models}), 
$\varepsilon_{\rm v}=0.03$ and $\varepsilon_{\rm v}=0.2$. 
}
\label{fig:shali}
\end{figure}

The values obtained from the numerical simulations
of the specific angular momentum
should be compared to the values that follow from this
Eq.~(\ref{eq:specmomang}) to conclude which of the above mentioned
views --- low or high specific angular momentum of the accreted
material --- is most probably correct.
Ruffert \& Anzer (1995) find that the numerical value is 0.72 
times\footnote{this value is the rms, while in Table~\ref{tab:models}
we list the averages}
the analytical estimate, which would indicate a large value for the
accreted specific angular momentum.
Although the simulations yield the values of all three components of
the angular momentum, the interesting component is the one pointing in
$z$-direction. 
Thus I will implicitely assume $j_{\rm z}$ when discussing properties
like fluctuations, magnitudes, etc.~that the simulations produce.
From the symmetry of the boundary conditions the average of the $x$
and $y$ components of the angular momentum should be zero, although
their fluctuations can be quite large.
Numerically I obtain the mass and angular momentum accretion rates as
a function of time, $\dot{M}(t)$ and $\dot{J}(t)$.
From these functions I calculate the instantaneous specific angular
momentum $j(t)=\dot{J}(t)/\dot{M}(t)$ as function of time, which I
will plot and from which I calculate the temporal mean $l$ listed in
Table~\ref{tab:models}.

One can numerically approximate the integrals (U.~Anzer, personal
communication) of the mass flux and angular momentum over the
deformed cross section of the accretion cylinder, to obtain the
coefficients $f$ in the relations Eq.~(\ref{eq:accmass}) 
and~Eq.~(\ref{eq:specmomang}):
\begin{equation}
   \dot{M} = f_{\rm m}(\varepsilon_{\rm v})
             \pi R^2_{\rm A} \rho_0 v_0  \quad,
  \label{eq:coeffmass}
\end{equation}
\begin{equation}
  \jmath_{\rm z} = f_{\rm j} (\varepsilon_{\rm v})
      \varepsilon_{\rm v}  v_0 R_{\rm A}  
  \label{eq:coeffspec}\quad.
\end{equation}
Here, I will only consider the effect of a velocity gradient.
The unitless functions $f$ are a function of $\varepsilon_{\rm v}$ and 
the functional relation of $v_{{\rm x}\infty}[\varepsilon_{\rm v}]$,
i.e.~whether $v_{{\rm x}\infty}$ depends purely linearly on
$\varepsilon_{\rm v}$ or as in Eq.~(\ref{eq:vgrad}) via the ``tanh''-term.
Figure~\ref{fig:shali} shows the values of the functions $f$ for the
mass and specific angular momentum and for both the linear and
``tanh'' case.
In the linear case there is no solution for 
$\varepsilon_{\rm v}\ga0.15$ (U.~Anzer, personal communication)
so the curves end at that point. 
Since $f_{\rm m}\approx1$ and is practically constant for 
$\varepsilon_{\rm v}\la0.1$, the Eq.~(\ref{eq:accmass}) is a good
approximation in this range.
If the prescription is correct that everything in the accretion
cylinder is accreted, we expect to see an increase of the mass
accretion rate by a factor of roughly 1.8 in the models with a fairly
large gradient of $\varepsilon_{\rm v}=0.2$ 
(cf.~Table~\ref{tab:models}).
The same trends apply to the specific angular momentum: 
its coefficient $f_{\rm j}$ remains relatively constant 
$f_{\rm j}\approx 1.5$ in the range $\varepsilon_{\rm v}\la0.1$,
and becomes a factor of 2 larger for $\varepsilon_{\rm v}\approx0.2$.
In the case including ``tanh'', the coefficent $f_{\rm j}$ decreases
again for $\varepsilon_{\rm v}\ga0.25$ because the gradient is so
steep that the ``tanh''-cutoff acts at very short distances.
So the short lever arm that enters into the angular momentum wins.

\begin{figure*}
 \begin{tabular}{cc}
  \epsfxsize=8.8cm \epsfclipon \epsffile{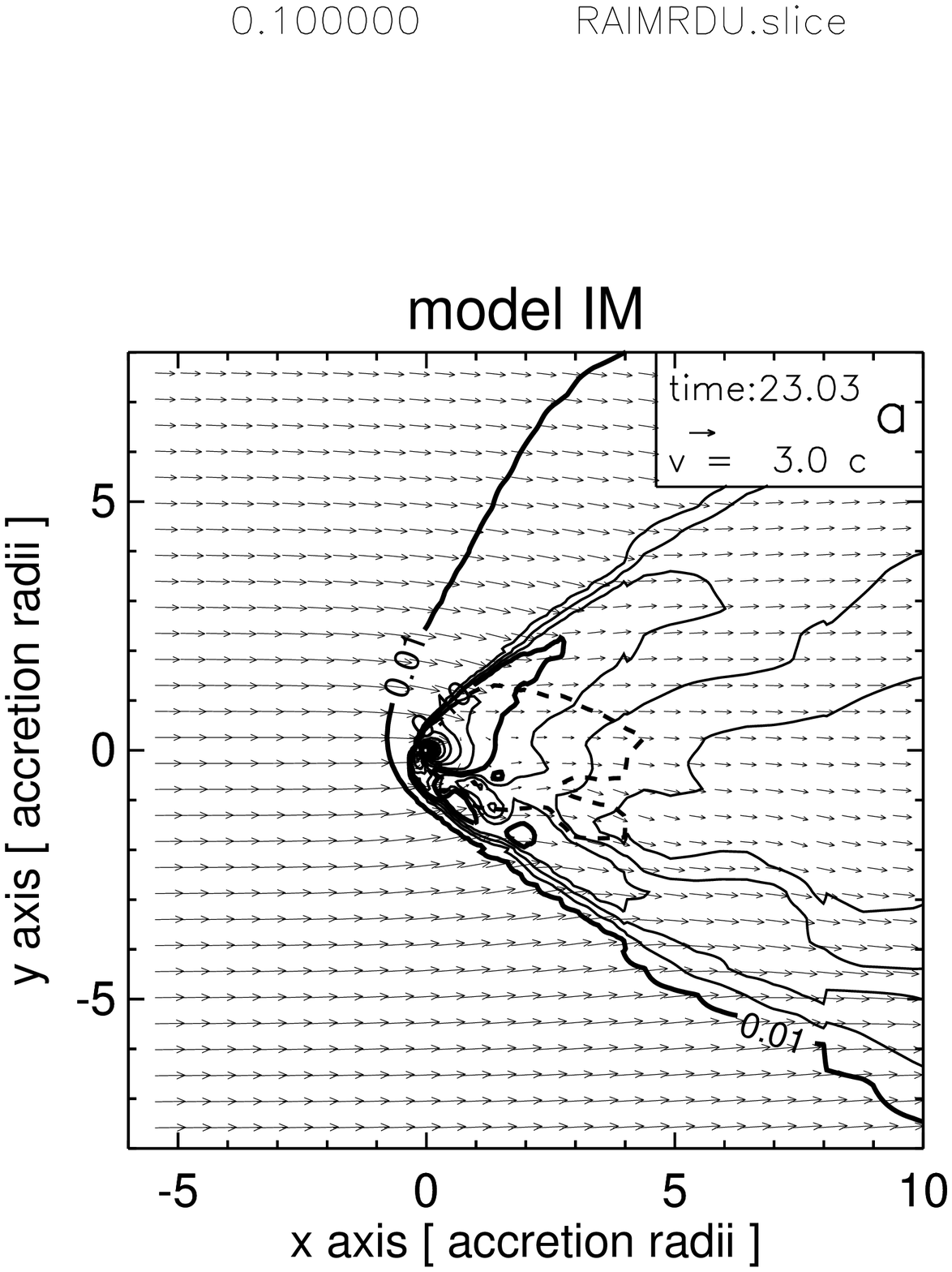} &
  \epsfxsize=8.8cm \epsfclipon \epsffile{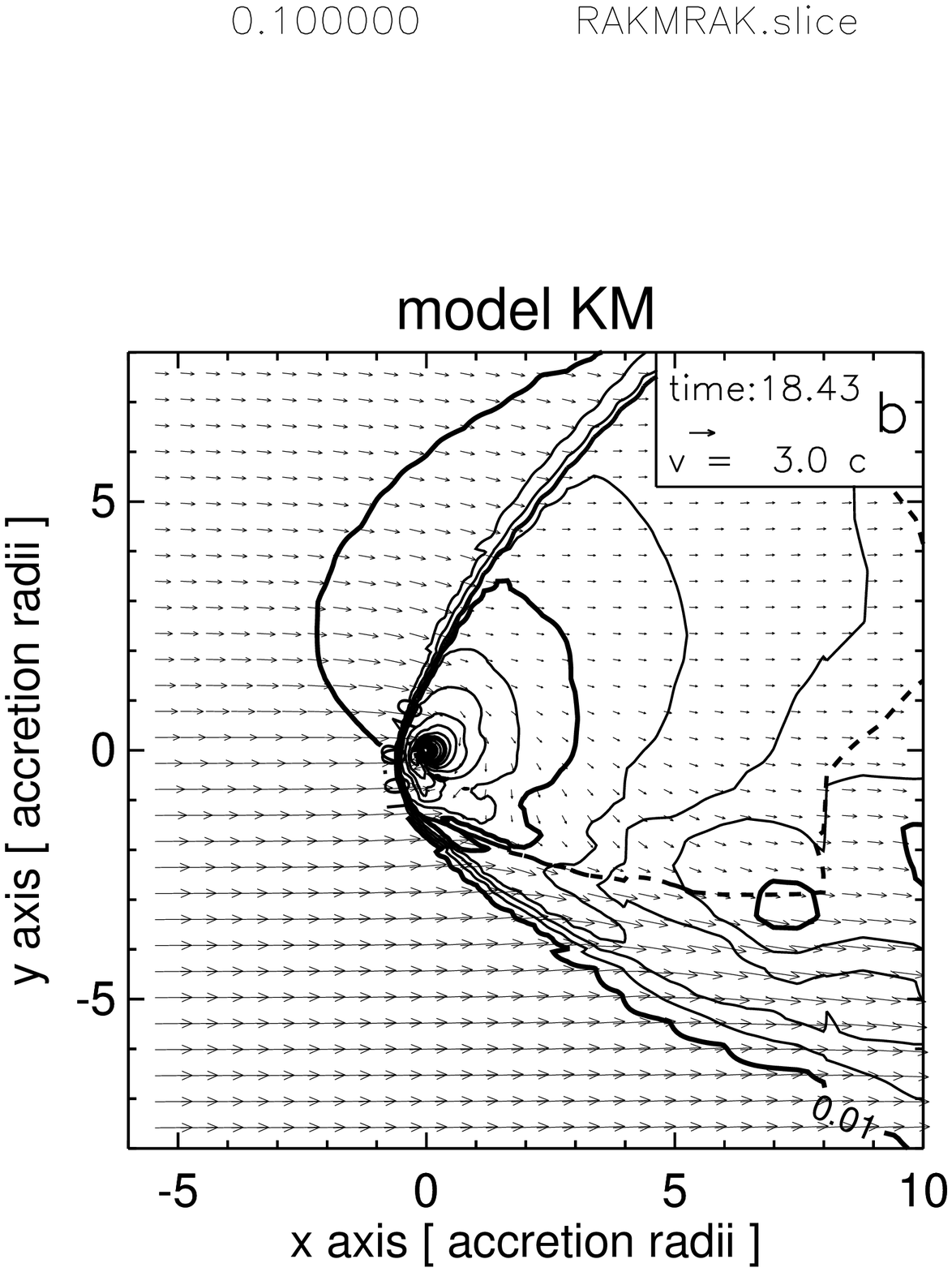} \\
  \epsfxsize=8.8cm \epsfclipon \epsffile{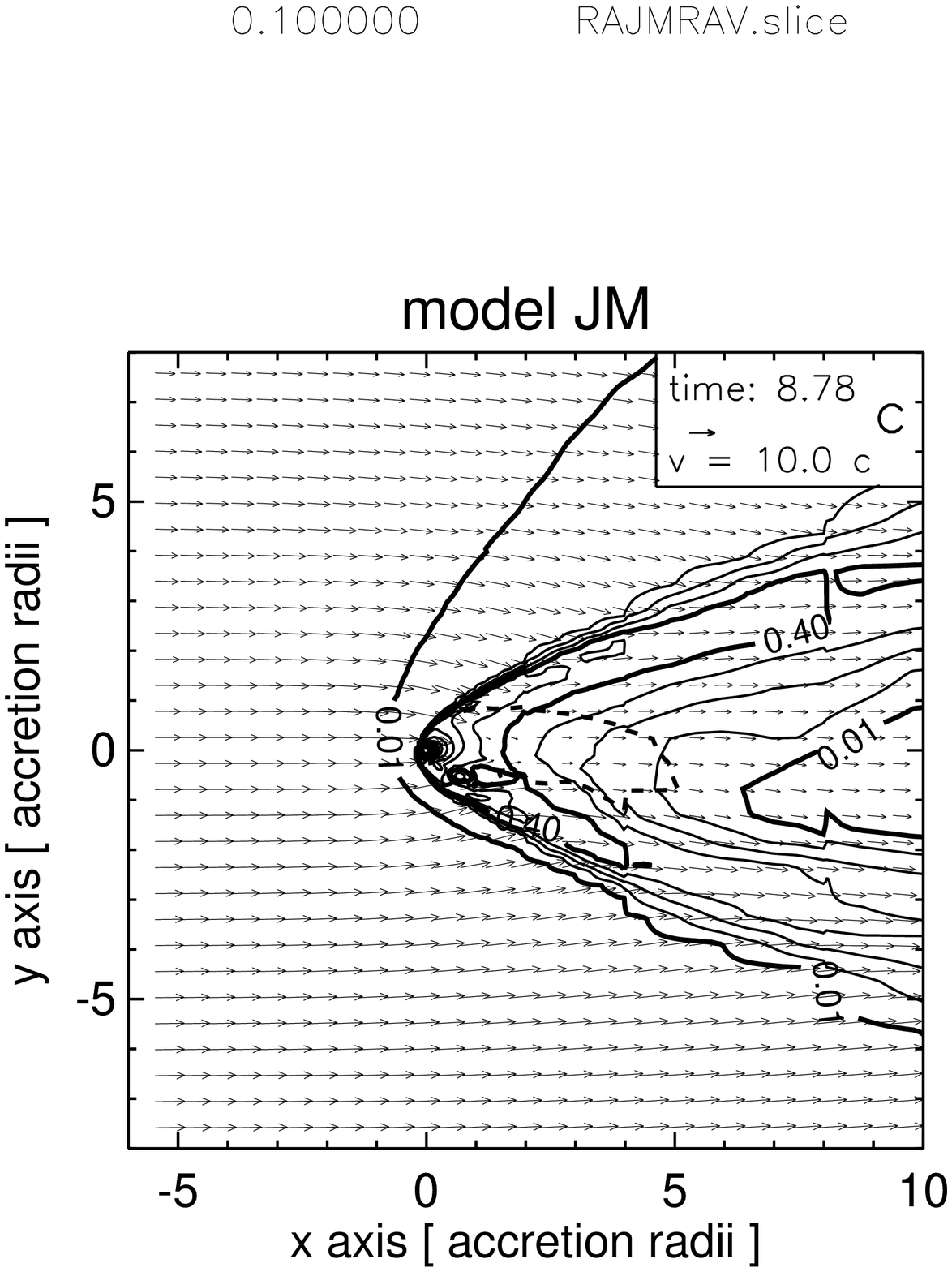} &
  \epsfxsize=8.8cm \epsfclipon \epsffile{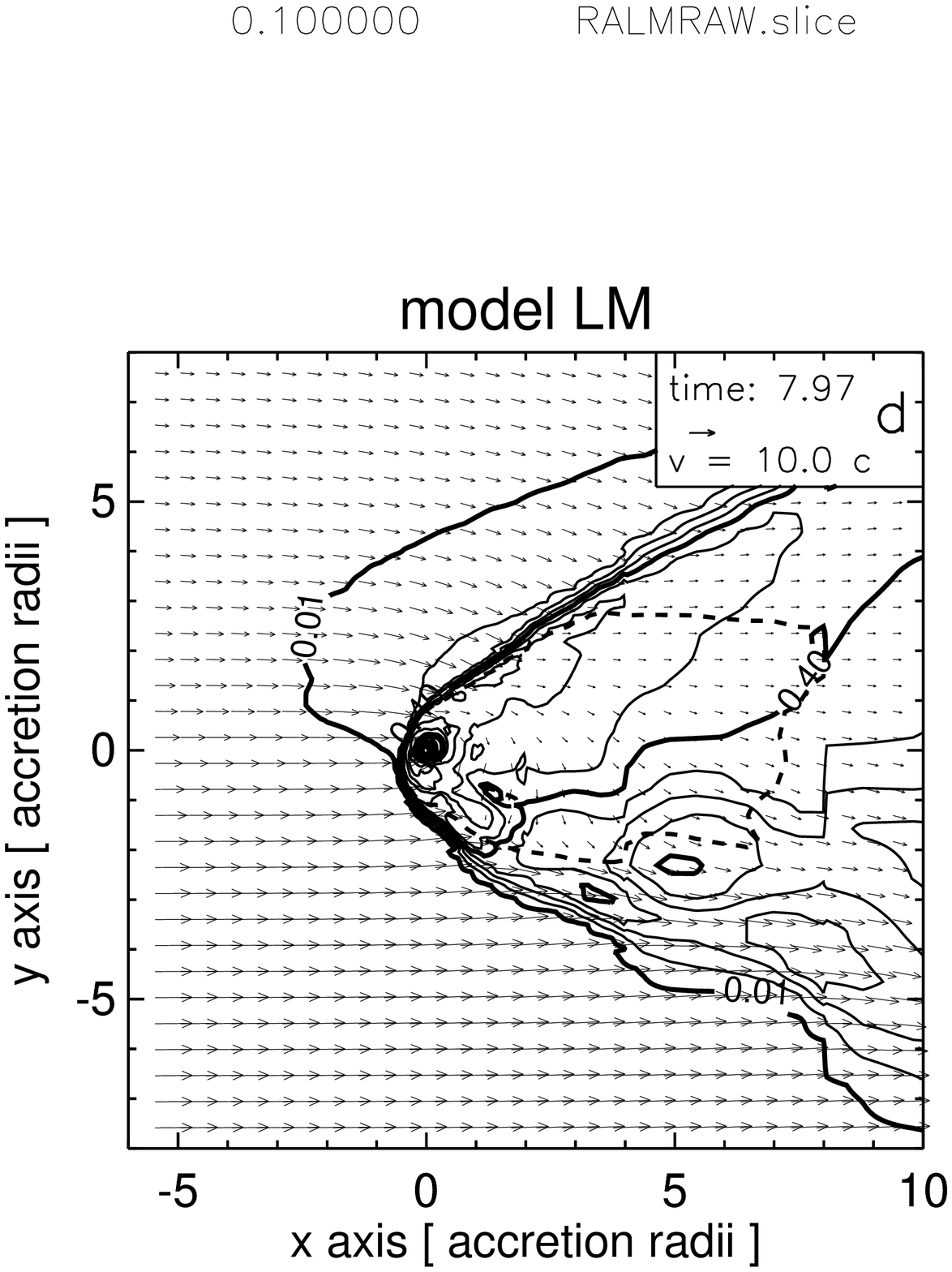} \\
  \epsfxsize=8.8cm \epsfclipon \epsffile{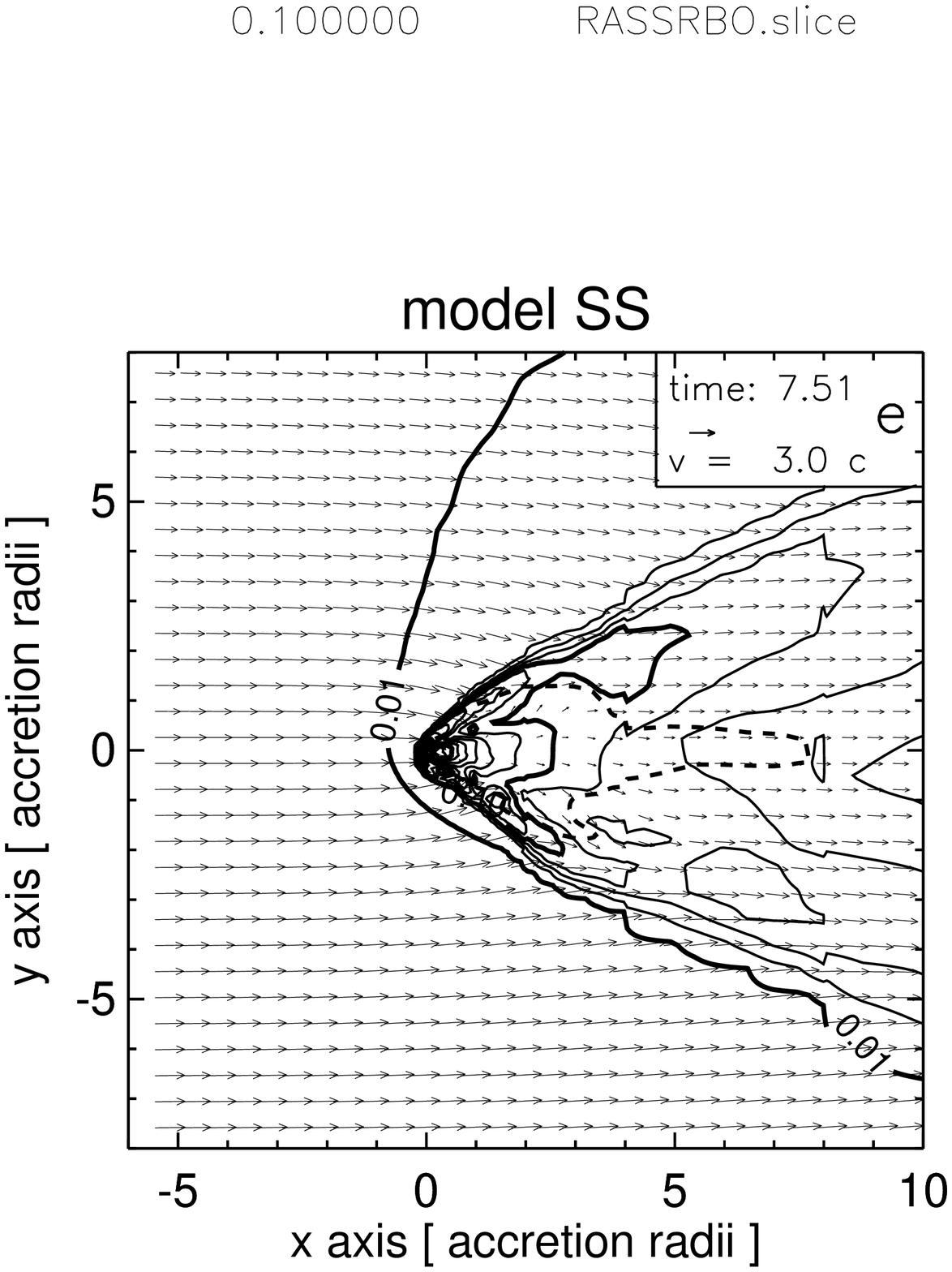} &
\raisebox{8cm}{\parbox[t]{8.8cm}{
\caption[]{\label{fig:big}
Contour plots showing snapshots of the density together
with the flow pattern at large distances from the accretor.
The contour lines are spaced logarithmically in intervals of 0.1~dex.
The bold contour levels are sometimes labeled with their respective
values (0.01 and 0.4).
%The dark shades of gray indicate a high density.
The dashed contour delimits supersonic from subsonic regions.
The time of the snapshot together with the velocity scale is given in
the legend in the upper right hand corner of each panel.
}}}
 \end{tabular}
%\picplace{20.cm}
\end{figure*}

\subsection{Models\label{sec:models}}

The combination of parameters that I varied, together with some results are
summarised in Table~\ref{tab:models}.
The first letter in the model designation indicates the Mach number
and the strength of the gradient: 
I, J, and S have $\varepsilon_{\rm v}=0.03$, while 
K, L, and R have $\varepsilon_{\rm v}=0.2$.
The second letter specifies the size of the accretor:
M (medium) and S (small) stand for
accretor radii of 0.1~$R_{\rm A}$, and~0.02~$R_{\rm A}$, respectively.
I basically simulated models with all possible combinations of
two relative wind flow speeds (Mach numbers of~3 and~10), two gradient
strengths (3\% and~20\%) and two different accretor sizes (0.02 and
0.1~accretion radii), all with an adiabtic index of 5/3.
The exeptional models are~ST and~SS --- in which I used an adiabatic
index of 4/3 --- and model~RL which has a very large accretor radius
and a very slow relative flow velocity.
The grids are nested to a depth $g$ such that the radius of the
accretor $R_\star$ spans several zones on the finest grid and
the softening parameter $\epsilon$ is then chosen to be a few zones
less than the number of zones that the accretor spans.
Model~IT and~ST are physically identical to models~IS and~SS,
respectively.
However, models~IT and~ST are numerically better resolved, because they
are nested one grid level finer.
Model~IM is identical to the model presented in Ruffert \& Anzer (1995),
i.e.~the velocity gradient is chosen as specified in Eq.~(2) of 
Ruffert \& Anzer (1995) without the ``tanh''-term of Eq.~(\ref{eq:vgrad}).
This term is included in model~IM*, however, since the results were
nearly indistinguishable between models~IM and~IM*, model~IM* was
evolved for only roughly one third of the time of model~IM.

As far as computer resources permitted, I aimed at evolving the models
for at least as long as it takes the flow to move
from the boundary to the position of the accretor which is at the center
(crossing time scale).
This time is given by $L/2{\cal M}_\infty$ and ranges from about 1
to about 10 time units.
The actual time $t_{\rm f}$ that the model is run can be found in
Table~\ref{tab:models}.

The velocity distribution following the ``tanh''-prescription of
Eq.~(\ref{eq:vgrad}) has an inflection point and thus is
Kelvin-Helmholz unstable with an amplification time constant of
roughly $\tau\approx(5R_{\rm A})/(\varepsilon_{\rm v}{\cal M}c_\infty)$
(Drazin~1981).
During the time it takes matter to move from the boundary to the
accretor, $t\approx(16R_{\rm A})/({\cal M}c_\infty)$, random
perturbations can grow by about $t/\tau\approx3\varepsilon_{\rm v}$
which is smaller than unity even for the large gradients 
($\varepsilon_{\rm v}=0.2$) used
in the simulations and listed in Table~\ref{tab:models}.

The calculations are performed on a Cray-YMP~4/64 and a Cray J90 8/512.
They need about 12--16 MWords of main memory
and take approximately 40 CPU-hours per simulated time unit
(for the $\delta=1/64$ models and Mach 10;
the $\delta=1/256$ models take four times as long, etc.;
$\delta$ is the size of a zone on the finest grid, see Table~\ref{tab:models}).

\begin{figure*}
 \begin{tabular}{cc}
\raisebox{8cm}{\parbox[t]{8.8cm}{
\caption[]{\label{fig:cont3}
Contour plots showing snapshots of the density together
with the flow pattern in a plane containing the center of the accretor
for all models with a velocity gradient of~3\%.
The contour lines are spaced logarithmically in intervals of 0.1~dex.
The bold contour levels are sometimes labeled with their respective
values (0.01 and 1.0).
The dashed contour delimits supersonic from subsonic regions.
The time of the snapshot together with the velocity scale is given in
the legend in the upper right hand corner of each panel.
}}}
 \end{tabular}
%\picplace{20.cm}
\end{figure*}

\begin{figure*}
 \begin{tabular}{cc}
  \epsfxsize=8.8cm \epsfclipon \epsffile{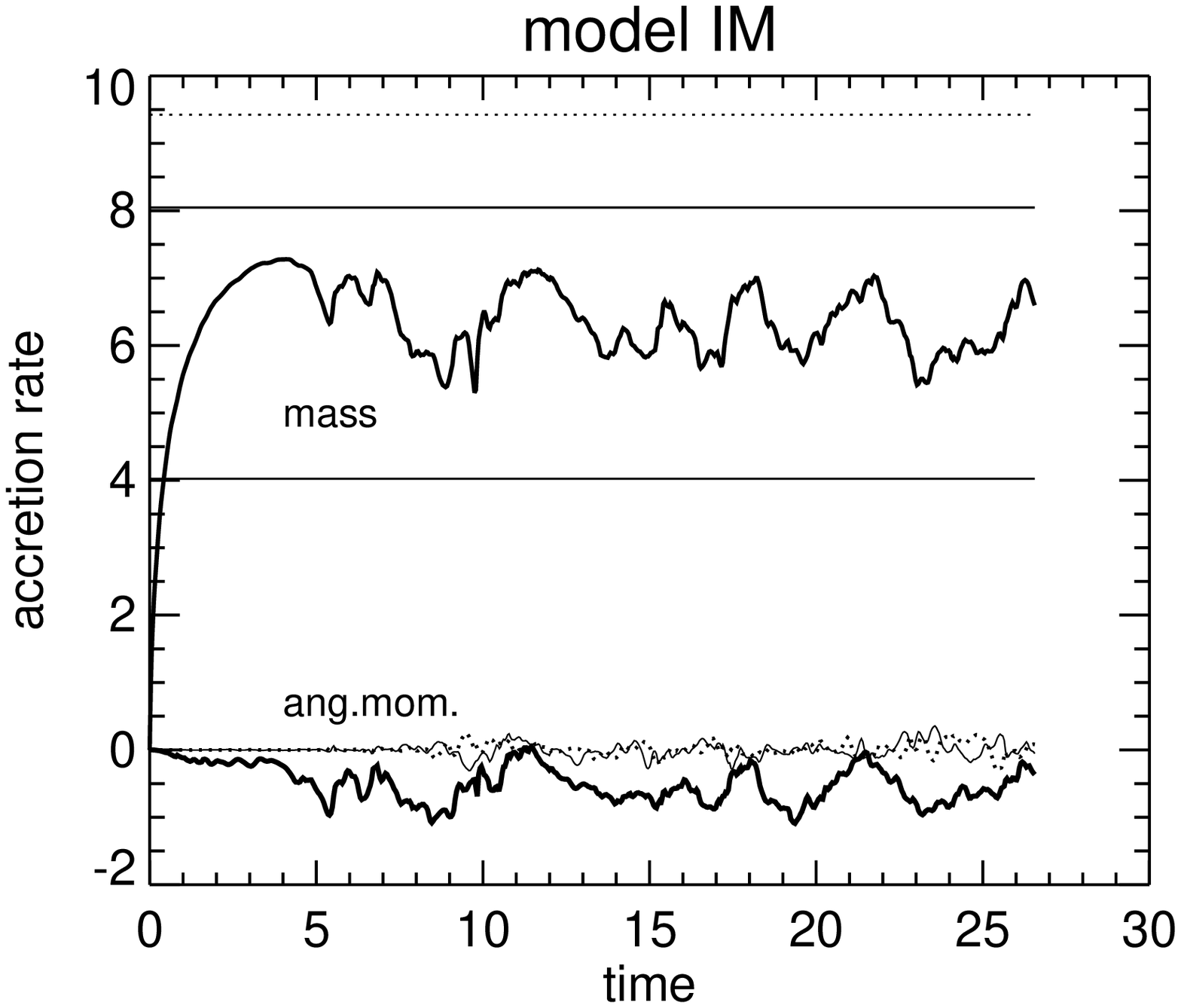} &
  \epsfxsize=8.8cm \epsfclipon \epsffile{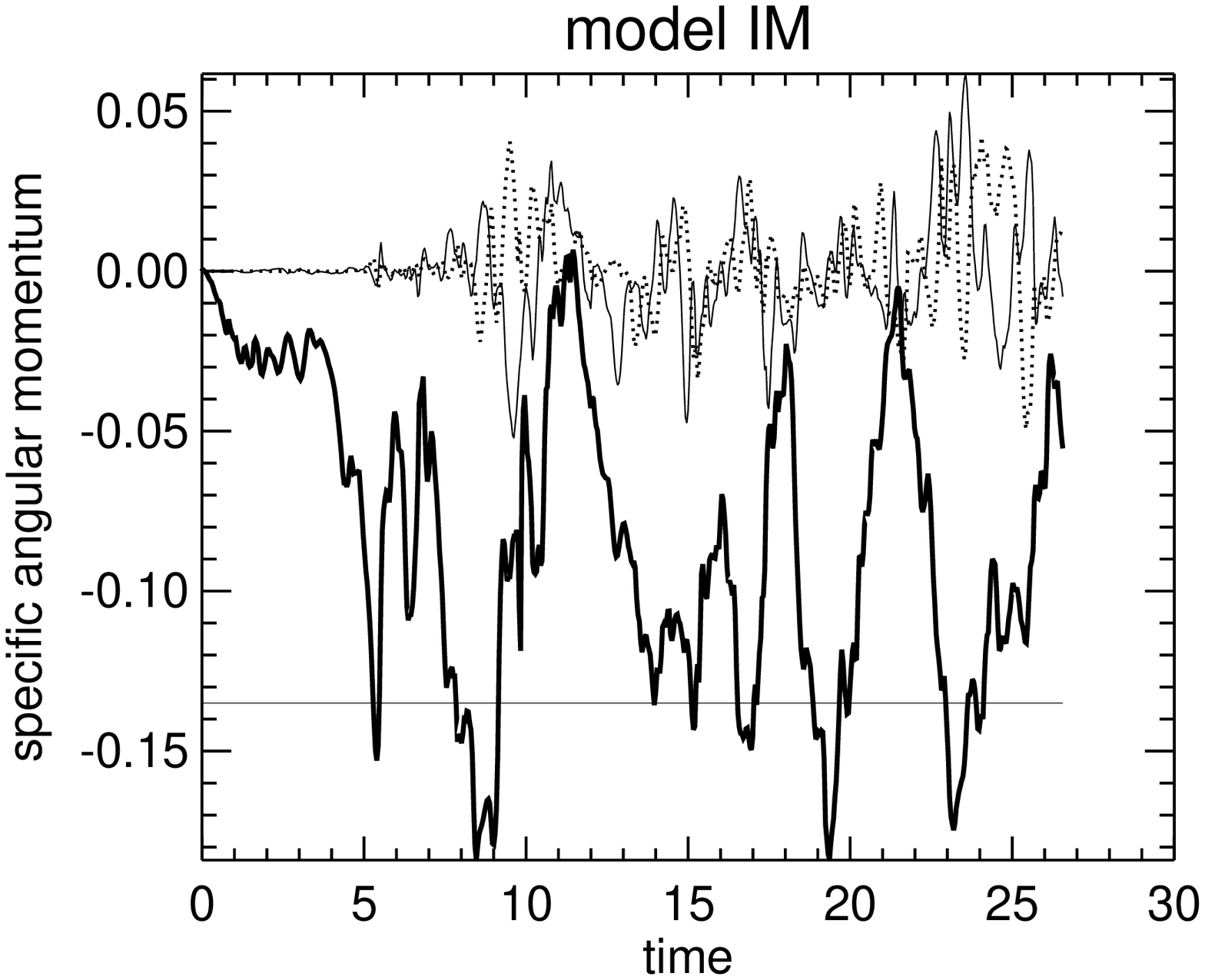} \\
  \epsfxsize=8.8cm \epsfclipon \epsffile{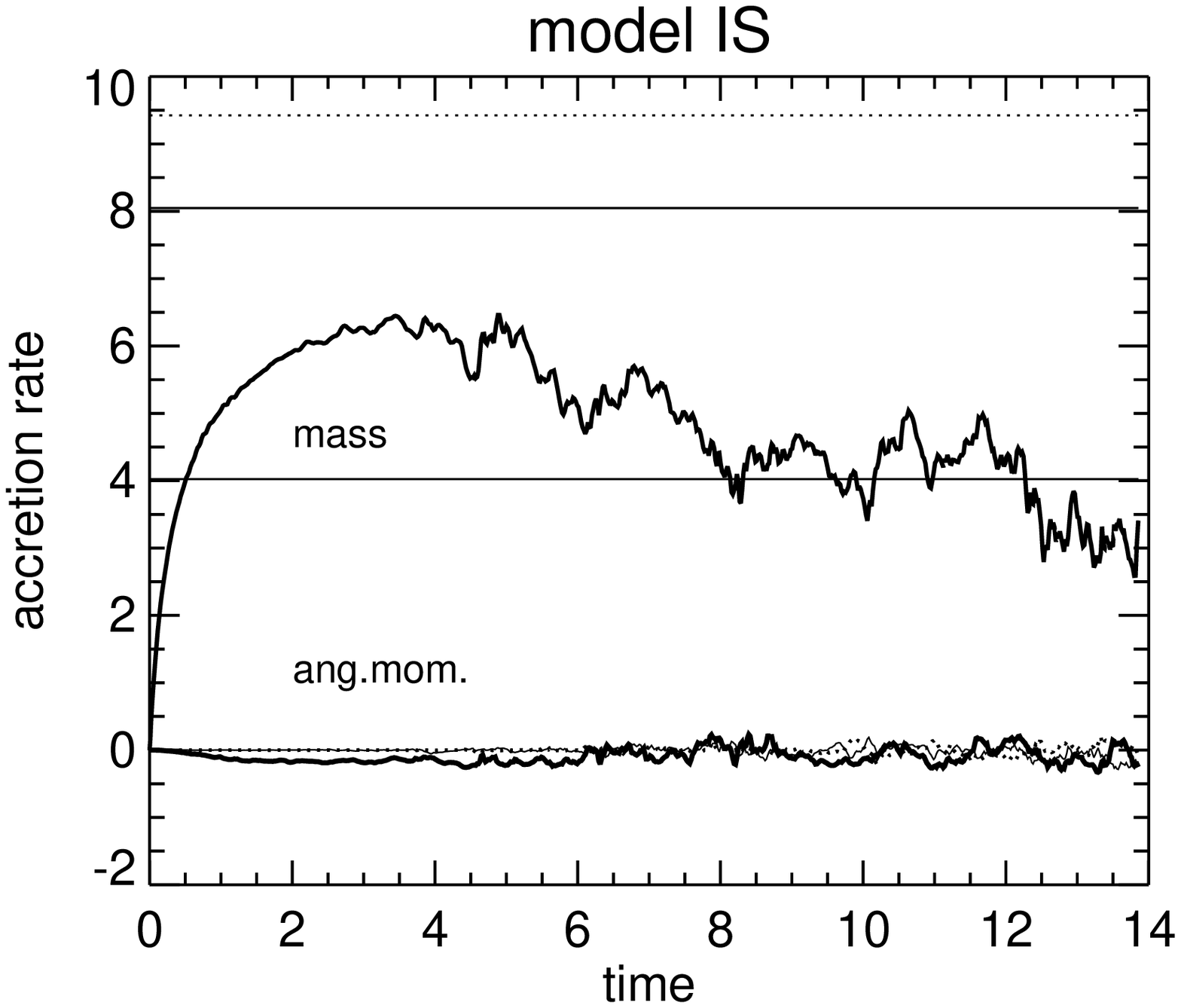} &
  \epsfxsize=8.8cm \epsfclipon \epsffile{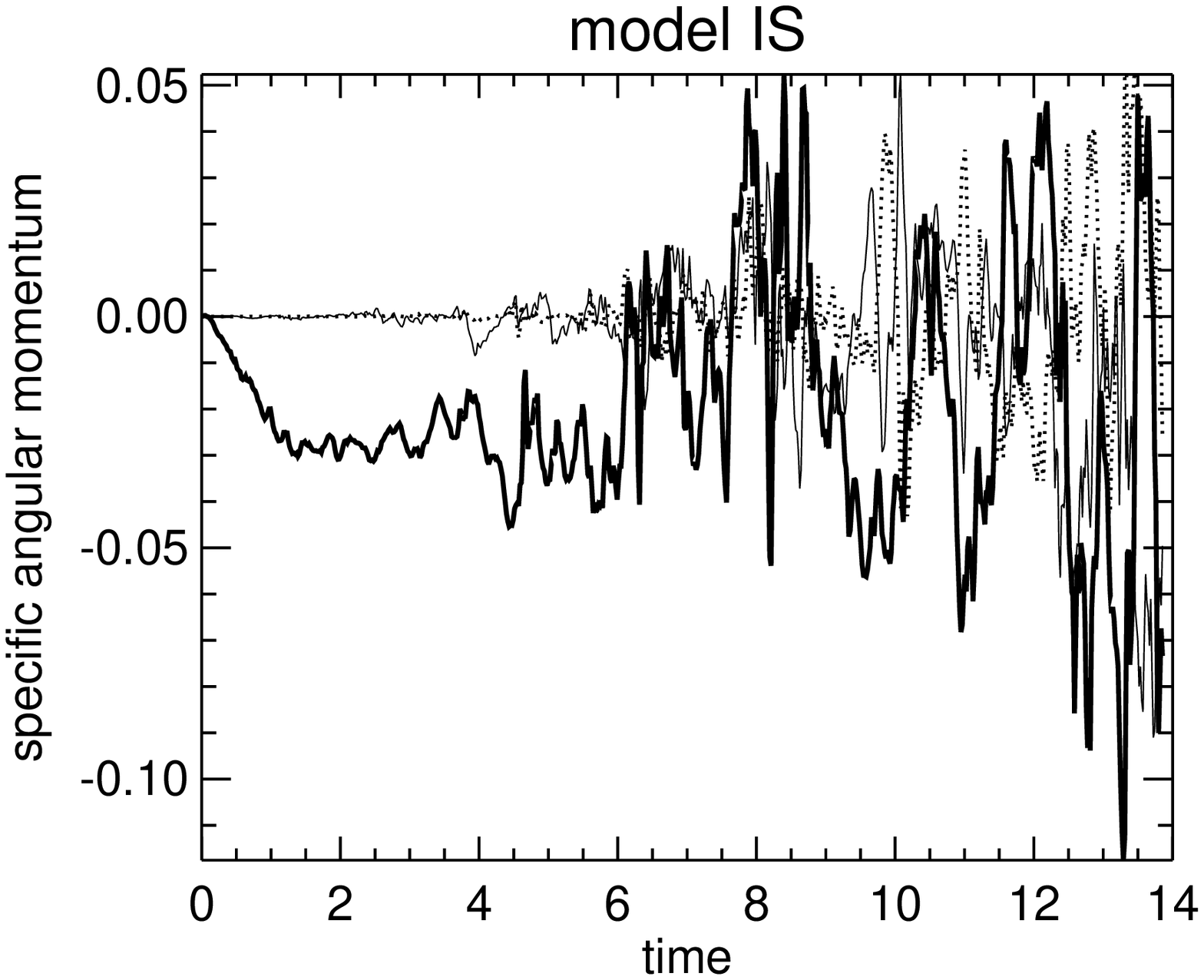} \\
  \epsfxsize=8.8cm \epsfclipon \epsffile{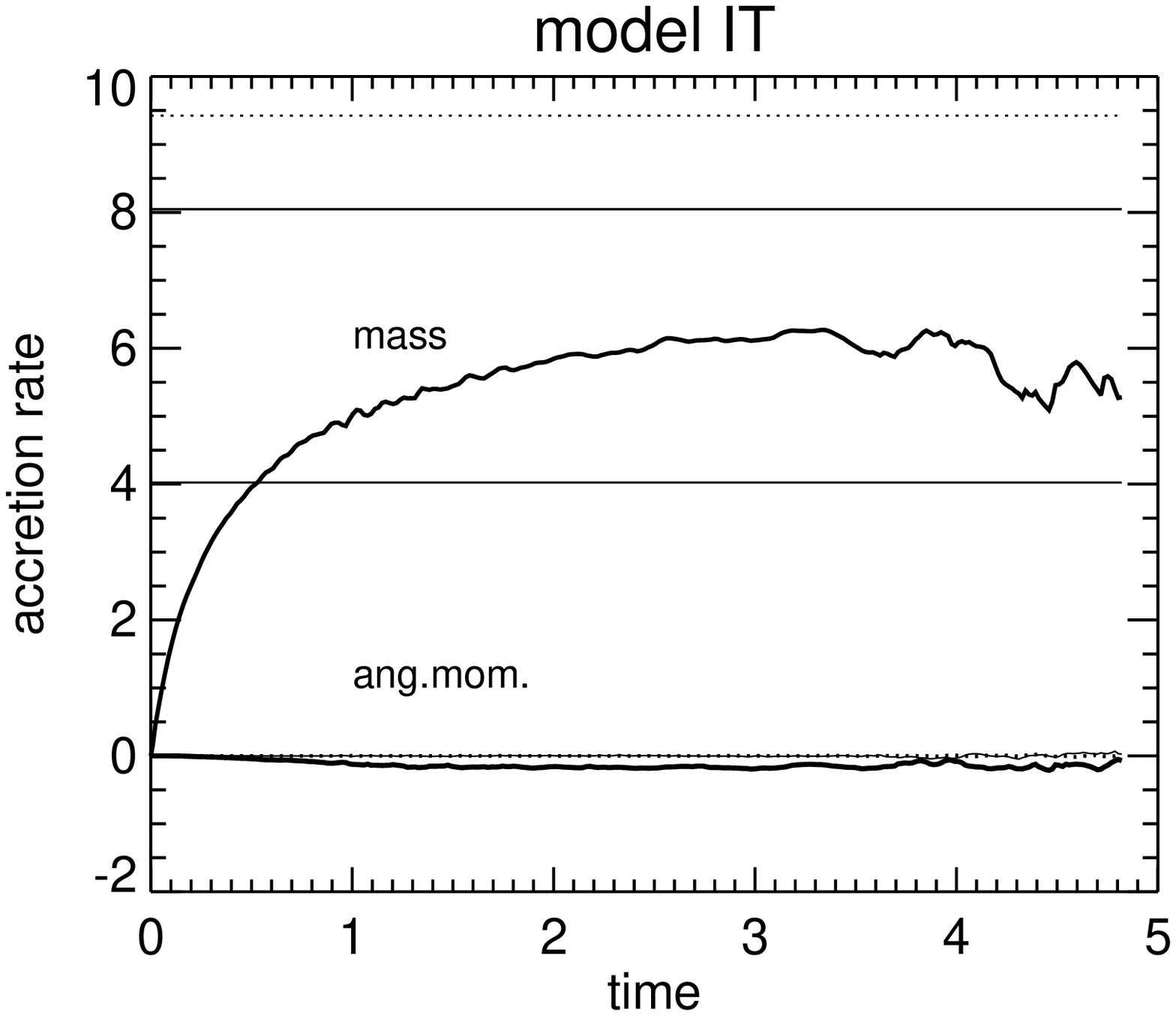} &
  \epsfxsize=8.8cm \epsfclipon \epsffile{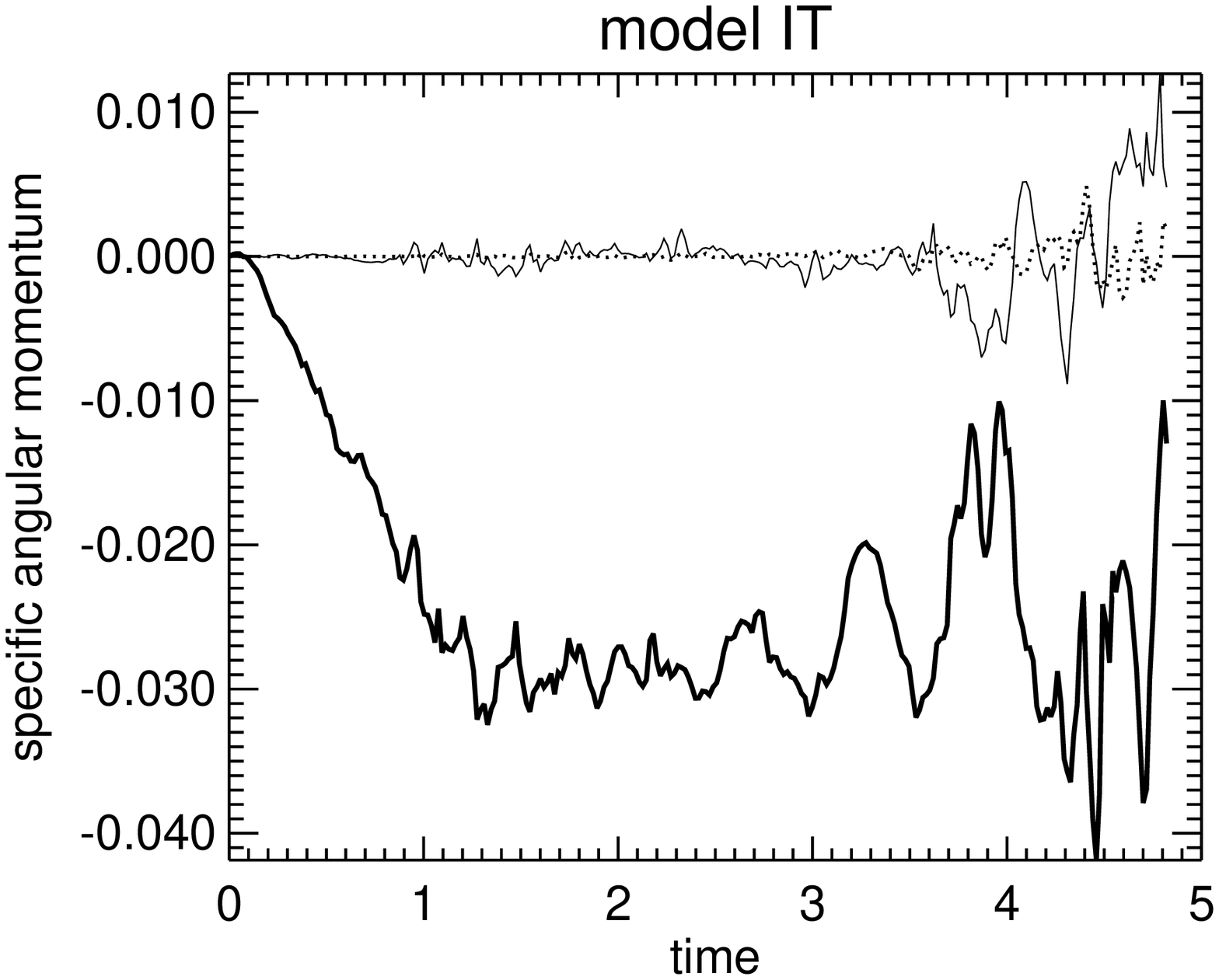}
 \end{tabular}
%\picplace{20.cm}
\caption[]{
The accretion rates of several quantities are plotted as a
function of time for the moderately supersonic (${\cal M}_\infty$=3)
models~IM, IS and~IT with a velocity gradient of~3\%.
The left panels contain the mass and angular momentum accretion rates,
the right panels the specific angular momentum of the matter
that is accreted.
In the left panels, the straight horizontal lines show the analytical
mass accretion rates: dotted is the Hoyle-Lyttleton rate
(Eq.~1 in Ruffert~1994a), 
solid is the Bondi-Hoyle approximation formula (Eq.~3 in Ruffert~1994a) 
and half that value.
The upper solid bold curve represents the
numerically calculated mass accretion rate.
The lower three curves of the left panels trace the x~(dotted),
y~(thin solid) and z~(bold solid) component of the angular momentum
accretion rate.
The same components apply to the right panels.
The horizontal line in the right Panel of model~IM shows the
specific angular momentum value as given by Eq.~(\ref{eq:specmomang}).
It is outside the range of the plot for models~IS and~IT.
}
\label{fig:valueI}
\end{figure*}

\begin{figure*}
 \begin{tabular}{cc}
  \epsfxsize=8.8cm \epsfclipon \epsffile{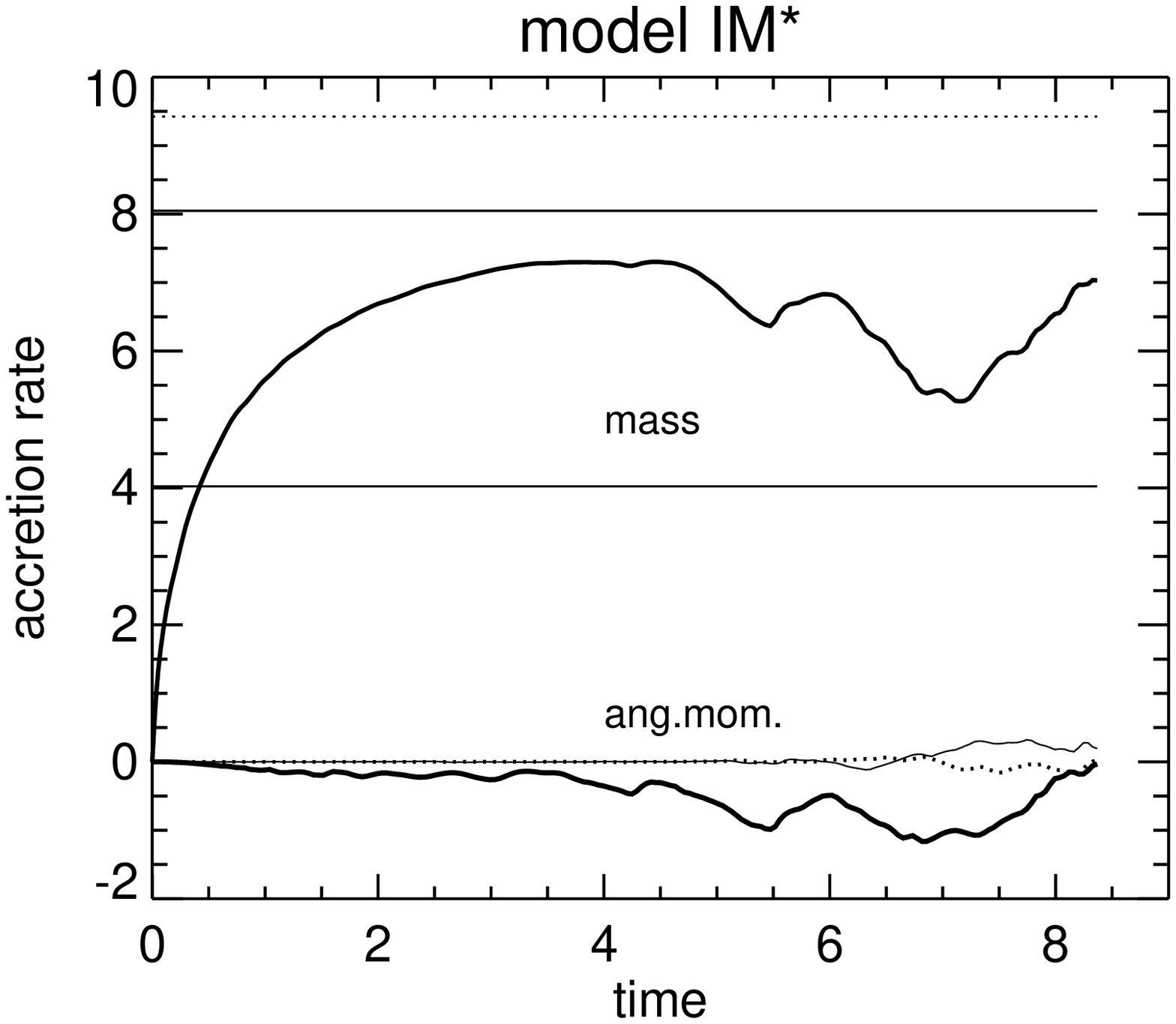} &
  \epsfxsize=8.8cm \epsfclipon \epsffile{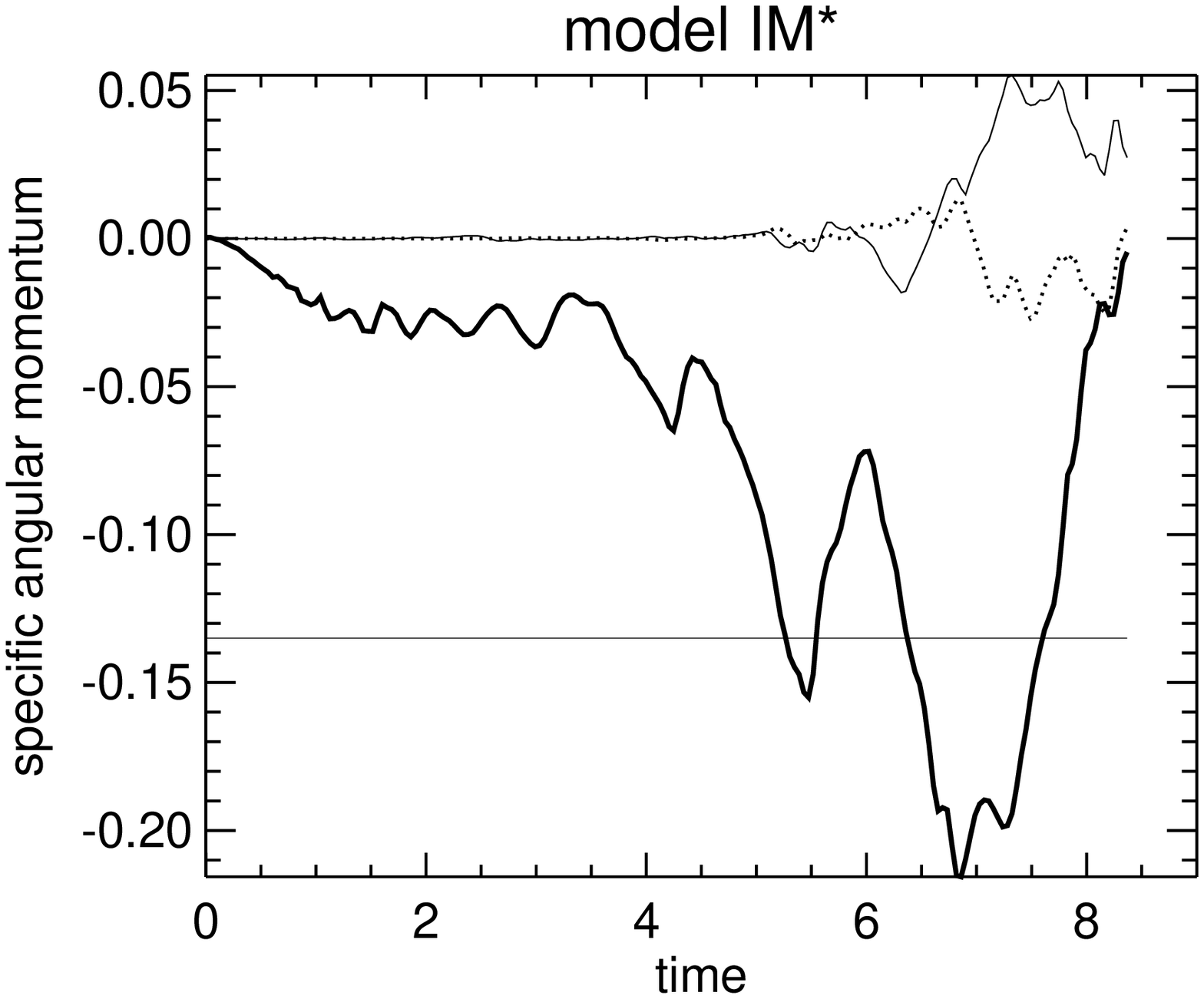}
 \end{tabular}
%\picplace{20.cm}
\caption[]{
The accretion rates of several quantities are plotted as a
function of time for model~IM*.
The left panel contains the mass and angular momentum accretion rates,
the right panel the specific angular momentum of the matter
that is accreted.
In the left panel, the straight horizontal lines show the analytical
mass accretion rates: dotted is the Hoyle-Lyttleton rate
(Eq.~1 in Ruffert~1994a), 
solid is the Bondi-Hoyle approximation formula (Eq.~3 in Ruffert~1994a) 
and half that value.
The upper solid bold curve represents the
numerically calculated mass accretion rate.
The lower three curves of the left panels trace the x~(dotted),
y~(thin solid) and z~(bold solid) component of the angular momentum
accretion rate.
The same components apply to the right panel.
The horizontal line in the right Panel shows the 
specific angular momentum value as given by Eq.~(\ref{eq:specmomang}).
}
\label{fig:valueIs}
\end{figure*}

%\begin{figure}
%  \epsfxsize=8.8cm \epsfclipon \epsffile{shade.ps}
%%\picplace{20.cm}
%\caption[]{Gray scale plots showing the density distribution of
%model~GS in a plane containing the center of the accretor.
%Darker shades represent higher densities.
%The time of the snapshot is the same as in Fig.~\ref{fig:contGH}e.
%}
%\label{fig:GSshade}
%\end{figure}

\begin{figure*}
 \begin{tabular}{cc}
  \epsfxsize=8.8cm \epsfclipon \epsffile{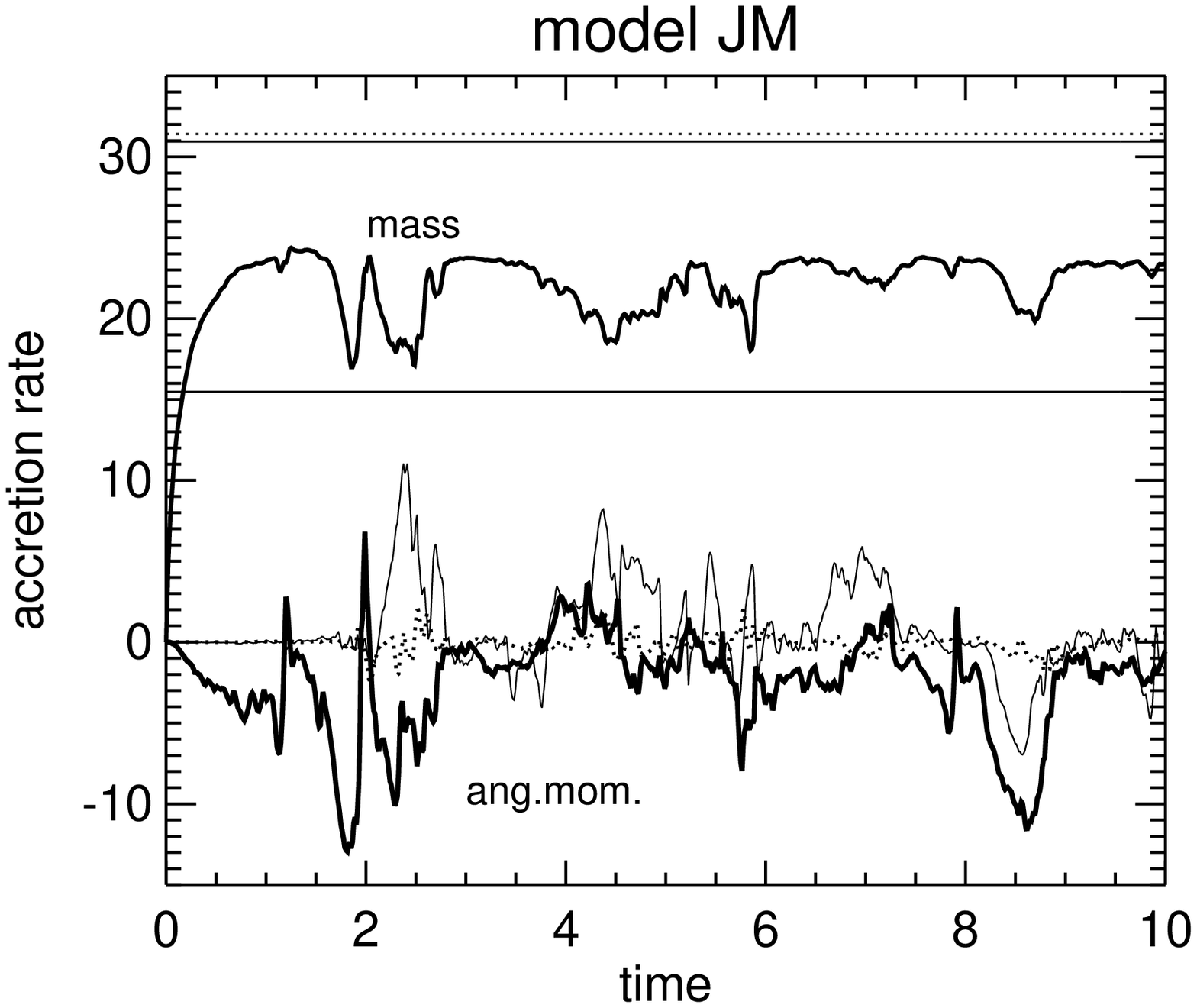} &
  \epsfxsize=8.8cm \epsfclipon \epsffile{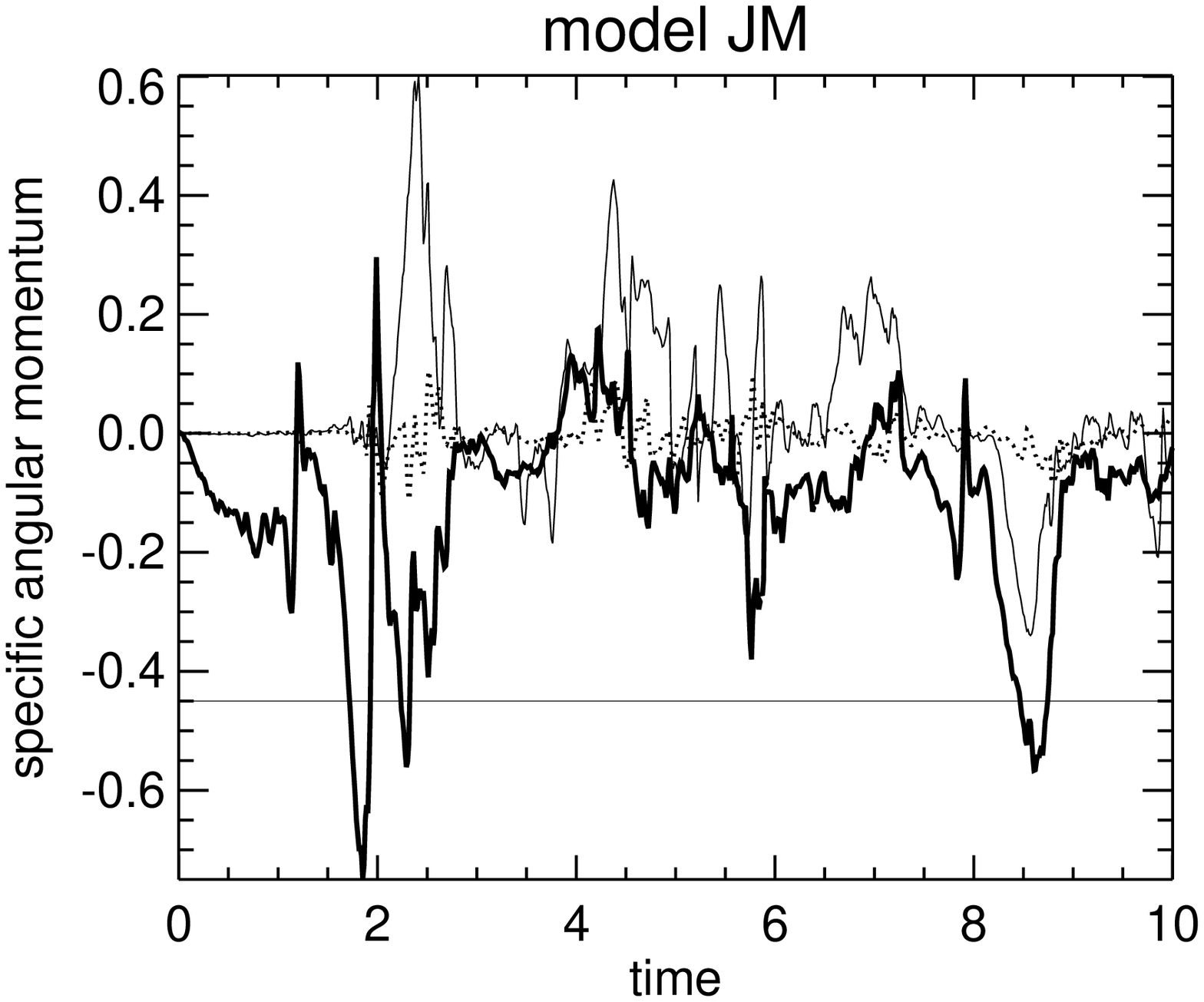} \\
  \epsfxsize=8.8cm \epsfclipon \epsffile{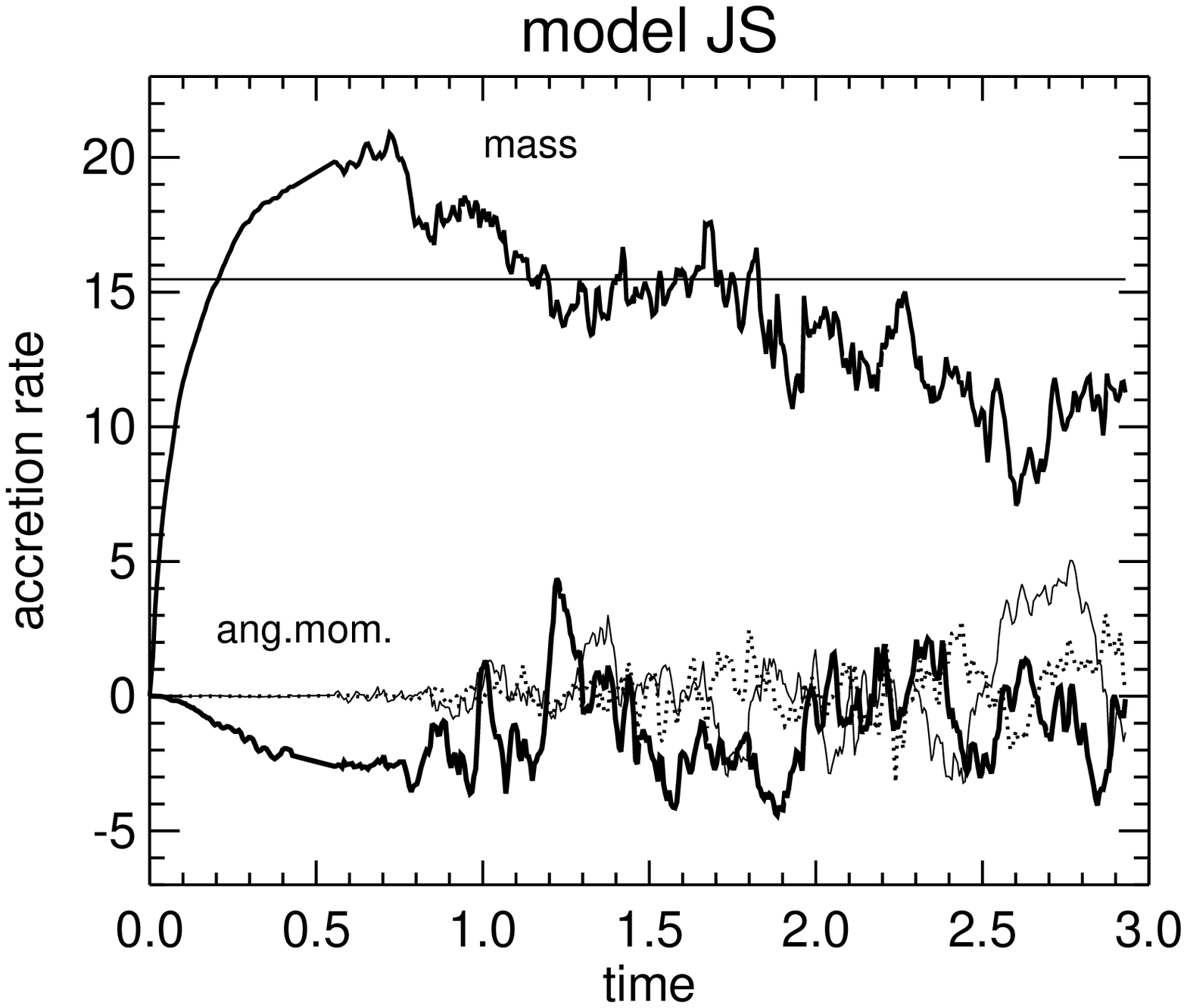} &
  \epsfxsize=8.8cm \epsfclipon \epsffile{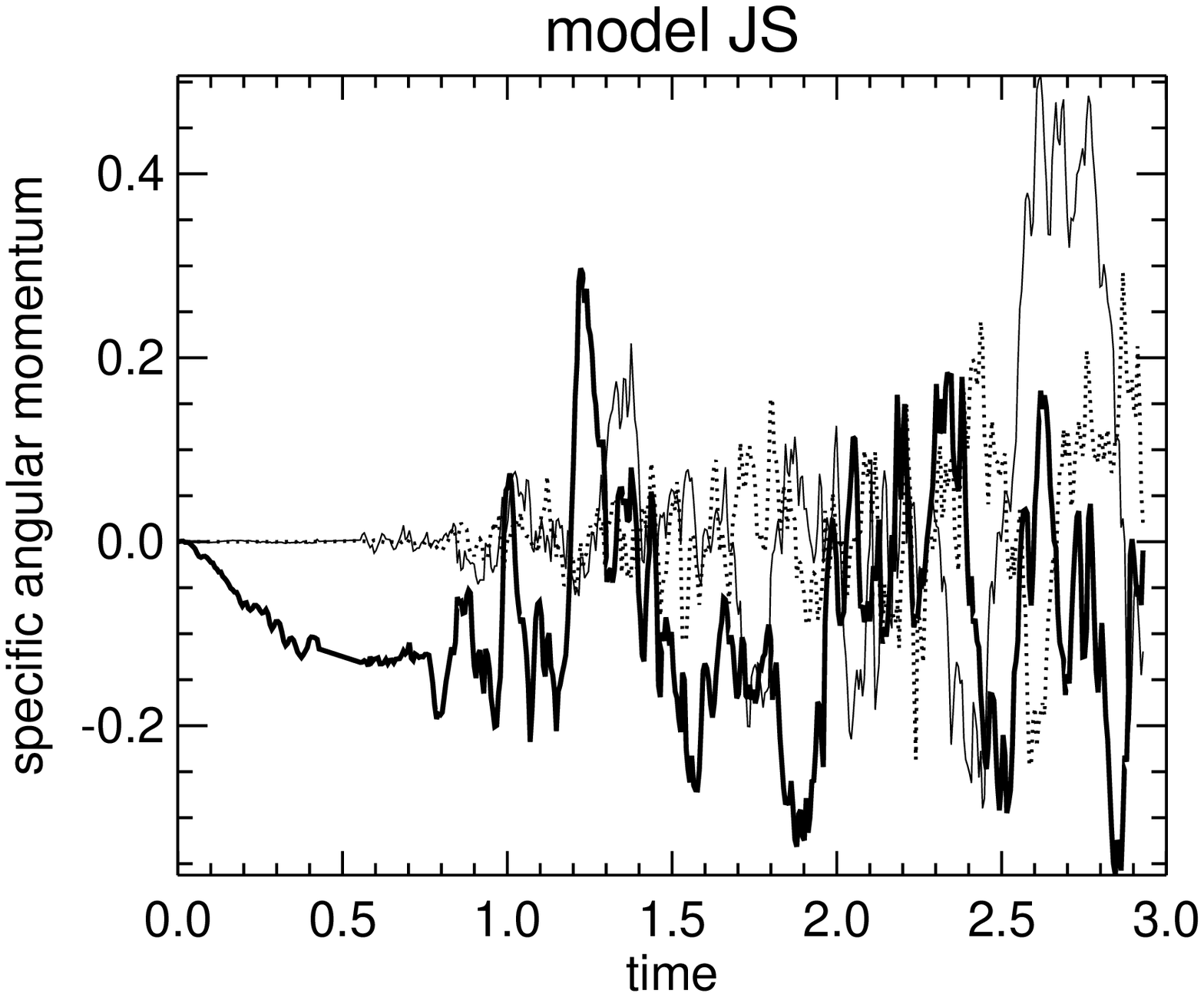}
 \end{tabular}
%\picplace{20.cm}
\caption[]{
The accretion rates of several quantities are plotted as a
function of time for the highly supersonic (${\cal M}_\infty$=10)
models~JM and~JS with a velocity gradient of~3\%.
The left panels contain the mass and angular momentum accretion rates,
the right panels the specific angular momentum of the matter
that is accreted.
In the left panels, the straight horizontal lines show the analytical
mass accretion rates: dotted is the Hoyle-Lyttleton rate
(Eq.~1 in Ruffert~1994a), 
solid is the Bondi-Hoyle approximation formula (Eq.~3 in Ruffert~1994a) 
and half that value.
The upper solid bold curve represents the
numerically calculated mass accretion rate.
The lower three curves of the left panels trace the x~(dotted),
y~(thin solid) and z~(bold solid) component of the angular momentum
accretion rate.
The same components apply to the right panels.
The horizontal line in the right Panel of model~JM shows the analytic
specific angular momentum value as given by Eq.~(\ref{eq:specmomang}).
It is outside the range of the plot for model~JS.
}
\label{fig:valueJ}
\end{figure*}

\section{Shape of the shock cone\label{sec:far}}

Figure~\ref{fig:big} shows the density distribution of five models 
towards the end of the simulations, emphasizing the distribution of
matter on scales between one and ten accretion radii.
The left Figs.~~\ref{fig:big}~a, c and~e, show models with a small
gradient in velocity (3\%), while the right Figs.~b and~d have large
gradients (20\%).
Similarly to 3D models without gradients (see Ruffert 1996, and
references therein), these new models do not exhibit the ``flip-flop''
flow visible in previous 2D simulations.
The shape of the shock cones shown in Fig.~\ref{fig:big} is fairly
constant in time and remains roughly conical, contrary to the 2D flows,
whose cones shifted strongly from side to side.
One notices the following features when inspecting Fig.~\ref{fig:big}.
The mass is distributed in a hollow shock cone (as has been reported
previously) for the models with a small gradient, i.e. the density is
maximal just behind the shock, while downstream from the accretor, the
density is minimal along the axis.
The asymmetry of the velocities in the incoming flow reflects itself
in higher density maxima along the cone on the side of the lower
velocities.
This density asymmetry is so pronounced in the models with strong
gradients (Figs.~\ref{fig:big}) that the ``hollow cone'' shape can
be recognized only with difficulty. 
The line of minimum density is very irregular and is shifted from the
$y=0$-axis by several accretion radii.
Already upstream of the shock a higher density is indicated by the
contour of value~0.01 being detached from the shock on the positive
$y$-axis side, while it is very close to the shock on the negative
$y$-axis side.
This density difference is easily explicable: on the side of smaller
velocities gravity can act relatively more strongly to divert the
flow.
Thus the effective local accretion radius is larger on the side of
smaller velocities and so a larger volume of matter can be
gravitationally focused by the accretor.
One also notices that the side of the cone with smaller densities is
more irregularly shaped than the high-density side.
The cavities and lumps produced by the fluctuating flow close to the
accretor (at distances closer than roughly one accretion radius) 
can propagate more easily downstream on the side of the cone with 
lower densities.
Since the velocity enters the accretion radius Eq.~{\ref{eq:accrad})
via a square, one might wonder, whether the velocity is so small, that
the local accretion radius is comparable to the distance of the
accretor to the boundary of the computational box which is
approximately at 16$R_{\rm A}$. 
This is not the case, since inserting $v_{\rm min}=v_0/2$ (from
Eq.~(\ref{eq:vgrad})) into Eq.~{\ref{eq:accrad}) one obtains 
4$R_{\rm A}$, which is a factor of 4 smaller than the distance to the
computational boundary.

\section{Results of models with 3\% velocity gradient\label{sec:descr1}}

\subsection{Moderately supersonic accretors, Mach 3}

The model denoted by IM in this work is identical to the simulation
presented in Ruffert \& Anzer (1995). 
Since for the parameters that I wanted to investigate in this paper a
slightly different prescription of the velocity gradient was necessary
(cf.~Sect.~\ref{sec:models}), a comparison is called for to check
how large the influence of the ``tanh''-term is for small gradients
$\varepsilon_{\rm v}=0.03$.
Models~IM and~IM* are identical except for the ``tanh''-term.
Fig.~\ref{fig:cont3}a shows a contour plot of the density in one
plane for model~IM, Fig.~\ref{fig:valueI} shows the accretion rates of
several quantities for model~IM, and Fig.~\ref{fig:valueIs} displays
the same quantities for model~IM*.
One notices that both the mass accretion rate as well as the
specific angular momentum accreted are similar in both models for the
time over which both have been calculated.
I stopped the simulation of model~IM* at $t\approx8.2$ because it
looked so alike to model~IM.
Thus I conclude that for small gradients $\varepsilon_{\rm v}=0.03$
the difference between the ``tanh''-prescription and the simple linear
dependence is indeed negligible.

Models~IS and~IT differed only in their numerical resolution: model~IT
was simulated with one grid level finer, while the accretor size was
very small ($R=0.02R_{\rm A}$) compared to models~IM and IM* (whose
radii were 5 times larger).
The density distribution of models~IS and~IT can be found in
Fig.~\ref{fig:cont3}, while their accretion rates are shown in
Fig.~\ref{fig:valueI}.
Because model~IT had one level of refinement more, the computational
cost was larger, so the time the simulation was run is roughly half of
the time of model~IS.
Until $t\approx5$ both models show the same features: the mass
accretion rate rises to roughly~6 units and starts fluctuating at
about $t\approx4$ time units.
The specific angular momentum rises continuously until
$t\approx1$ then fluctuates around a value of roughly -0.03 while
increasing the amplitudes of the fluctuations.
Thus the use of~9 nested grids seems sufficient, since the calculation
with~10 grids did not show any qualitative differences.

It is clear that model~IT has not been evolved for long enough to
obtain meaningful averages, since the fluctuation of the mass
accretion rate has hardly begun when the simulation is stopped
(Fig.~\ref{fig:valueI}).
Although model~IS was calculated for a longer time,
the continuously decreasing mass accretion rate of model~IS
(Fig.~\ref{fig:valueI}) indicates that a stationary state has not yet
been reached und so the simulation of this model should have been 
continued even further.
However. the high computational cost precluded this.
The fluctuations of the specific angular momentum, too, seem to
increase with time corroborating the the statement.
Thus the average quantities obtained from this model~IS cannot be very
exact.

Two uncertainties that could not be resolved by the single model~IM
presented in Ruffert \& Anzer (1995), can now be answered.
The first one pertained to the fact that the specific angular
momentum (visible in the top right panel of Fig.~\ref{fig:valueI})
seemed to reach but not cross the zero-line.
This seems to be a coincidence of the initial and boundary conditions,
since in model~IS (visible in the middle right panel of
Fig.~\ref{fig:valueI}) the fluctuations are indeed large enough to
change the sign of the specific angular momentum accreted.
We will return to this point in Sect.~\ref{sec:spec}.
The second uncertainty was whether in the generic case the values
of the specific angular momentum attained and exceeded the
analytically estimated ones given by Eq.~(\ref{eq:specmomang}).
Model~IS clearly does {\it not} attain these values by a large margin,
roughly a factor of 3 --- the analytic value for model~IS is the same
as for model~IM, since the accretor radius does not enter into 
Eq.~(\ref{eq:specmomang}).
The smaller accretor radius seems to allow only smaller values of
angular momentum to be accreted: if the lever arm (which is the radius
of the accretor) is smaller the velocities have to be an appropriate
amount larger (a factor of~5, roughly) to compensate.
This is obviously not the case: the arrows in the left panels of
Fig.~\ref{fig:cont3} close to the surface of the accretor have roughly
the same length.
The smaller accretor sizes also have the effect that the time scale of
the fluctuations of model~IS are shorter than the ones of model~IM
(compare the right panels of Fig.~\ref{fig:valueI}).

The corresponding plot to the top left panel in Fig.~\ref{fig:valueI}
of model~IM for the axisymmetric case can be found in the top left
panel of Fig.~16 in Ruffert \& Arnett~(1994).
For model~IS the closest would be the top left panel of Fig.~22 in
Ruffert \& Arnett~(1994). 
One can see, that while the amplitude of the fluctuations of the
the $z$-component of the accreted angular momentum is comparable, the
average of this component of the models with velocity gradients is
clearly non-zero.
Contrary to this, the $x$- and $y$-components fluctuate more strongly
in the models without gradients, but in all models their temporal
average is close to zero (see e.g.~Table~\ref{tab:models}).
The run, average and fluctuations of the mass accretion rate is
similar in all models. 

Due to the non-axisymmetric upstream boundary conditions it is not
surprising that the shape of the bow shock is not symmetric either.
There is an indication of this fact in the panels shown in
Fig.~\ref{fig:cont3} for the density distribution close to the
accretor, but it is very prominent when inspecting the shock cone
position further away from the accretor (see e.g.~Fig.~1 in Ruffert \&
Anzer~1995).
The temporal evolution shows the usual kinks and deformations of the
shock cone that were described in the previous papers (Ruffert~1996,
and references therein).

\begin{figure*}
 \begin{tabular}{cc}
 \end{tabular}
\caption[]{Contour plots showing snapshots of the density together
with the flow pattern in a plane containing the center of the accretor
for all models with a velocity gradient of~20\%.
The contour lines are spaced logarithmically in intervals of 0.1~dex.
The bold contour levels are labeled with their respective values 
(0.01 or~1.0).
The dashed contour delimits supersonic from subsonic regions.
The time of the snapshot together with the velocity scale is given in
the legend in the upper right hand corner of each panel.
}
%\picplace{20.cm}
\label{fig:cont20}
\end{figure*}

\begin{figure*}
 \begin{tabular}{cc}
  \epsfxsize=8.8cm \epsfclipon \epsffile{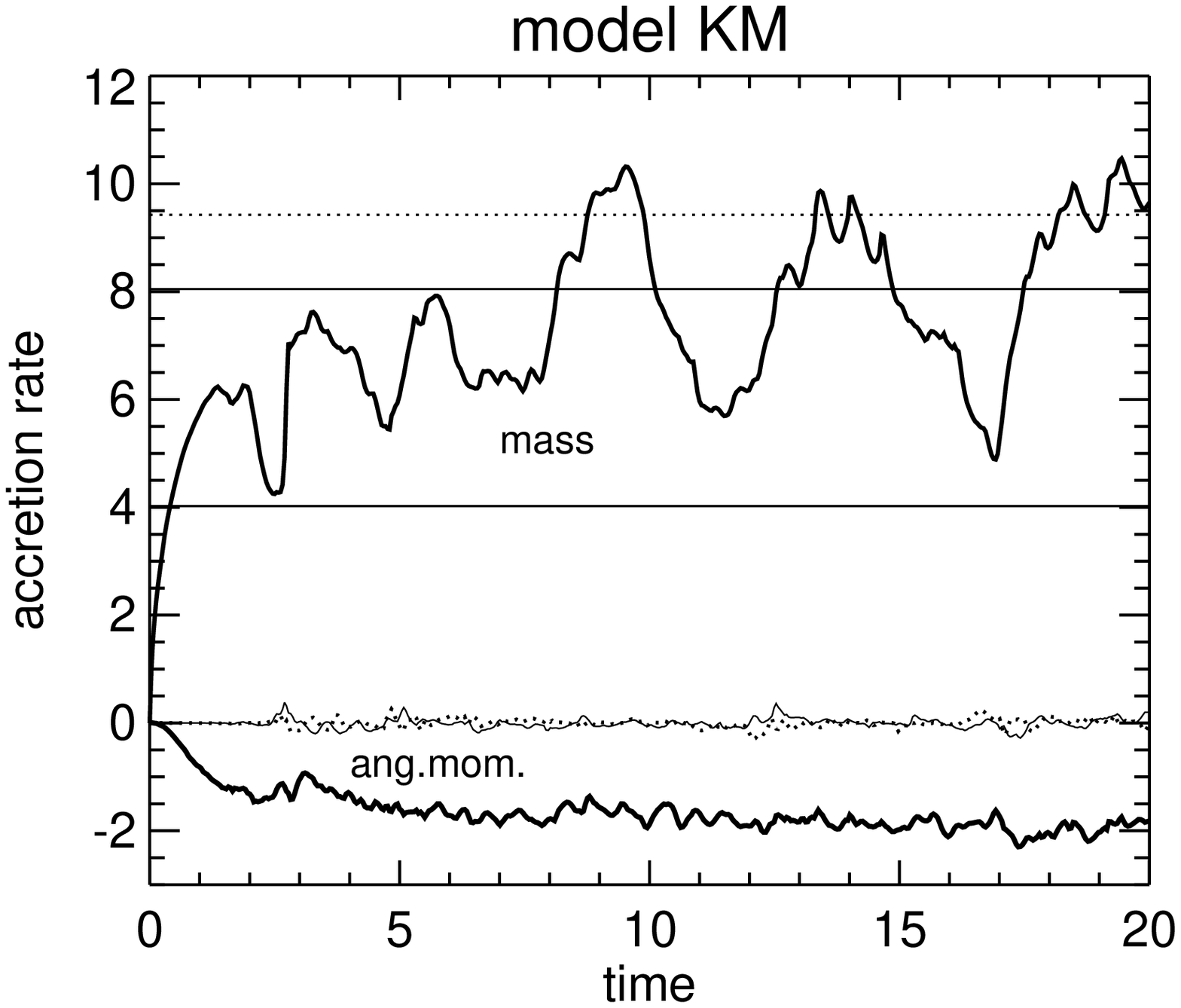} &
  \epsfxsize=8.8cm \epsfclipon \epsffile{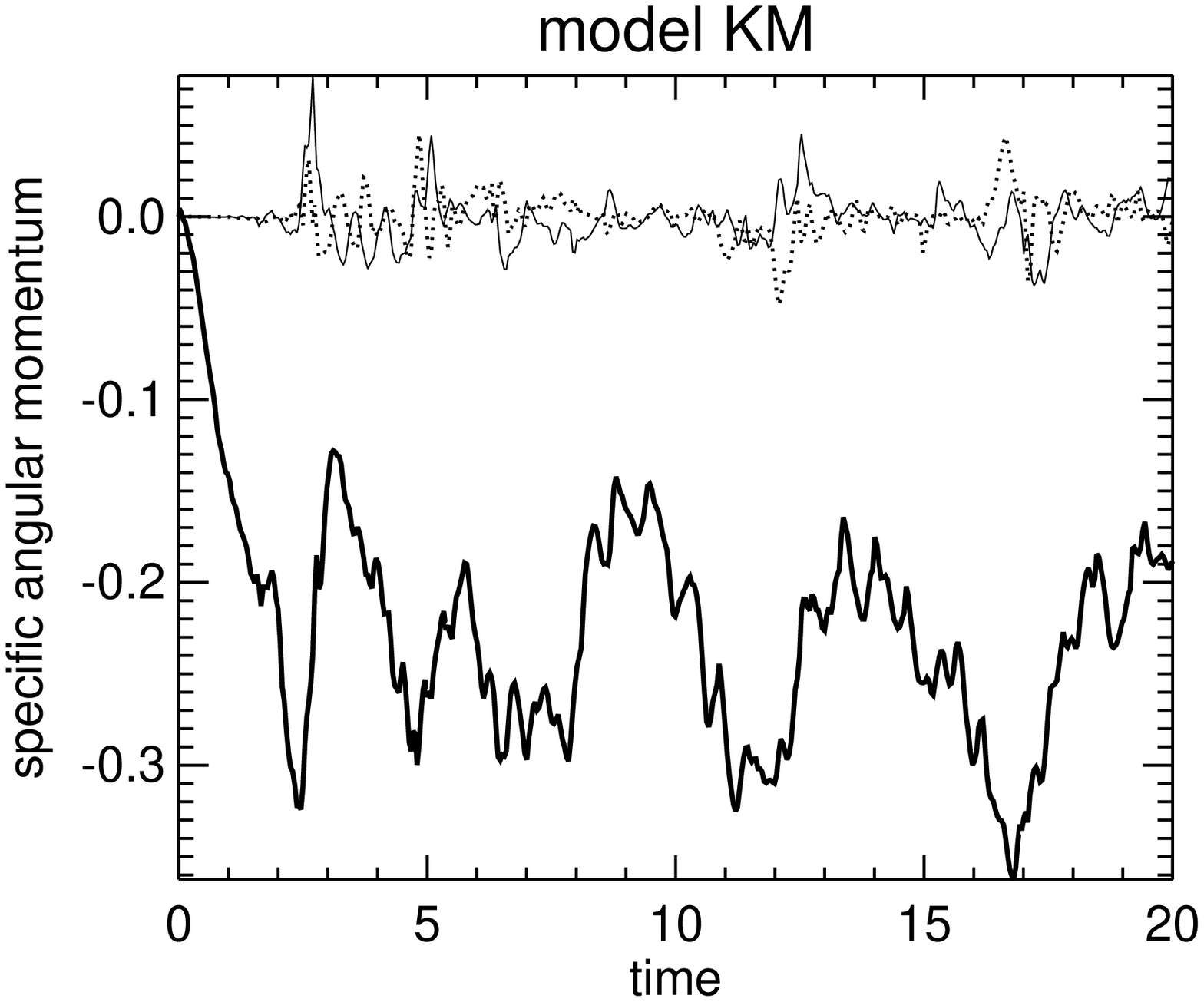} \\
  \epsfxsize=8.8cm \epsfclipon \epsffile{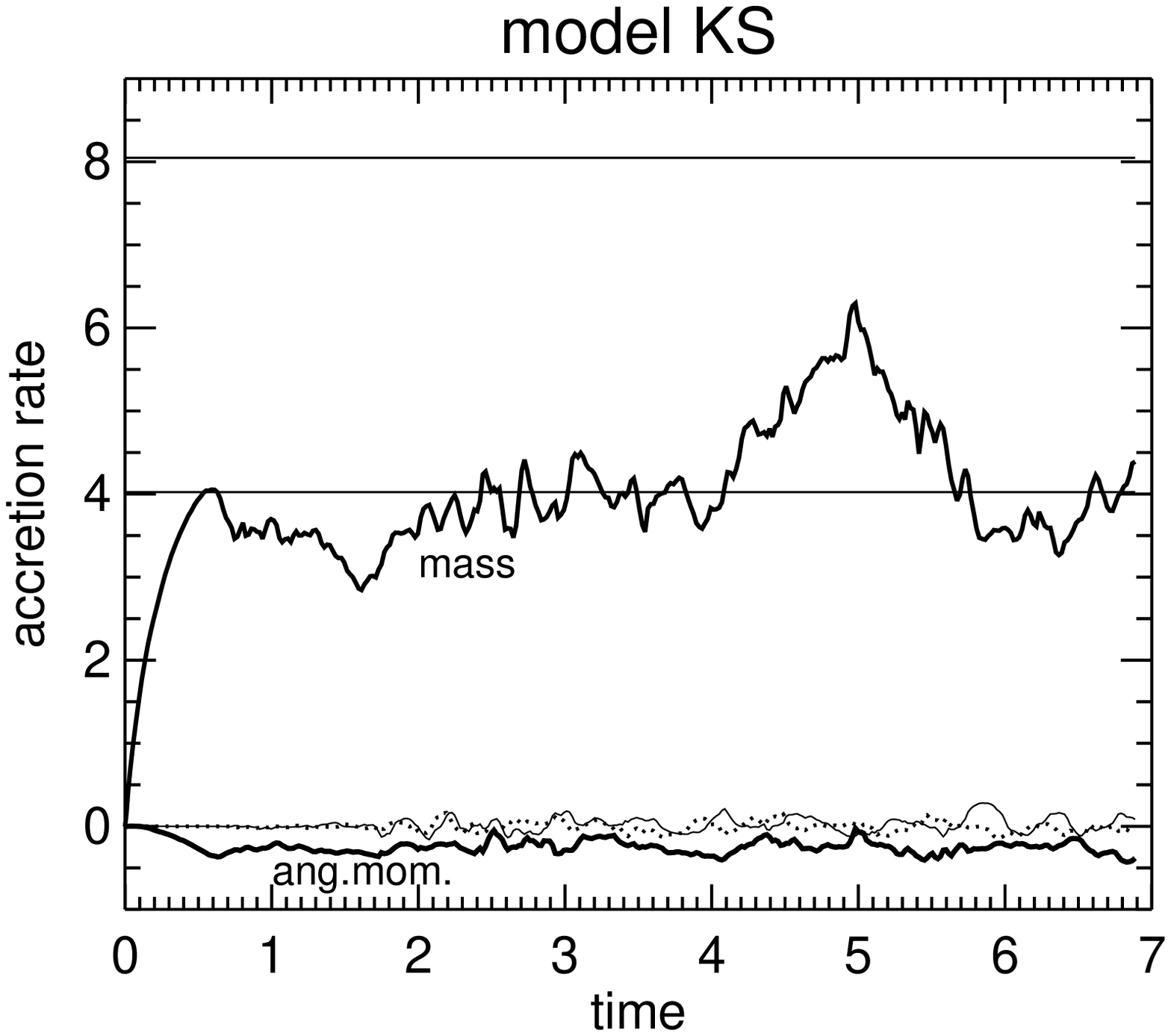} &
  \epsfxsize=8.8cm \epsfclipon \epsffile{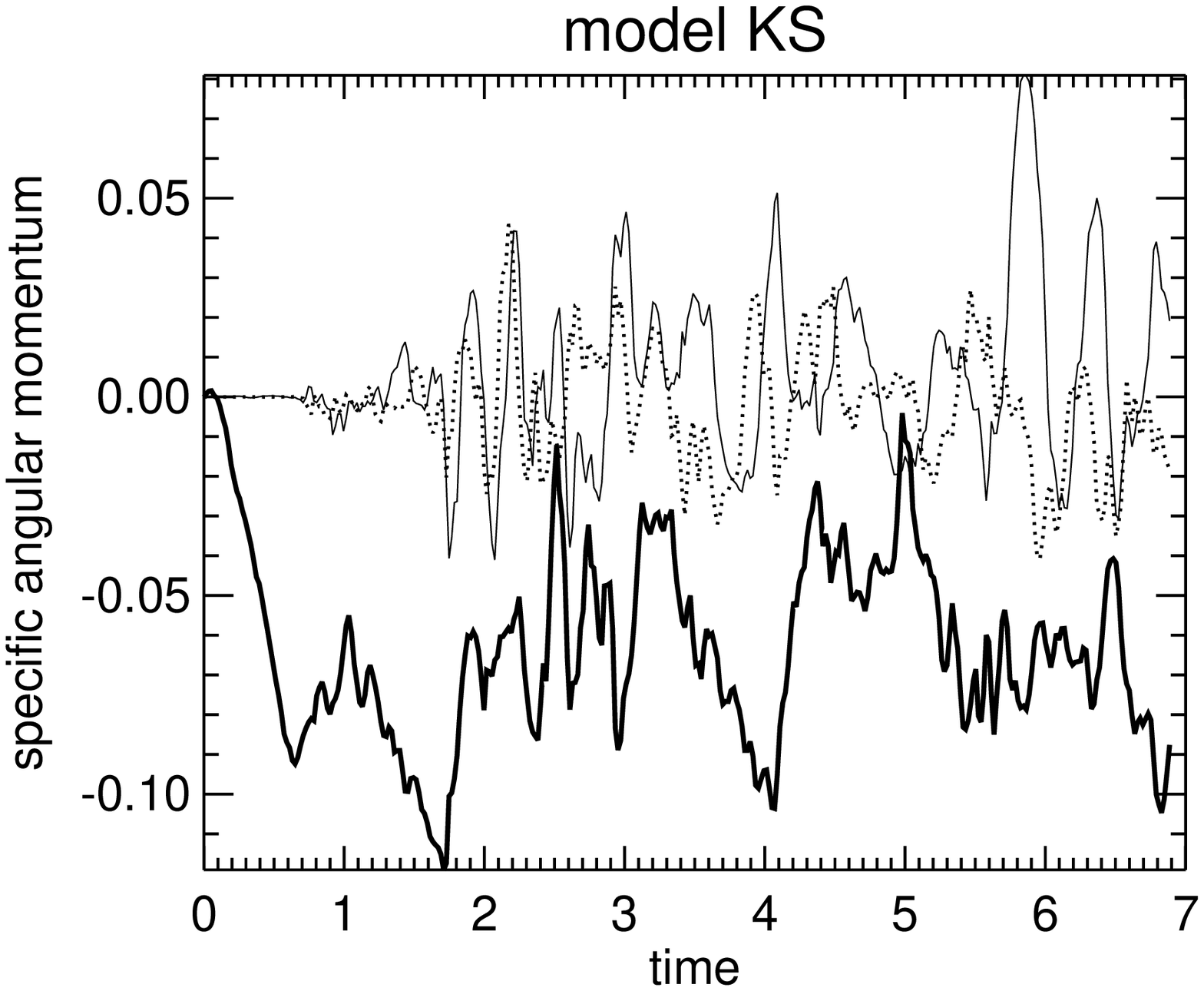}
 \end{tabular}
%\picplace{20.cm}
\caption[]{
The accretion rates of several quantities are plotted as a
function of time for the moderately supersonic (${\cal M}_\infty$=3)
models~KM and~KS with a velocity gradient of~20\%.
The left panels contain the mass and angular momentum accretion rates,
the right panels the specific angular momentum of the matter
that is accreted.
In the left panels, the straight horizontal lines show the analytical
mass accretion rates: dotted is the Hoyle-Lyttleton rate
(Eq.~1 in Ruffert~1994a), 
solid is the Bondi-Hoyle approximation formula (Eq.~3 in Ruffert~1994a) 
and half that value.
The upper solid bold curve represents the
numerically calculated mass accretion rate.
The lower three curves of the left panels trace the x~(dotted),
y~(thin solid) and z~(bold solid) component of the angular momentum
accretion rate.
The same components apply to the right panels.
The value (-0.9) of the specific angular momentum as given by
Eq.~(\ref{eq:specmomang}) is outside the range of the plot for both 
models. 
}
\label{fig:valueK}
\end{figure*}

\begin{figure*}
 \begin{tabular}{cc}
  \epsfxsize=8.8cm \epsfclipon \epsffile{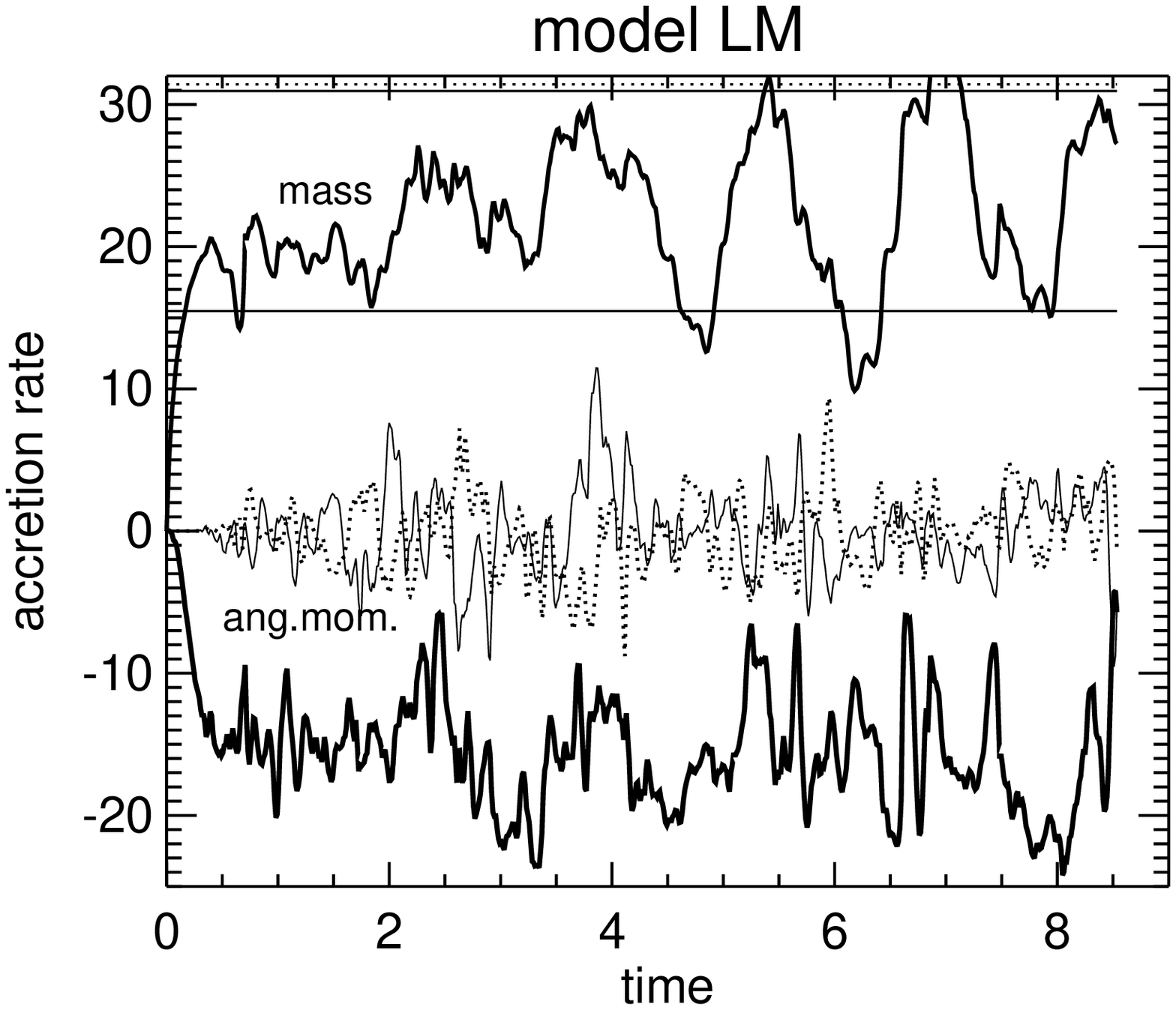} &
  \epsfxsize=8.8cm \epsfclipon \epsffile{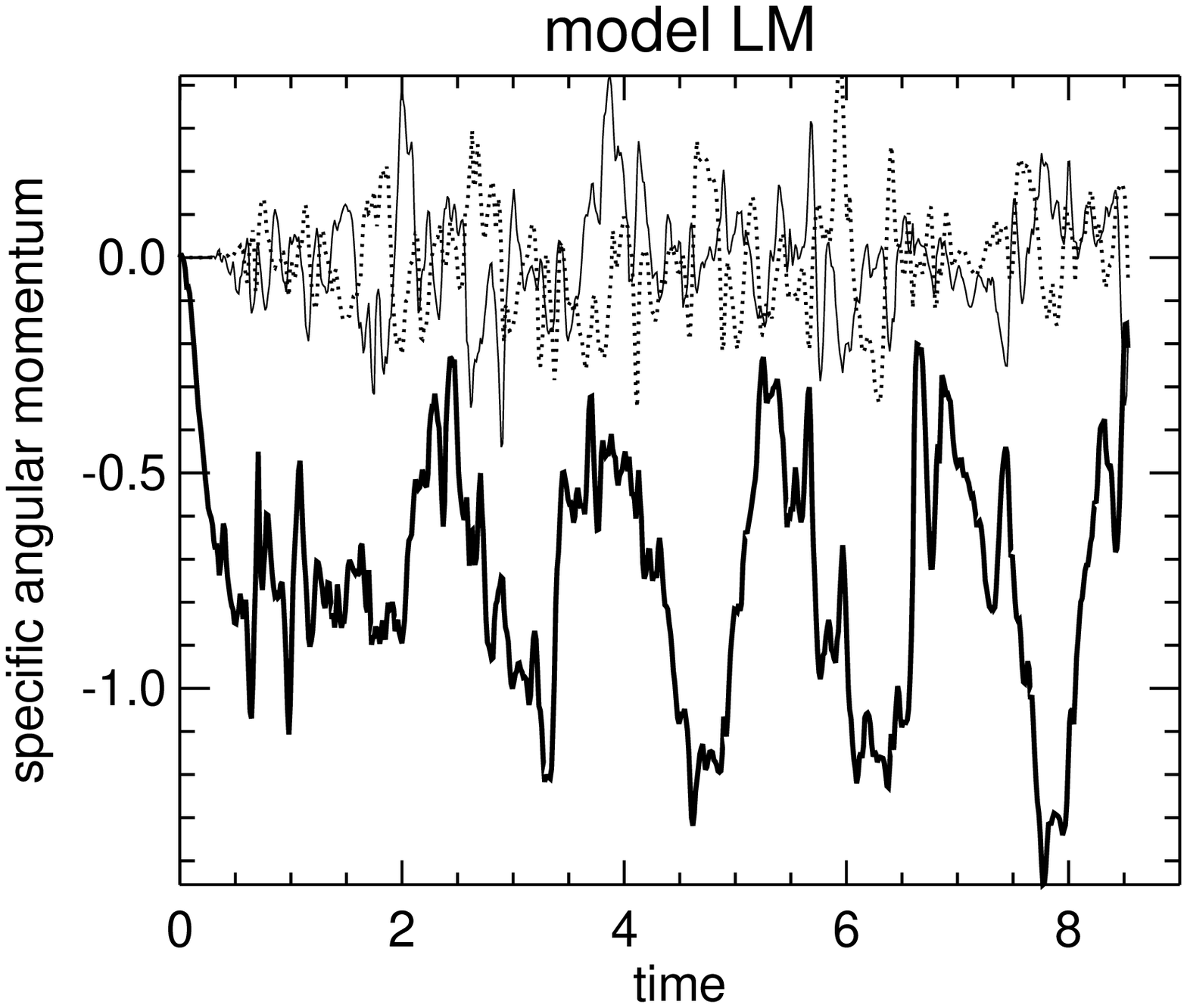} \\
  \epsfxsize=8.8cm \epsfclipon \epsffile{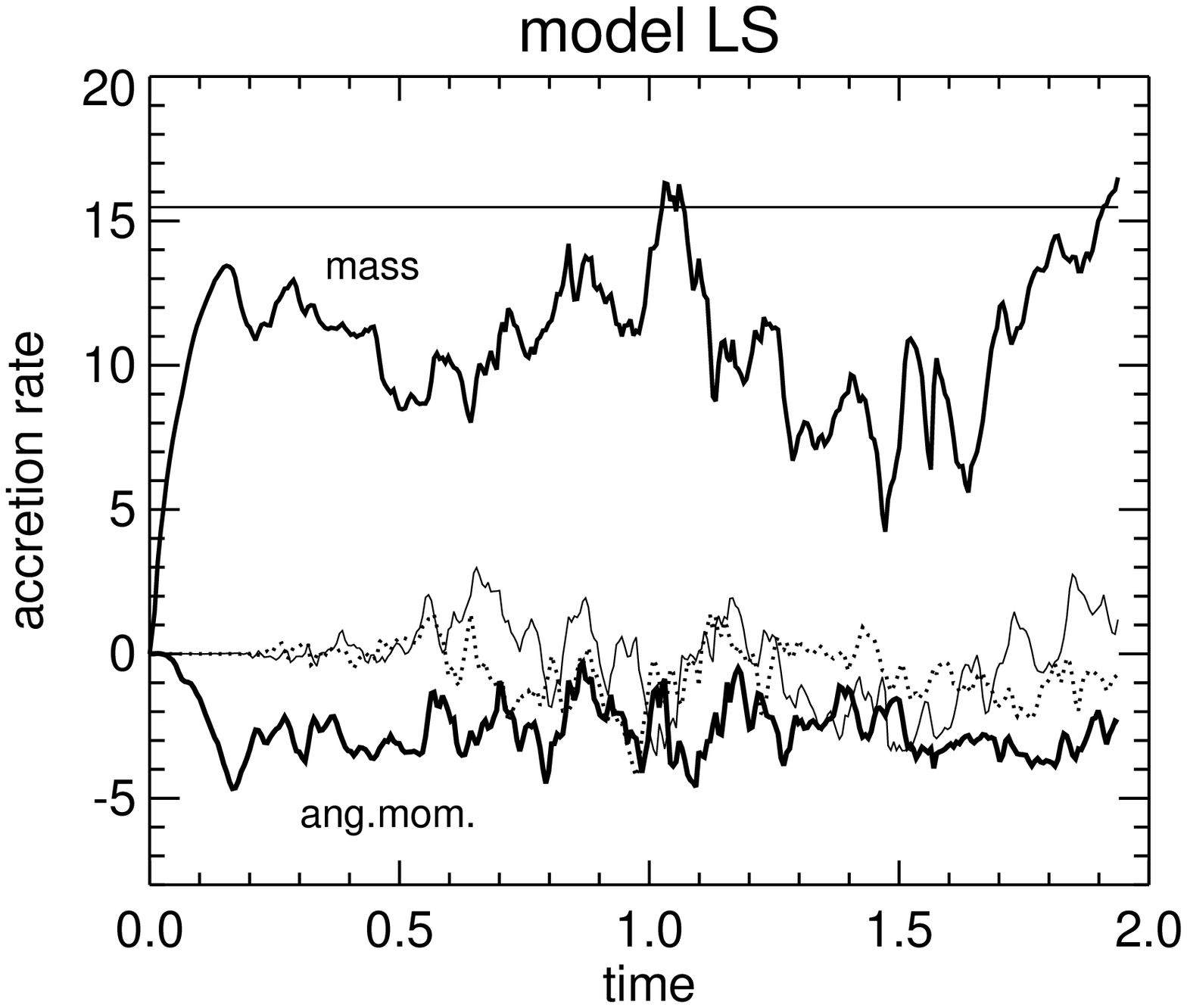} &
  \epsfxsize=8.8cm \epsfclipon \epsffile{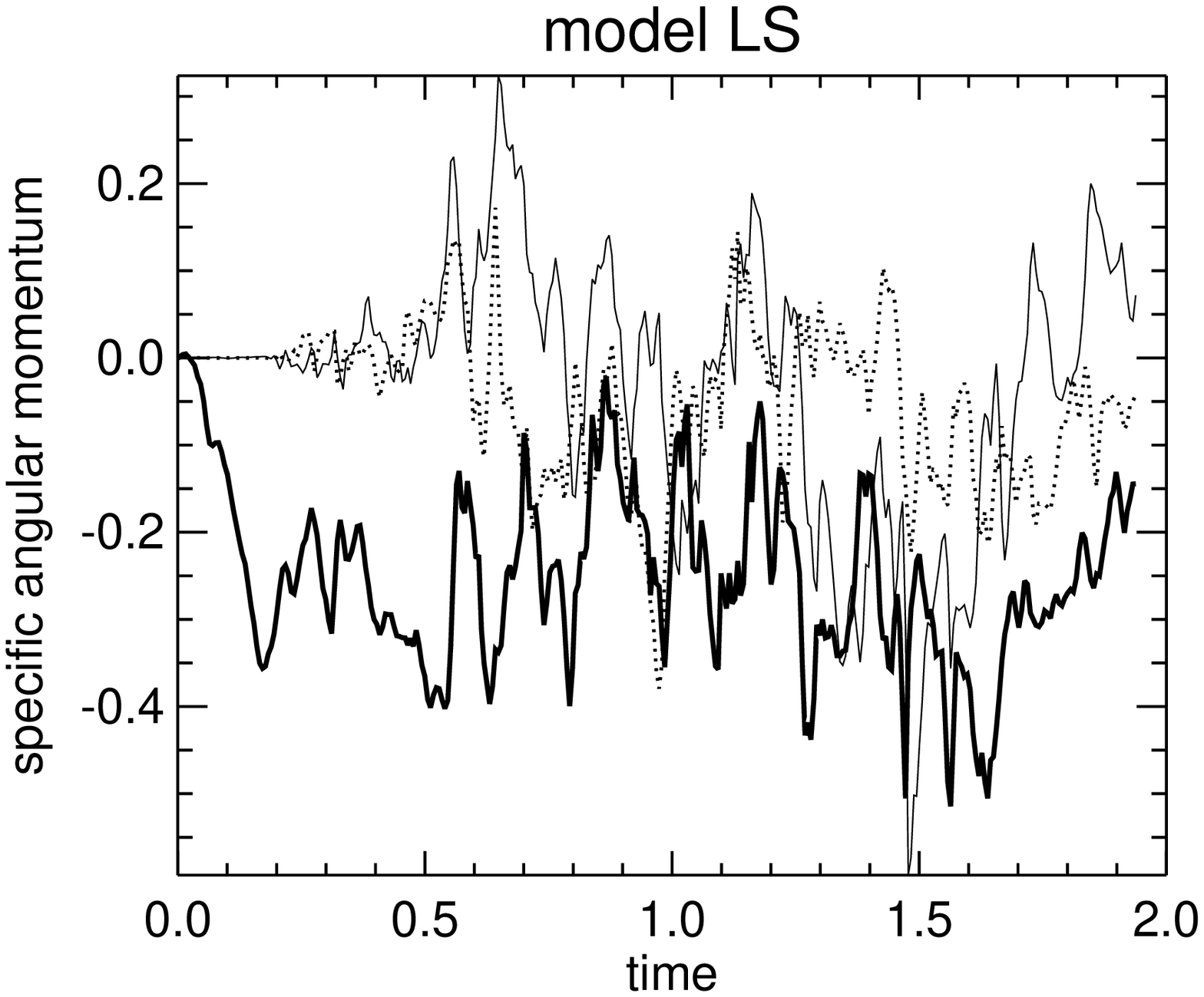}
 \end{tabular}
%\picplace{20.cm}
\caption[]{
The accretion rates of several quantities are plotted as a
function of time for the highly supersonic (${\cal M}_\infty$=10)
models~LM and~LS with a velocity gradient of~20\%.
The left panels contain the mass and angular momentum accretion rates,
the right panels the specific angular momentum of the matter
that is accreted.
In the left panels, the straight horizontal lines show the analytical
mass accretion rates: dotted is the Hoyle-Lyttleton rate
(Eq.~1 in Ruffert~1994a), 
solid is the Bondi-Hoyle approximation formula (Eq.~3 in Ruffert~1994a) 
and half that value.
The upper solid bold curve represents the
numerically calculated mass accretion rate.
The lower three curves of the left panels trace the x~(dotted),
y~(thin solid) and z~(bold solid) component of the angular momentum
accretion rate.
The same components apply to the right panels.
The value (-3.0) of the specific angular momentum as given by
Eq.~(\ref{eq:specmomang}) is outside the range of the plot for both 
models. 
}
\label{fig:valueL}
\end{figure*}

\subsection{Highly supersonic accretors, Mach 10}

The right panels of Fig.~\ref{fig:cont3} display the density contours
of models~JM and~JS, which are equivalent to models~IM and~IS, except
for the different flow speed upsteam: Mach~10 for the J-models
contrary to Mach~3 for the I-models.
The corresponding accretion rates can be found in
Fig.~\ref{fig:valueJ}.
The highly supersonic models~JM and~JS, too, do not converge to a
quiescent steady state but show an unstable fluctuating flow pattern.

In the same way as model~IM, also model~JM meets and exceeds the
analytically given value (Eq.~(\ref{eq:specmomang})) of the specific
angular momentum (top right panel Fig.~\ref{fig:valueJ}, horizontal
line), however only in rare, short bursts.
Thus, on average, the fraction of the specific angular momentum
to the analytic value is smaller than the value of the fraction of
model~IM (cf.~Table~\ref{tab:models} colum $l_{\rm z}$).
However, similarly to the difference between models~IM and~IS, the
specific angular momentum of model~JS is smaller still.
In fact, when looking at the bottom right panel of
Fig.~\ref{fig:valueJ}, the curve for the $j_{\rm z}$ component
overlaps very strongly with the curves of the other components.
There is, however, still clearly a systematic shift of $j_{\rm z}$ not
visible in~$j_{\rm x}$ and~$j_{\rm y}$ which fluctuate around zero.

The left two panels in Fig.~\ref{fig:valueJ} (models~JM and~JS)
can be compared to the two top panels of Fig.~9 in Ruffert (1994b)
which show the results of the equivalent models without gradients (FM
and FS, respectively).
The magnitude and fluctuation amplitude of the mass accretion rate are
similar which is also the case for the angular momentum accretion
rates (note different scale of $y$-axis in figures).
Since the mass accretion rate of model~JS seems to be steadily
declining (although also fluctuating, bottom left panel in
Fig.~\ref{fig:valueJ}), this model at the time we stopped the
simulation is probably still influenced by transients generated
through the initial conditions.
Thus the average values of e.g.~the specific angular momentum, have to
be used with caution.

\section{Results of models with 20\% velocity gradient\label{sec:descr2}}

Models~KM, KS, LM and LS break the axisymmetry very strongly since
they have as boundary condition a 20\% velocity gradient over one
accretion radius.
This asymmetry induces a strong clockwise vortex around the accretor
as can be seen in Fig.~\ref{fig:cont20}.
The sign of the velocity gradient has been chosen in such a way that
upstream of the accretor the higher velocities are on the negative
$y$-side of the $xy$-plane (lower half of the contour plots).
However, the numeric simulations confirm the sign of the analytically
estimated accreted angular momentum (Eq.~(\ref{eq:specmomang})): the
vector points into the plane of the plot, which corresponds to a
clockwise rotation in these contour plots. 
A note of caution is necessary, however: recall that
Eq.~(\ref{eq:specmomang}) was derived only to lowest order
in~$\varepsilon_{\rm v}$ and~$\varepsilon_\rho$, thus assuming
that~$\varepsilon_{\rm v}$ and~$\varepsilon_\rho$ are small
compared to unity.
Obviously this is questionable with the choice 
$\varepsilon_{\rm v}=0.2$ for the models presented in this section.
It is still interesting to see by how much the analytic estimates
deviate from the numerical models is this extreme case.

\subsection{Moderately supersonic accretors, Mach~3}

The accretion rates of several quantities for models~KS and~KM can be
found in Fig.~\ref{fig:valueK}.
Contrary to model~IM, model~KM does not exhibit such strong
fluctuations, thus it never even comes close (by a factor 3) to the
analytically predicted value (-0.9) of Eq.~(\ref{eq:specmomang}), 
nor does it ever come close to the zero line.
The large scale motion of the vortex around the accretor is fairly
stable in time, which explains why the fluctuations relative to the
mean of the specific angular momentum are smaller in model~KM than in
model~IM.

The fluctuations of the mass accretion rate (left panels of
Fig.~\ref{fig:valueK} of model~KM seems to increase with time
indicating that this model is still evolving in time and has not yet
reached a steady mean state.
These panels should be compared to the analogous panels (left in
Fig.~\ref{fig:valueI} for the models~IM and~IS.
One notices that both the mean and the amplitude of the
fluctuations is larger in model~KM than in model~IM
Although the mean mass accretion rate of model~KS is lower than the
mean in model~IS this might be a transient effect, since toward the
end of model~IS the mass accretion rate is of the same magnitude as in
model~KS. 
Thus for the larger accretors (M-models) the vortex allows more mass
to be accreted, while for the small accretors (S-models) the
difference is negligible.
Both models~KM and~KS have one difference in common compared to
model~IM and~IS respectively: the unstable flow, which manifests
itself e.g.~via a fluctuating mass accretion rate, begins much faster
for the models with a large gradient, models~KM and~KS.
Also the large-scale fluctuations of the specific angular momentum
appear at roughly $t\approx1$ in models~KM and~KS, while it takes
until $t\approx4$ in models~IM and~IS.

\subsection{Highly supersonic accretors, Mach 10}

The accretion rates of several quantities for the highly supersonic
models~LM and~LS are shown in Fig.~\ref{fig:valueL}.
The fluctuation amplitude of the mass accretion rate of model~LM is
larger than both the amplitudes of model~JM (top left panel in
Fig.~\ref{fig:valueJ}) and of model~FM (top left panel of Fig.~9 in
Ruffert~1994), however the mean seems roughly equal.
On the whole model~LM looks more unstable and active than the
models~JM and~FM with small or no gradients.
The mass accretion rate of model~LS does not seem to decline
constantly during the first two time units as was the case for
model~JS.

The trend that the model with the smaller accretor (in this case
model~LS compared to model~LM) has the $z$-component shifted closer to
the zero line is repeated also for the highly supersonic models with
large gradients. 
The fluctuations around the mean of the $z$-component has roughly the
same amplitude and frequency as the fluctuations of the $x$ and
$y$-components around zero.

\begin{figure*}
 \begin{tabular}{cc}
\raisebox{8cm}{\parbox[t]{8.8cm}{
\caption[]{Contour plots showing snapshots of the density together
with the flow pattern in a plane containing the center of the accretor
for models~SS, ST and~RL.
The contour lines are spaced logarithmically in intervals of 0.1~dex.
The bold contour levels are labeled with their respective values.
The dashed contour delimits supersonic from subsonic regions.
The time of the snapshot together with the velocity scale is given in
the legend in the upper right hand corner of each panel.\label{fig:contm}
}}}
 \end{tabular}
%\picplace{20.cm}
\end{figure*}

\begin{figure*}
 \begin{tabular}{cc}
  \epsfxsize=8.8cm \epsfclipon \epsffile{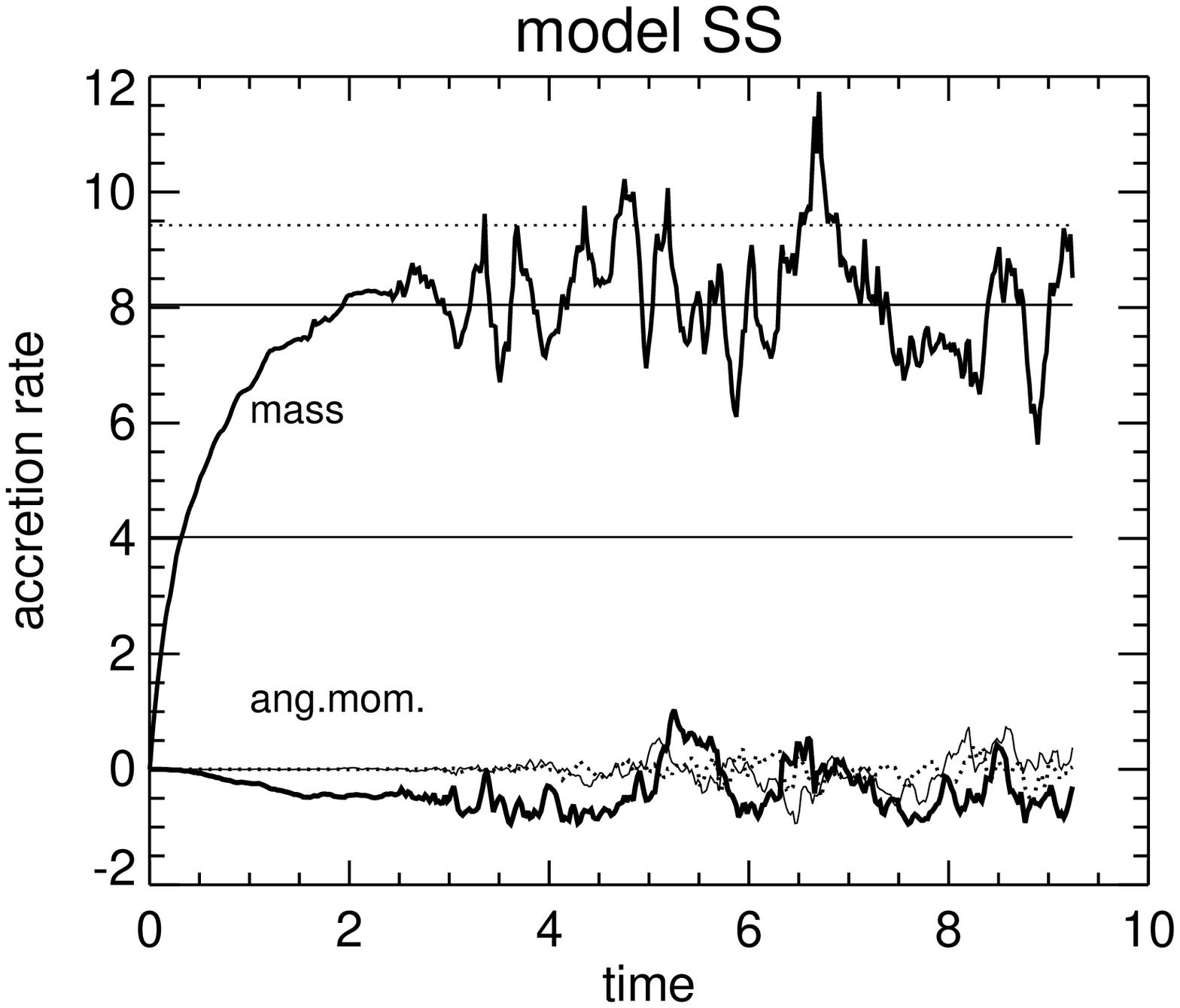} &
  \epsfxsize=8.8cm \epsfclipon \epsffile{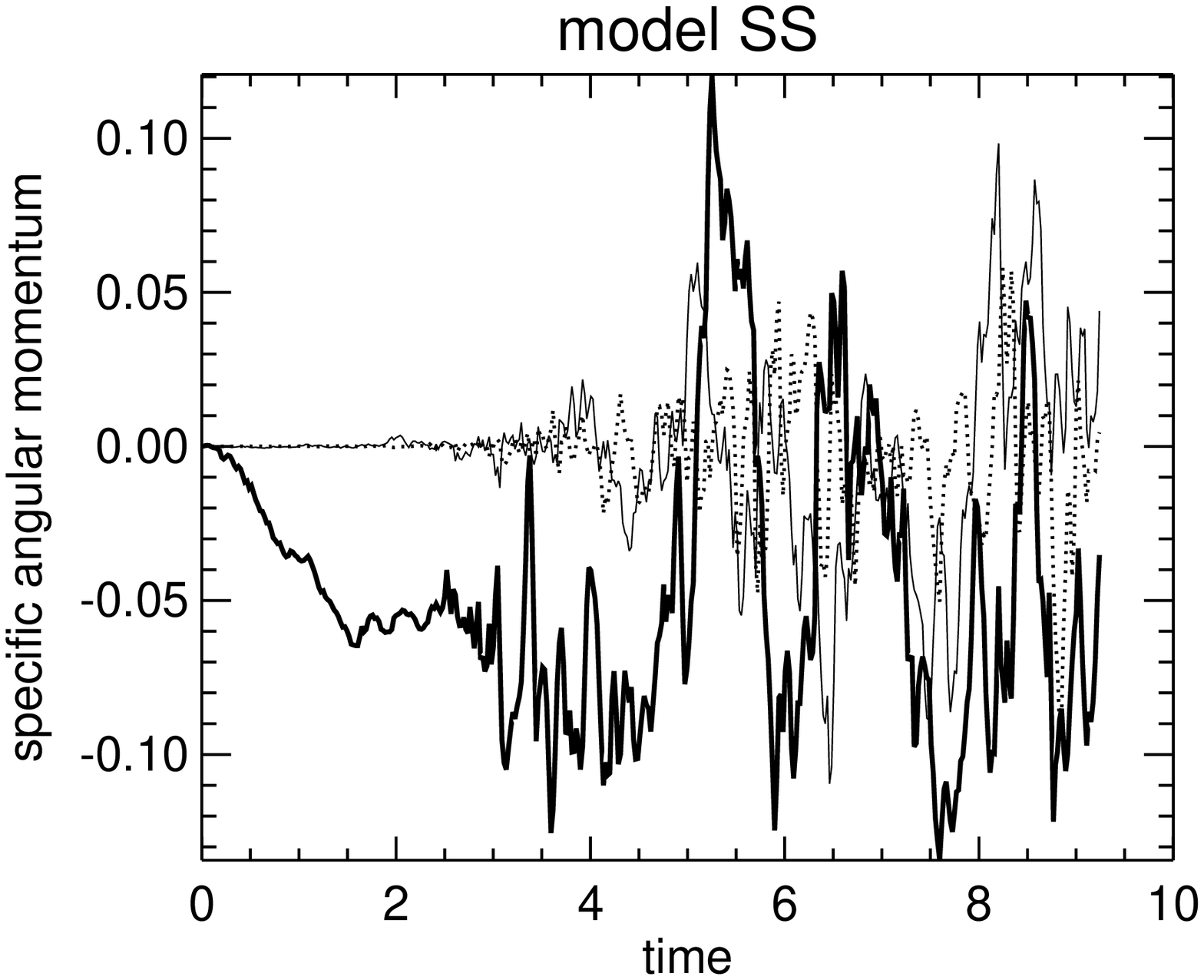} \\
  \epsfxsize=8.8cm \epsfclipon \epsffile{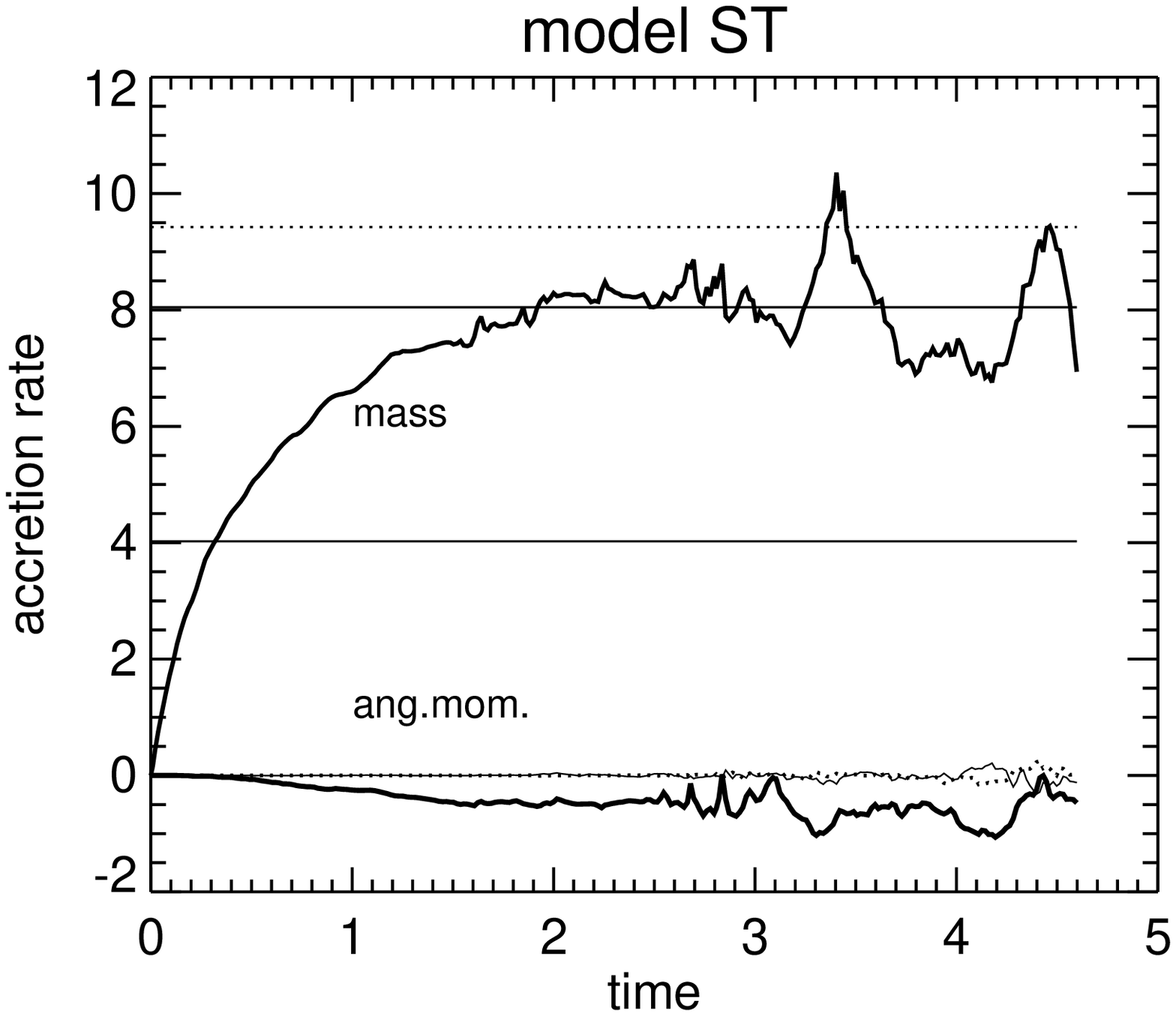} &
  \epsfxsize=8.8cm \epsfclipon \epsffile{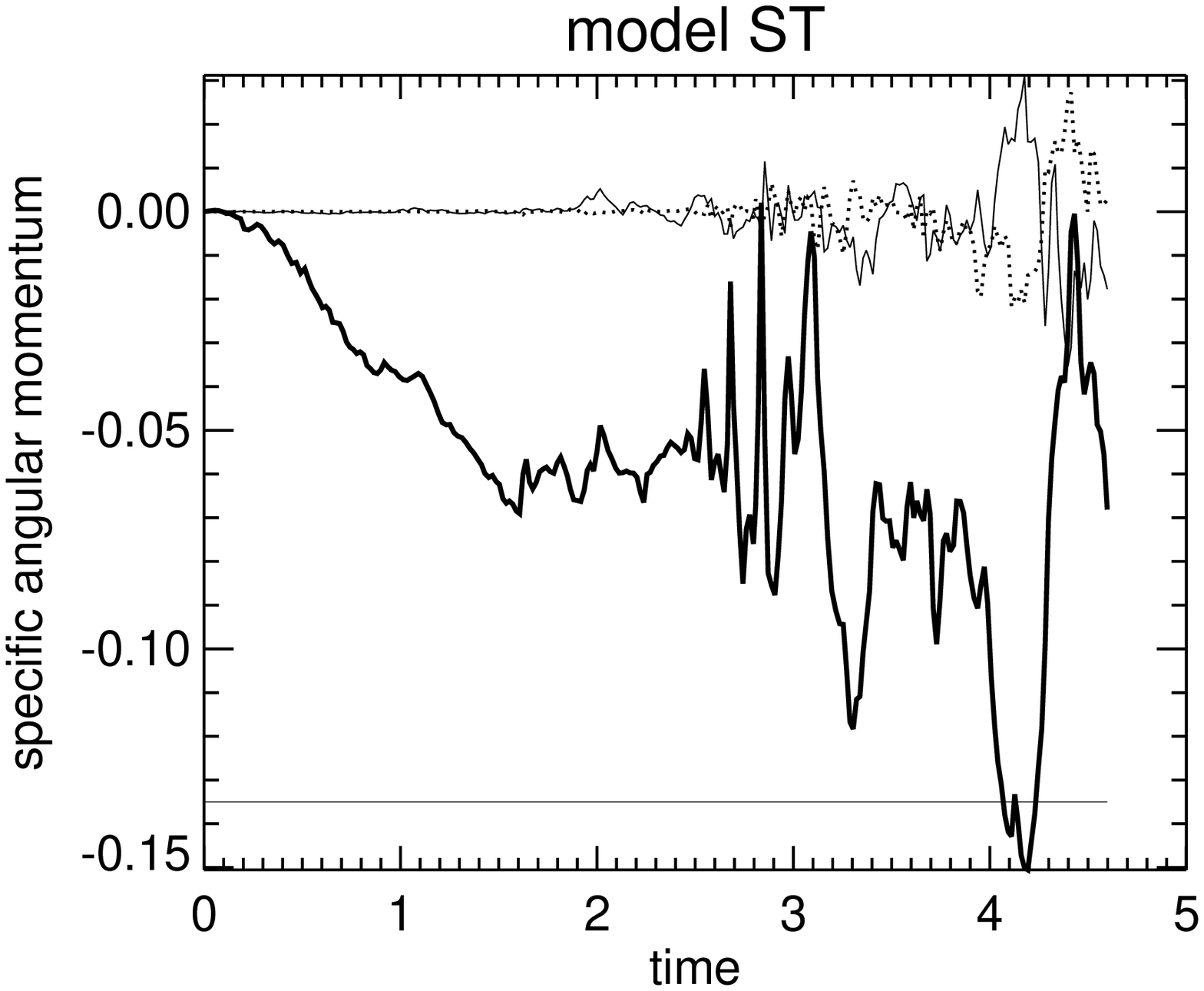} \\
  \epsfxsize=8.8cm \epsfclipon \epsffile{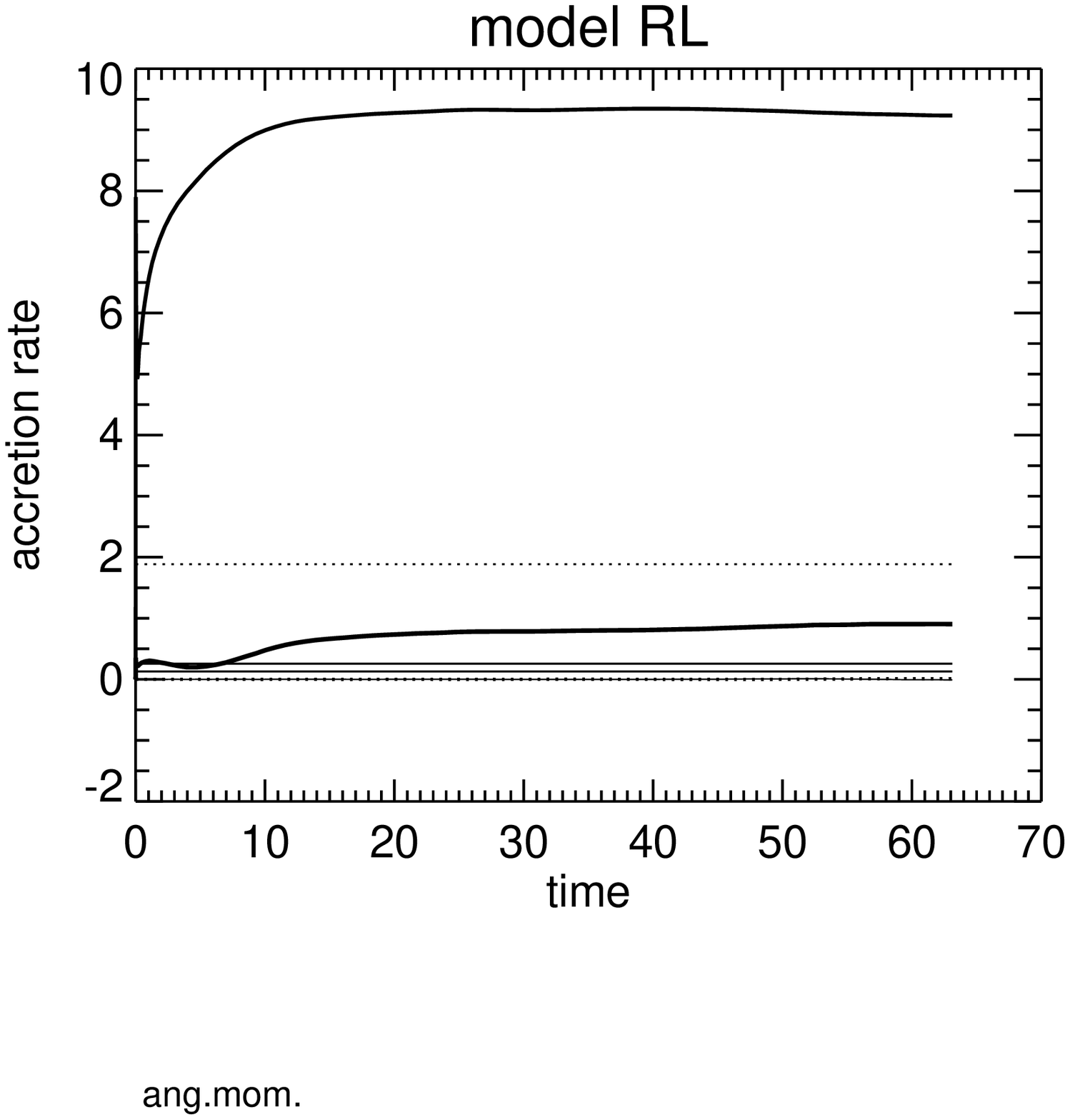} &
  \epsfxsize=8.8cm \epsfclipon \epsffile{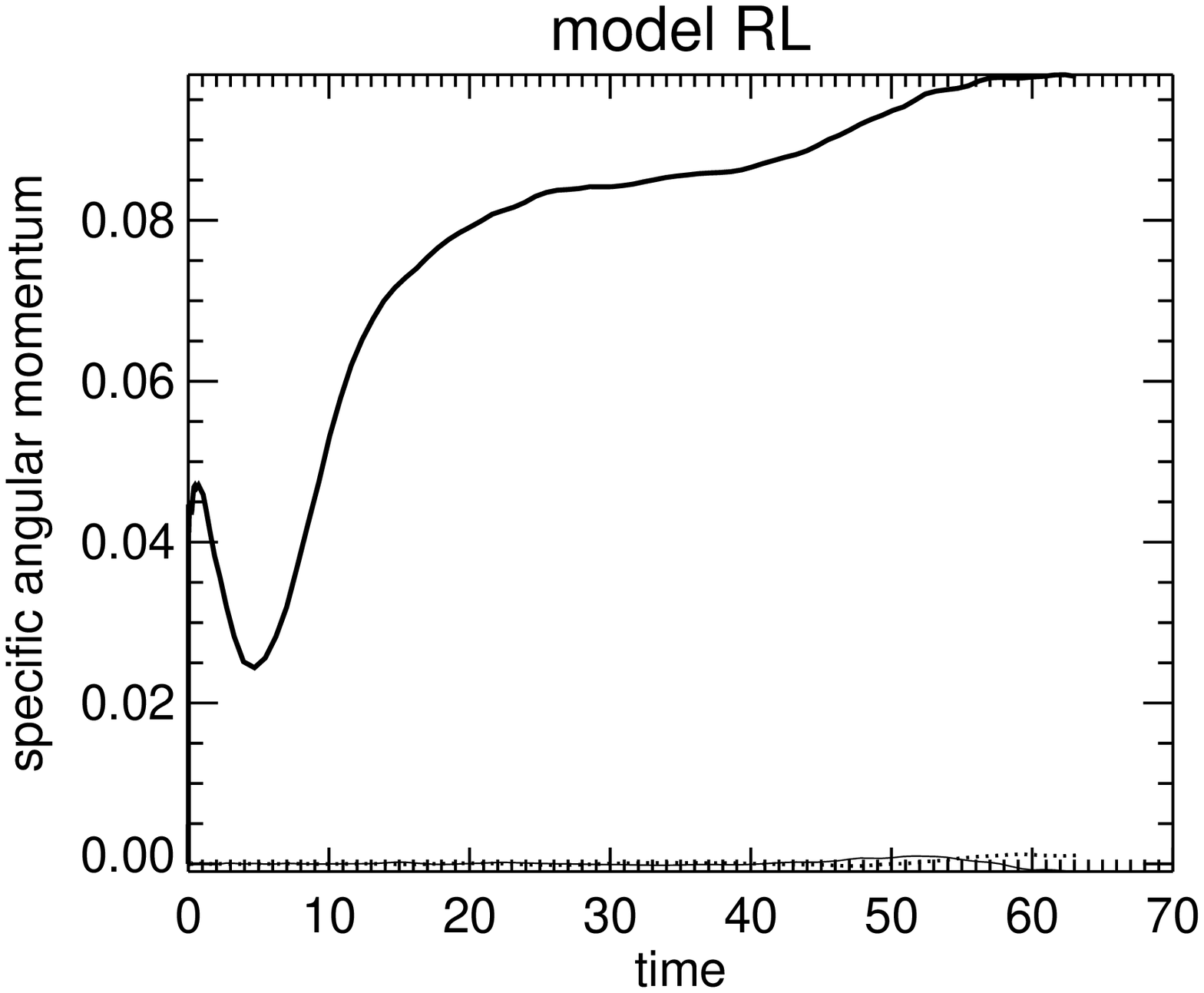}
 \end{tabular}
%\picplace{20.cm}
\caption[]{
The accretion rates of several quantities are plotted as a
function of time for models~SS, ST and~RL.
The left panels contain the mass and angular momentum accretion rates,
the right panels the specific angular momentum of the matter
that is accreted.
In the left panels, the straight horizontal lines show the analytical
mass accretion rates: dotted is the Hoyle-Lyttleton rate
(Eq.~1 in Ruffert~1994a), 
solid is the Bondi-Hoyle approximation formula (Eq.~3 in Ruffert~1994a) 
and half that value.
The upper solid bold curve represents the
numerically calculated mass accretion rate.
The lower three curves of the left panels trace the x~(dotted),
y~(thin solid) and z~(bold solid) component of the angular momentum
accretion rate.
The same components apply to the right panels.
The horizontal line in the right Panel of model~ST shows the 
specific angular momentum value as given by Eq.~(\ref{eq:specmomang}).
It is outside the range of the plot for models~SS and~RL.
}
\label{fig:valueS}
\end{figure*}

\section{Results of models with index 4/3\label{sec:descr3}}

As a check for the numerics two models were run with the same physical
initial and boundary conditions, but differing in resolution of the
accretor: model~SS used 9~grids, model~ST used 10~grids
(cf.~Table~\ref{tab:models} for the other parameters).
The two left panels of Fig.~\ref{fig:contm} show contour plots of the
density distribution, while the acretion rates of mass and angular
momentum can be found in Fig.~\ref{fig:valueS}.
Especially the temporal evolution of the accretion rates confirm that
also for $\gamma=4/3$ using grids to 9~levels deep is sufficient:
within the time in which the two models~SS and~ST overlap, the mass
accretion rate and the angular momentum accretion agree to within a
few percent.

Models~SS and~ST (Fig.~\ref{fig:valueS}) are equivalent to
models~IS and~IT, (Fig.~\ref{fig:valueI}) so a comparison should show
what effects are due to the different adiabatic index $\gamma$.
Additionally, model~SS can be compared to model~CS shown in Fig.~6 of 
Ruffert~(1995).
Both the mean and the amplitude of the mass accretion rate of model~SS
are similar to the ones of model~CS, and even the angular momentum
accretion fluctuations do not show any striking difference.
When comparing model~SS to model~IS one notices that the mass
accretion rate proceeds in a much more unstable way in model~SS.
The rate of mass accretion as well as the specific angular
momentum of the $z$-component are larger in model~SS.
So much so, that in a short burst model~ST actually reaches the
analytically estimated value (middle right panel of
Fig.~\ref{fig:valueS}) contrary to model~IS or~IT.
Thus we conclude that the models with $\gamma=4/3$ accrete in a less
stable way than their $\gamma=5/3$ equivalents, which is the opposite
from what had been observed in the simulations without gradients
(Ruffert 1995 and references therein).
More $\gamma=4/3$ models are necessary before any more systematic
statements can be made.

\begin{figure}
  \epsfxsize=8.8cm \epsfclipon \epsffile{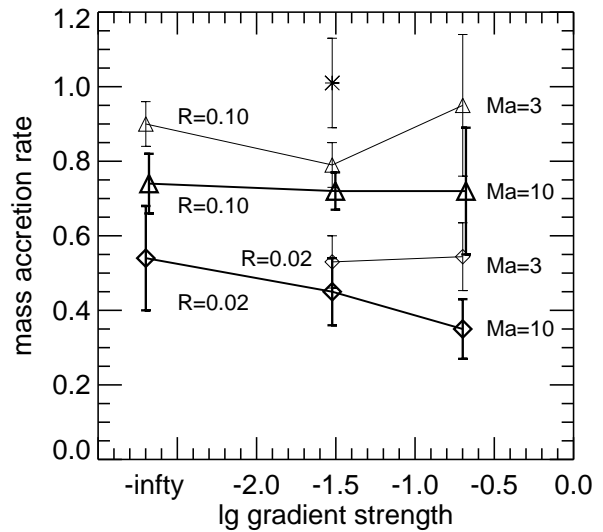}
%\picplace{6.3cm}
\caption[]{Mass accretion rates (units: $\dot{M}_{\rm BH}$) are shown
  as a function of the strength of the velocity gradient: 20\% and 3\%
  are the results from this work, while the values for models without
  gradient (at the x-axis position ``-infty'') are taken from
  Ruffert~(1994a) and Ruffert \& Arnett~(1994).
  Diamonds ($\diamond$) denote models in which the accretor has a
  radius of $0.02~R_{\rm A}$, triangles ($\triangle$) models with
  $0.1~R_{\rm A}$.
  The large bold symbols belong to models with a speed of 
  ${\cal M}_\infty=10$,
  while the smaller symbols belong to models with ${\cal M}_\infty=3$.
  The accretor radius and Mach number are also written near each set
  of points.
  All models have $\gamma=5/3$, except for the ``star'' which denotes
  model~SS. 
  The error bars extending from the symbols indicate one standard
  deviation from the mean ($S$ in Table~\ref{tab:models}).
  Some points were slightly shifted horizontally to be able to discern the 
  error bars.
}
\label{fig:massacc}
\end{figure}

\begin{figure}
\epsfxsize=8.8cm \epsfclipon \epsffile{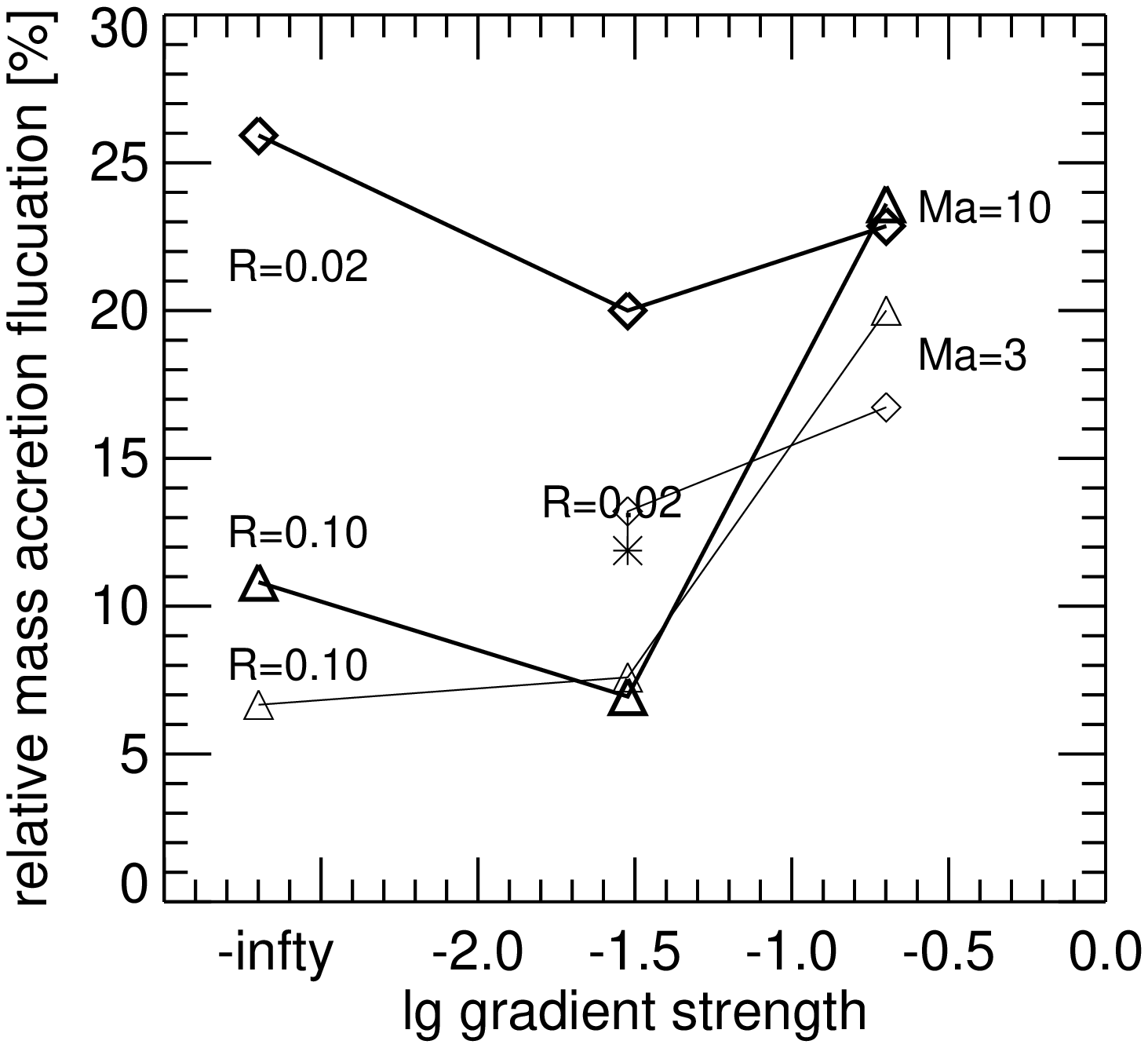}
%\picplace{8.0cm}
\caption[]{
The relative mass fluctuations,
i.e.~ the standard deviation $S$ divided by the average mass accretion
rate $\overline{\dot{M}}$ (cf.~Table~\ref{tab:models}), is shown as
a function of the strength of the velocity gradient: 20\% and 3\% are
the results from this work, while the values for models without
gradient (at the x-axis position ``-infty'') are taken from
Ruffert~(1994a) and Ruffert \& Arnett~(1994).
Diamonds ($\diamond$) denote models in which the accretor has a
radius of $0.02~R_{\rm A}$, triangles ($\triangle$) models with
$0.1~R_{\rm A}$.
The large bold symbols belong to models with a speed of 
${\cal M}_\infty=10$,
while the smaller symbols belong to models with ${\cal M}_\infty=3$.
All models have $\gamma=5/3$, except for model~SS (with $\gamma=4/3$)
denoted by a star~($*$).
}
\label{fig:relfluct}
\end{figure}

\section{Results of model~RL\label{sec:descr4}}

Model~RL is extreme in two aspects: firstly the radius of the accretor
is one accretion radius in size and thus very large, and secondly the
flow speed upstream of the accretor is subsonic, Mach~0.6.
Additionally the gradient was chosen to be large (0.2).
These conditions, although extreme, might, however, be present in
stellar binary systems with a wind from a nova (K.~Schenker, personal
communication), thus the model is not totally academic, but one has to
take care in relating the physical size of the stars to the radius of
the accretor used in the simulations.
The contour plot showing the density distribution together with the
instantaneous velocities is shown in Fig.~\ref{fig:contm}b, while the
corresponding accretion rates are shown in the bottom panels of
Fig.~\ref{fig:valueS}.
The large scale of the contour plot shows the velocity gradient
clearly.

After $t\approx10$ time units model~RL has practically reached a
stationary state in which mass is accreted at a constant rate
(bottom left panel in Fig.~\ref{fig:valueS}.
This behaviour as well as the value of the mass accretion rate is very
similar to model~SL in Ruffert~(1994b; top left panel of Fig.~2).
The small fluctuations of the angular momentum visible in model~SL are
initial transients due to a small (3\%) perturbation of the initial
density distribution. 
I did not perturb the density in any model presented in this paper.
This model~RL is the only model presented in this work that reaches a
constant state, which is probably due to the subsonic flow, sa has
been reported elsewhere for models without gradients (e.g.~Ruffert 1995).

The main difference concerning accreted quantities between model~RL
and all other models in this paper is the sign of the accreted angular 
momentum: it is positive (cf.~Fig~\ref{fig:valueS}), while the
$z$-component of the angular momentum accreted of all other models is
negative (see also Table~\ref{tab:models}).
This is probably due to the fact that an accretor with such a large
radius accretes mainly via its large geometric cross section, and not
following the classic Bondi-Hoyle-Lyttleton scenario involving
gravitational focussing, accretion mainly from the downstream side, etc.
(the low Mach number might also play a role).
Thus the matter is directly absorbed from the incoming stream without
first being ``processed'' though a shock, taking with it the angular
momentum it had upstream: this produces a anti-clockwise rotating
vortex around the accretor (contrary to the clockwise vortex mentioned
in Sect.~\ref{sec:descr2}).

\section{Analysis of Results\label{sec:analy}}

\subsection{Mass accretion rate\label{sec:massacc}}

The mass accretion rates obtained in this work for all models
except~IT, RL, and~ST are collected in Fig.~\ref{fig:massacc} 
together with the amplitude of their fluctuations (one standard
deviation).
This figure indicates that, to first order, the mass accretion rate
is independent of $\varepsilon_{\rm v}$, even for the large 
$\varepsilon_{\rm v}=0.2$.
Although the mean mass accretion rate does vary slightly
with Mach number and
accretor radius, even in units of $\dot{M}_{\rm BH}$, the variation of
the rates across different velocity gradients $\varepsilon_{\rm v}$
remains within the fluctuations of the unstable flow.
This is only in partial agreement with Fig.~\ref{fig:shali}.
Up to $\varepsilon_{\rm v}=0.1$ no variation is expected from 
Fig.~\ref{fig:shali} if all mass in the accretion cylinder is actually
accreted (which is an assumption that enters when deriving
Eq.~(\ref{eq:accmass}), etc.).
However, since the mass accretion rate seems to remain unchanged even
for the models with $\varepsilon_{\rm v}=0.2$ (contrary to 
Fig.~\ref{fig:shali}), we conclude that for these large velocity
gradients not all matter in the accretion cylinder is accreted
any longer.

The relative mass fluctuations, i.e.~ the standard deviation $S$
divided by the average mass accretion rate $\overline{\dot{M}}$ 
(cf.~Table~\ref{tab:models}), have been collected in
Fig.~\ref{fig:relfluct}. 
Although some models cluster around a relative fluctuation of 5\%--10\%
while others are around 15\%--25\%, it is not clear which combination
of parameters is responsable for this division.

Two general statements can be made.
When, starting from axisymmetric models, slightly increasing in the
velocity gradient to a few percent, the relative accretion rate
fluctuation either remains unchanged or decreases.
In the axisymmetric case, any eddy will produce a fluctuation of the
accretion rates. 
As long as the velocity gradient is small, the vortex generated by the
the incoming angular momentum around the accretor is of the same
strength as the eddies and it might be able to stabilize the flow
around the accretor.
When further increasing the velocity gradient the relative
fluctuations increase strongly.
So the stabilizing effect is lost indicating that the vortex itself
contributes to the eddies and the fluctuations.

\begin{figure}
\epsfxsize=8.8cm \epsfclipon \epsffile{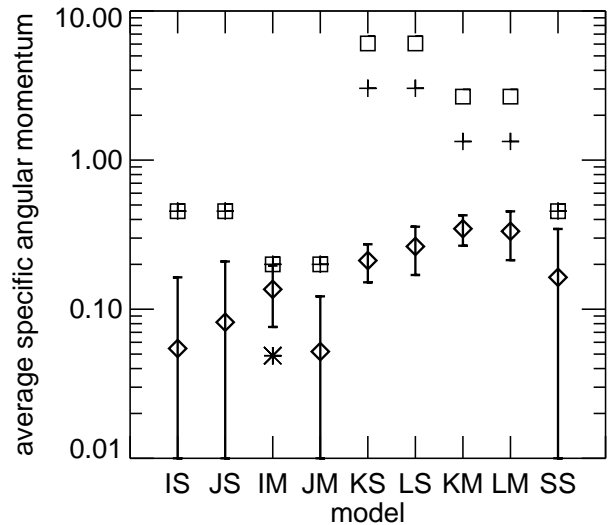}
%\picplace{6.3cm}
\caption[]{The average specific angular momentum (units: $l_{\rm s}$,
Kepler velocity vortex at surface of accretor,  as given by
Eq.~(\ref{eq:kepler})) is shown for most models by diamond
symbols ($l_{\rm z}$ in Table~\ref{tab:models}).
The ``error bars'' extending from the symbols indicate one standard
deviation from the mean ($\sigma_{\rm z}$ in Table~\ref{tab:models}).
The long error bars extending to the bottom axis are an indication
that the fluctuations of the respective model are so large, that the
specific angular momentum changes sign from time to time.
The plus signs above the diamonds indicate the specific angular
momentum $j_{\rm z}$ according to the Shapiro \& Lightman (1976)
prescription, Eq.~(\ref{eq:specmomang}), while the squares denote the
values $\jmath_{\rm z}$ taken from the semi-numerical estimate,
Eq.~(\ref{eq:coeffspec}) and Fig.~\ref{fig:shali}.
All models have $\gamma=5/3$, except for model~SS ($\gamma=4/3$).
The star ($*$) at the position of IM is the value taken from Ishii
et al.~(1993) (see text in Sect.~\ref{sec:works})
}
\label{fig:angacc}
\end{figure}

\subsection{Specific angular momentum\label{sec:spec}}

Assuming a vortex flowing with Kepler velocity $V$ just above the
accretor's surface with radius $R_\star$, the specific angular
momentum of such a vortex is 
\begin{equation}
   l_{\rm s} = R_\star V = \sqrt{R_\star/R_{\rm A}}
           {\cal M}_\infty  R_{\rm A} c_\infty / \sqrt{2}   \quad .
 \label{eq:kepler}
\end{equation}
The term $\sqrt{2}$ has erroneously been omitted in my previous works
(e.g.~Ruffert, 1996).
Although for short periods of time the specific angular momentum
can exceed $l_{\rm s}$, it is difficult to imagine how accreted matter 
can {\it on average} (temporal) exceed this value.
This implies that smaller objects (smaller $R_\star$) can accrete 
only smaller specific angular momenta, which goes to zero like 
$\sqrt{R_\star}$.

In Fig.~\ref{fig:angacc}, I plot for several models the numerically
obtained quantities $l_{\rm z}$ along with the amplitude of the 
fluctuations (one standard deviation, $\sigma_{\rm z}$), 
which can be found in Table~\ref{tab:models}.
These are plotted in units of $l_{\rm s}$ (Eq.~(\ref{eq:kepler})).
Additionally, above the diamonds denoting the above mentioned ratio 
$l_{\rm z}/l_{\rm s}$, I plot, using plus-signs, the values that one
expects for the analytically estimated quantity $j_{\rm z}$ from
Eq.~\ref{eq:specmomang} and denoting by squares, $\jmath_{\rm z}$ from
the semi-numerical estimate, Eq.~(\ref{eq:coeffspec}).

Several trends can easily be noticed in Fig.~\ref{fig:angacc}.
Model~JM seems to be well below the general trend, indicating that the 
simulation was not evolved for long enough; I will not include this
model in the following discussion.
The four ``K'' and~``L'' models form a fairly homogeneous group
accreting roughly 0.3 of the Kepler specific angular momentum.
For the ``I''- and~``J''-models this fraction is roughly 0.1, thus
confirming that for models with smaller gradients, the vortex around
the accretor is less pronounced.
These two groups vary less among themselves than the variation one
would expect if the analytical estimates Eq.~(\ref{eq:specmomang})
(plus signs) or Eq.~(\ref{eq:coeffspec}) (squares) were valid.
Thus, when estimating the specific angular momentum one should be
guided by the Kepler-values Eq.~(\ref{eq:kepler}).
When applying Eq.~(\ref{eq:accmass}) one should bear in mind the
allowable parameter range: small gradients, supersonic flow and small
accretors. 
A good counter example is model~RL, which exhibits a constant state
and the sign of the accreted angular momentum is opposite to 
Eq.~(\ref{eq:specmomang}).

The smaller lever arm acting in models with smaller accretors is
included in Eq.~(\ref{eq:kepler}). 
Still, the ``S''-cases have a slightly smaller value of 
$l_{\rm z}/l_{\rm s}$ compared to the ``M''-cases.
The reduction is, however, not uniform for all models; the models with
large gradients show reductions of at most a factor of 2, while the
other models have a factor of 3.
So small accretors impede high specific angular momentum accretion in
some additional way than only via their smaller lever arm.
At the distance of the surface (= radius of the accretor),
matter seems to move in eddies at some fraction of the Kepler-speeds.
This fraction is dependent more on the velocity gradient 
$\varepsilon_{\rm v}$ than on the size of the accretor or the Mach
number. 
Recall, that from Fig.~\ref{fig:shali} one would expect the models
with large gradients (``K'' and ``L'') to accrete specific angular
momenta a factor of 2 larger than the analogous models with small
gradients (``I'' and ``J'').
This is in contradiction to what is shown in Fig.~\ref{fig:angacc},
confirming again that the assumption of accreting everything from the
accretion cylinder is not correct for large gradients.

When the fluctuations of the specific angular momentum are larger than
its mean, then the accreted angular momentum can change sign,
indicating a reversal of the rotation direction of a disk around the
accretor. 
This will most easily be attained for models with a small gradient,
since the fluctuations need not be large in these cases.
These models have their ``error bars''
extending completely down to the $x$-axis in Fig.~\ref{fig:angacc}.
For one of these models, model~SS, Fig.~\ref{fig:disk} shows the
reversal of the disk surrounding the accretor.
The ``normal'' rotation direction is shown in Fig.~\ref{fig:disk}e for
model~KS which has a large gradient $\varepsilon_{\rm v}=0.2$.
From the bottom right panel in Fig.~\ref{fig:valueK} one can see that
the specific angular momentum never changes sign indicating a
strong regular flow around the accretor.
This is visible in Fig.~\ref{fig:disk}e: the disk (region of azimuthal
flow) extends to a distance of at least 0.3~$R_{\rm A}$ and is,
however, slightly eccentric. 
In contrast, the disk around the models with small gradient,
e.g.~model~SS, is smaller: roughly 0.1~$R_{\rm A}$.
Figs.~\ref{fig:disk}a to~d show this small disk alternating between
clockwise and anti-clockwise rotation.
Shocks appear when matter originating outside the disk is accreted in
the opposite direction of the current disk rotation, e.g.~shortly
after the disk has been counterrotating (in Fig.~\ref{fig:disk}a),
matter from downstream forces the disk to rotate in anti-clockwise
direction (in Fig.~\ref{fig:disk}b).
In this figure a shock is visible at ($x\approx-0.05$,~$y\approx-0.05$):
note the velocity discontinuity, the change from supersonic to
subsonic (dashed line) and the increase in density (darker shades)
when going with the flow in anti-clockwise direction at 
($x\approx-0.05$,~$y\approx-0.05$).

\begin{figure*}
 \begin{tabular}{cc}
\raisebox{8cm}{\parbox[t]{8.8cm}{
\caption[]{\label{fig:disk}
A closeup view of the matter and velocity distribution around the
accretor for two models with small accretor ($R=0.02\,R_{\rm A}$).
%The density distribution is shown with contours and shades of gray:
%darker tones indicate higher density.
Arrows are overlayed to show the instantaneous velocity of matter.
The contour lines are spaced logarithmically in intervals of 0.5~dex
for model~SS and~0.2~dex for model~KS.
The bold contour levels are sometimes labeled with their respective
values (0.01 and 1.0).
The dashed contour delimits supersonic from subsonic regions.
The time of the snapshot together with the velocity scale is given in
the legend in the upper right hand corner of each panel.
}}}
 \end{tabular}
%\picplace{20.cm}
\end{figure*}

Shocks imply generation of entropy, thus if shocks appear more often
when the direction of rotation of the disk is reversed one would
expect that matter with higher specific entropy is accreted during
phases when the specific angular momentum of the matter changes sign.
To find such a possible correlation, I draw in Fig.~\ref{fig:ecorrel}
a dot connecting the two quantities for every second time step of the
numerical simulation.
This was done only for the four models~JM, JS, IS, and~SS which
exhibited a change in sign of the specific angular momentum (see
e.g.~Fig~\ref{fig:angacc}).
The two models~JM and~IS (Fig.~\ref{fig:ecorrel}a and~c) have an only
small intrinsic scatter of the specific entropy accreted
(roughly~10\%) indicating that the fluctuations visible in
Figs.~\ref{fig:valueI} and~\ref{fig:valueJ} do not generate equally
fluctuating shock structures.
For model~JM the entropy does not at all seem to correlate with the
angular momentum, while for model~IS a marginal indication exists
that the specific entropy is slightly higher for larger (=more
positive) specific angular momenta.
On the other hand, the highest entropies appear at the most negative
momenta. 
In contrast, models~JS and~SS exhibit a relatively large scatter of the
specific entropy (roughly 40\% and 25\%, respectively) and fairly
clear correlations.
In model~JS (Fig.~\ref{fig:ecorrel}b) high entropy material is
accreted preferentially when the angular momentum is small (around zero).
This result confirms what has been described at the beginning of this
paragraph. 
The correlation in Model~SS (Fig.~\ref{fig:ecorrel}d) is different
again: no material with low specific entropy (less than say 4.9)
is accreted when the specific angular momentum is positive.
Thus the disk seems to be in constant turmoil (with many shocks) when
the disk rotates in anti-clockwise direction (which is contrary to
the ``normal'' direction, cf.~beginning of this Sect.~\ref{sec:spec}).
However, when the disk rotates in clockwise direction both high and low
entropy material (more or less than 4.9) is present.

\begin{figure*}
 \begin{tabular}{cc}
  \epsfxsize=8.8cm \epsfclipon \epsffile{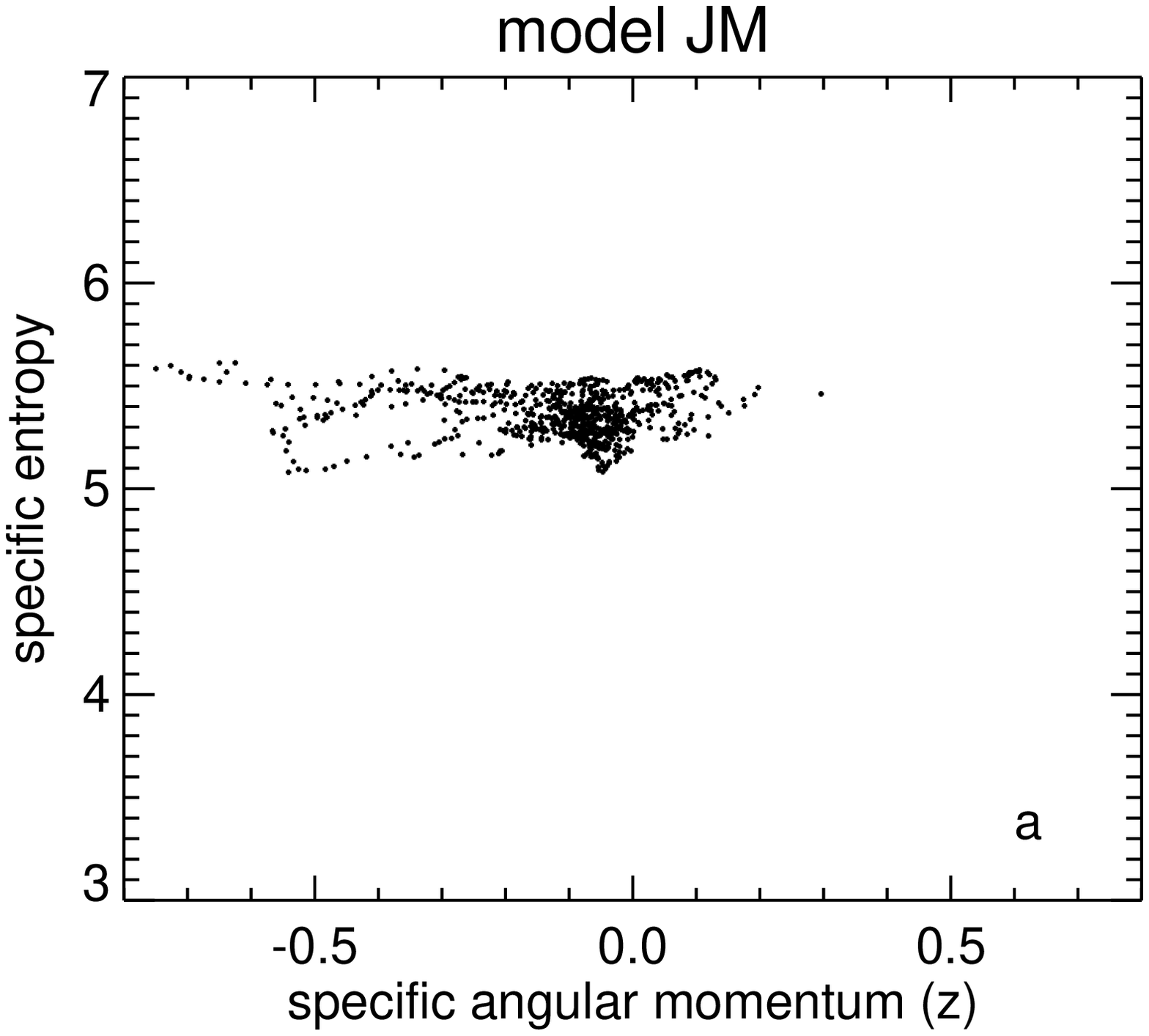} &
  \epsfxsize=8.8cm \epsfclipon \epsffile{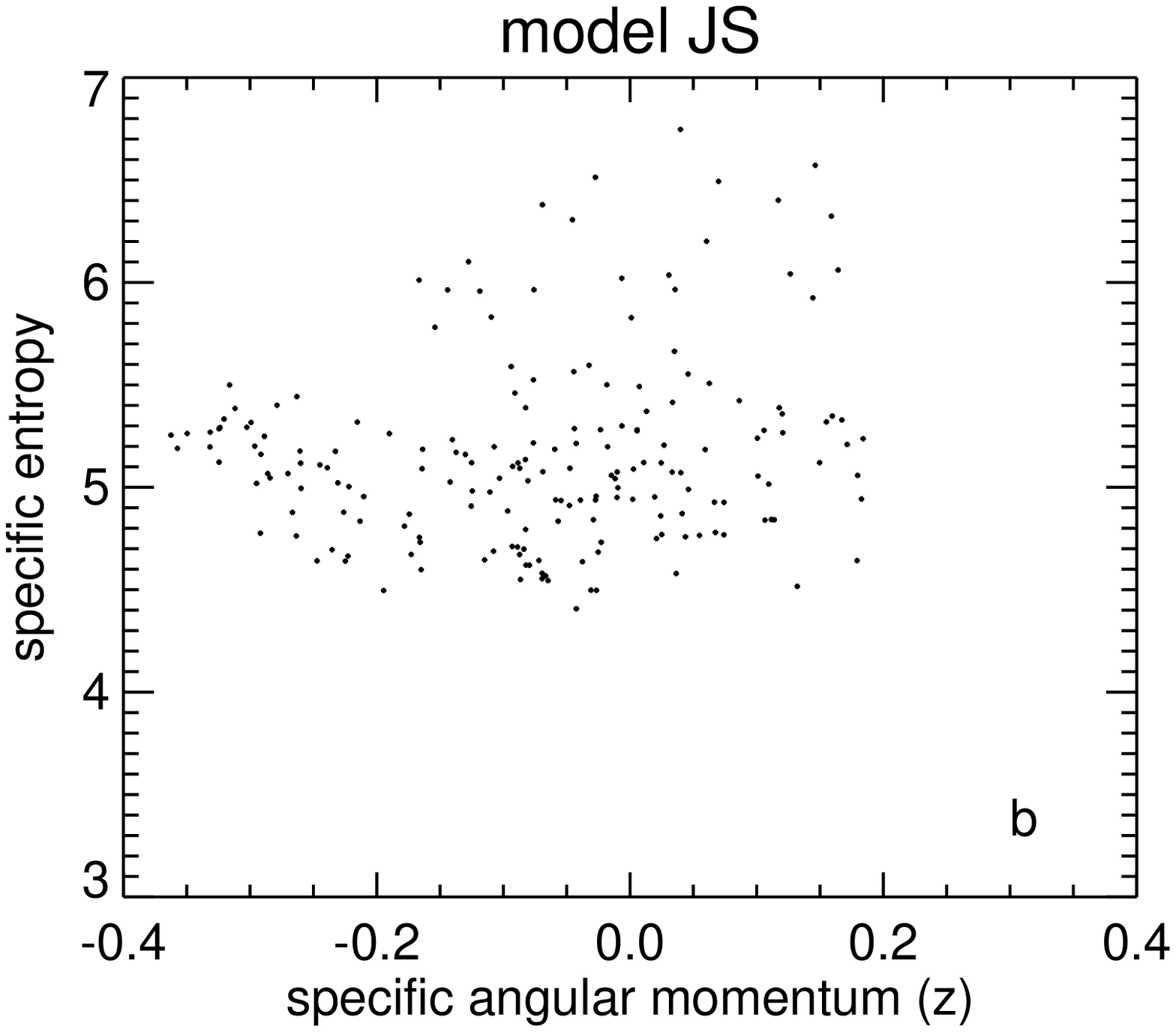} \\
  \epsfxsize=8.8cm \epsfclipon \epsffile{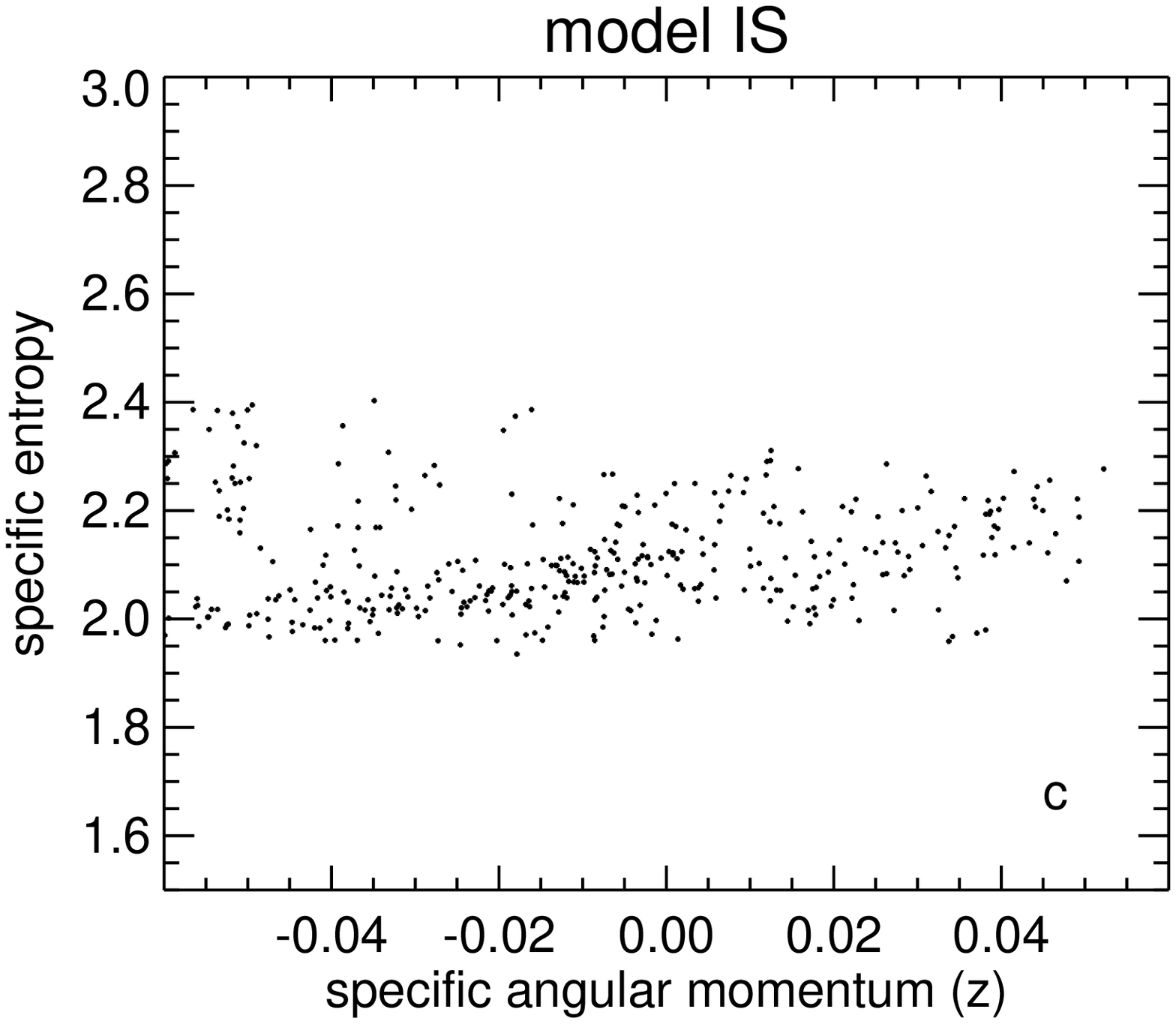} &
  \epsfxsize=8.8cm \epsfclipon \epsffile{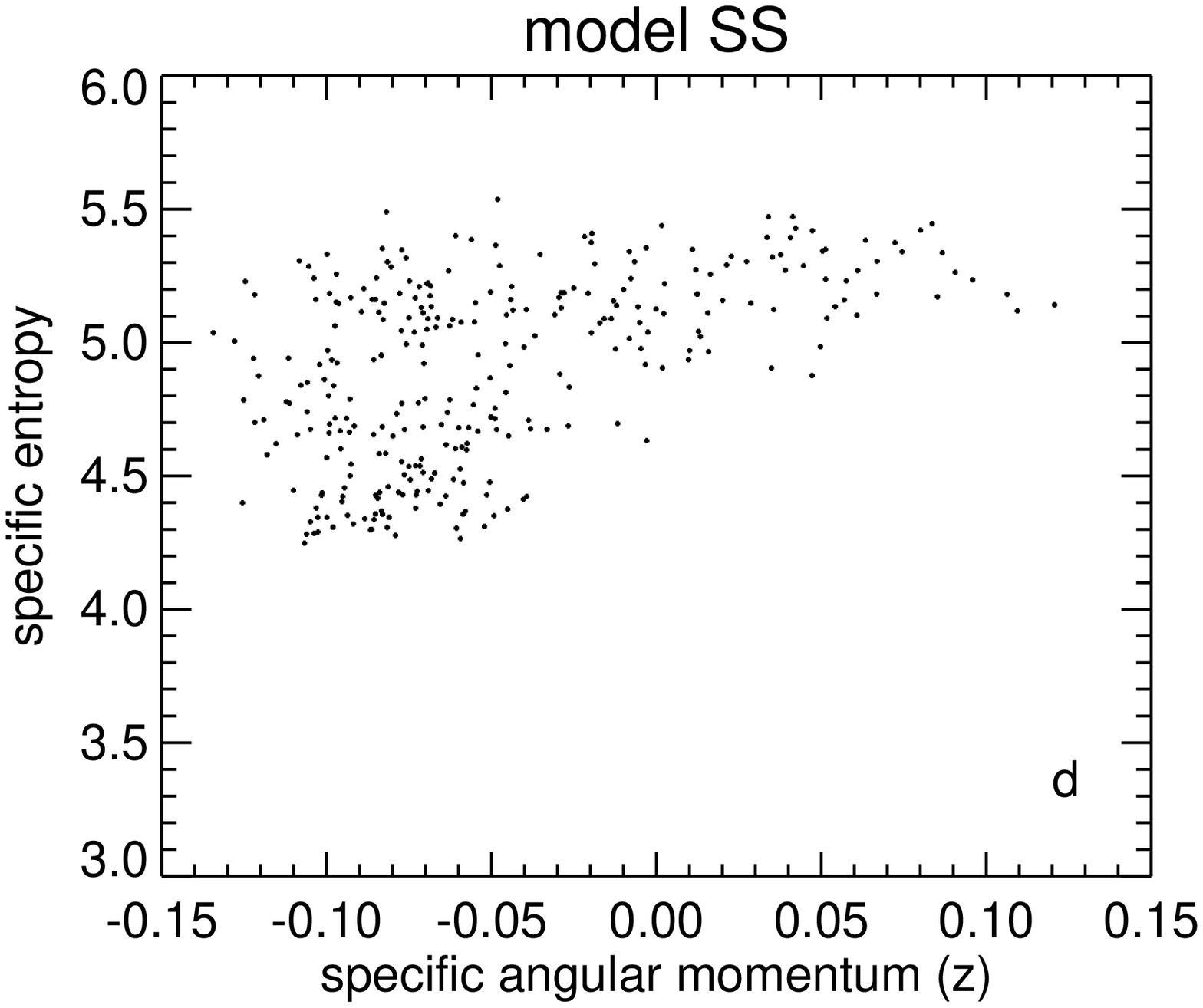} 
 \end{tabular}
\caption[]{\label{fig:ecorrel}
The specific entropy of accreted matter  is plotted versus the specific 
angular momentum of this matter for models~JM, JS, IS and~SS.
Each dot displays the two quantities at one moment in time.
The times for which the dots are plotted are $t\!\ga\!1$, $t!\ga\!2$,
$t\!\ga\!6$, and $t\!\ga\!3$, respectively.
}
%\picplace{20.cm}
\end{figure*}

\begin{figure*}
 \begin{tabular}{cc}
  \epsfxsize=8.8cm \epsfclipon \epsffile{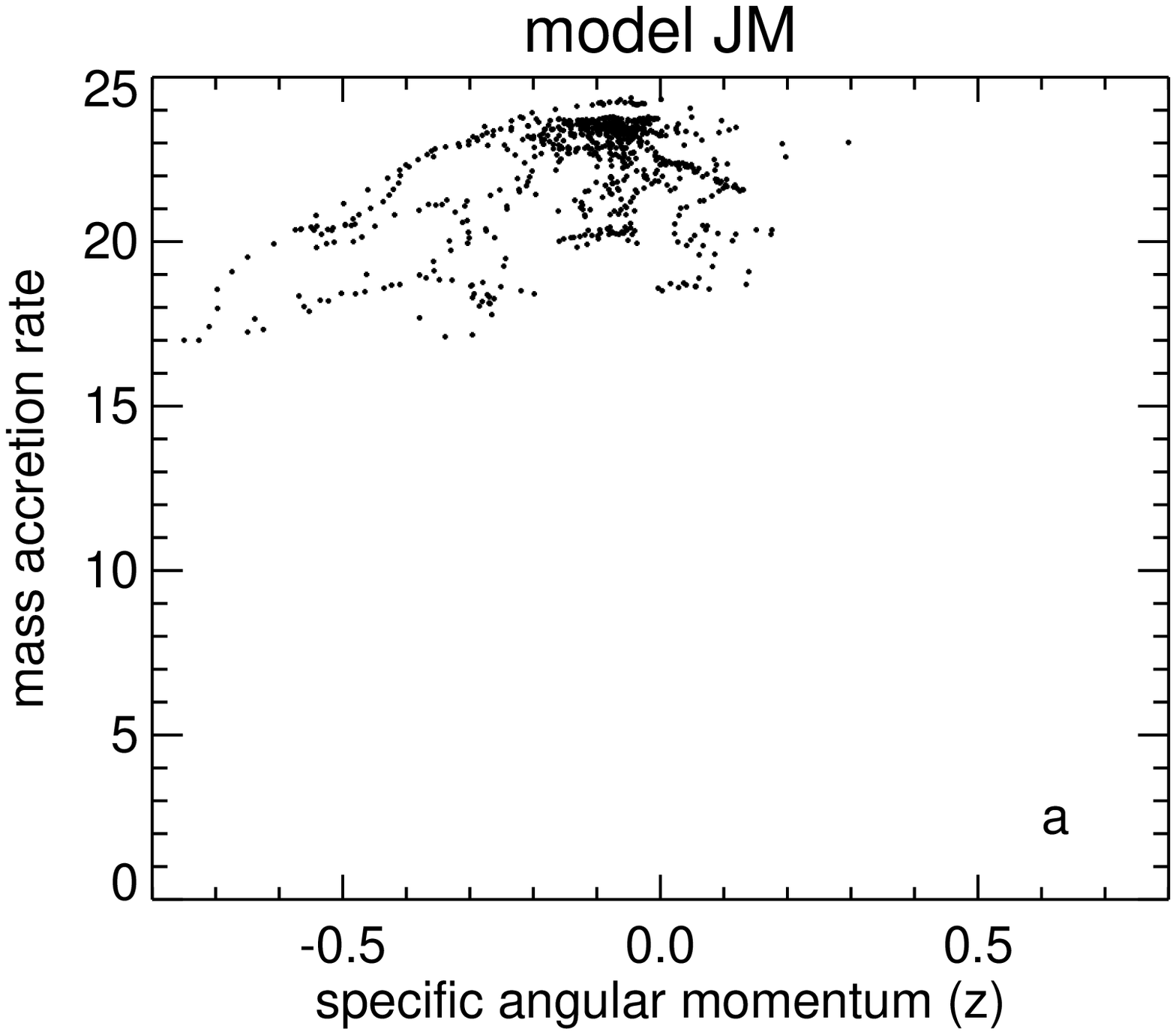} &
  \epsfxsize=8.8cm \epsfclipon \epsffile{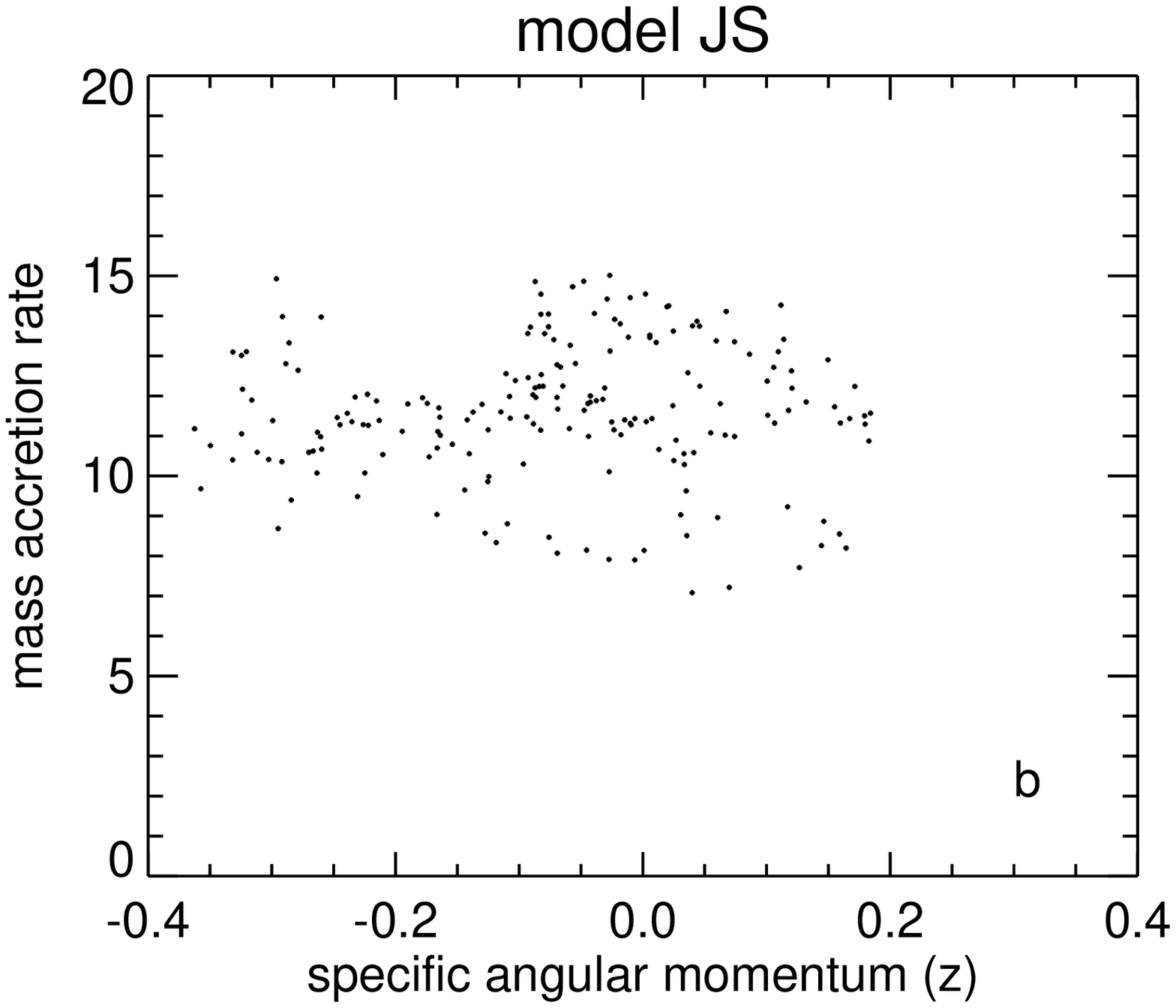} \\
  \epsfxsize=8.8cm \epsfclipon \epsffile{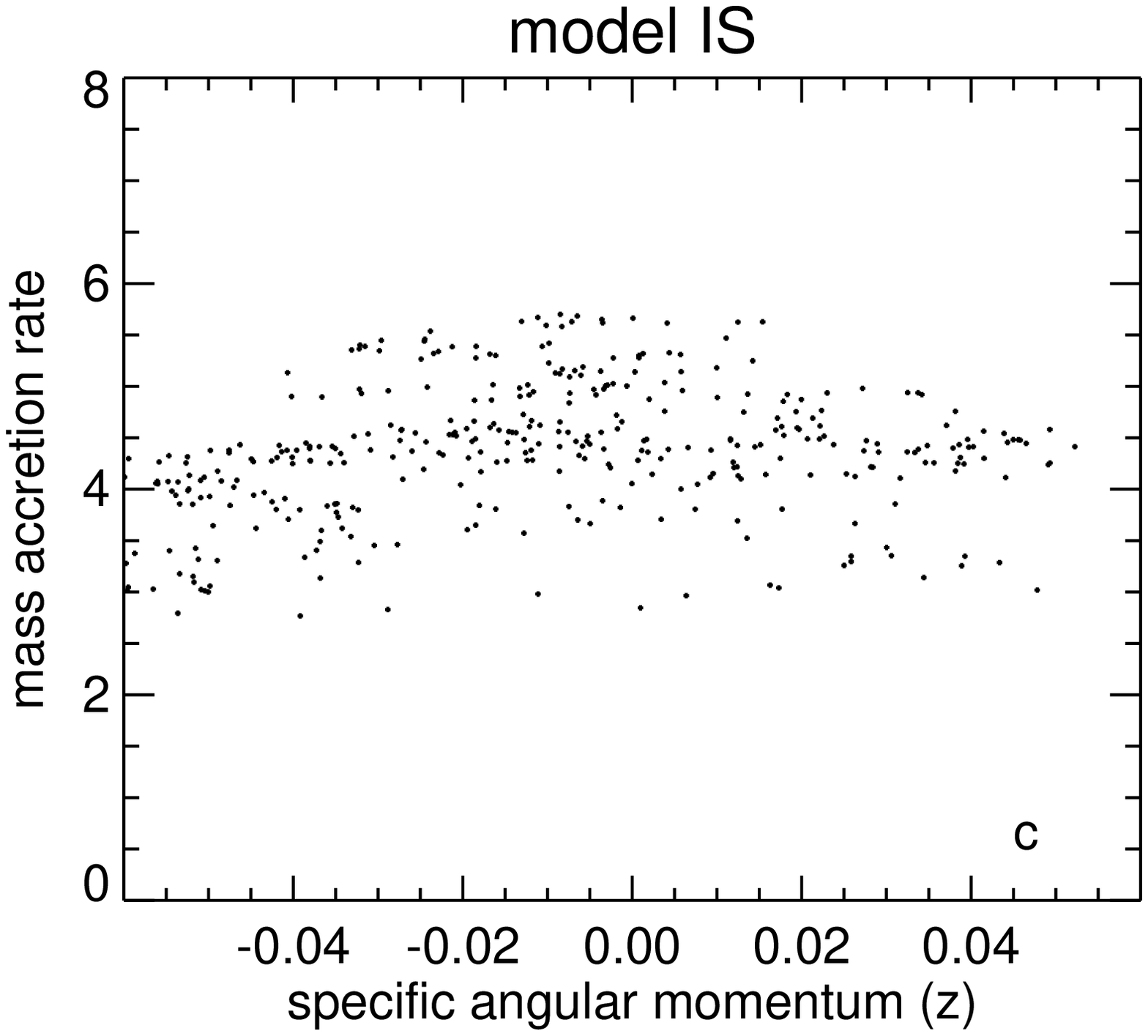} &
  \epsfxsize=8.8cm \epsfclipon \epsffile{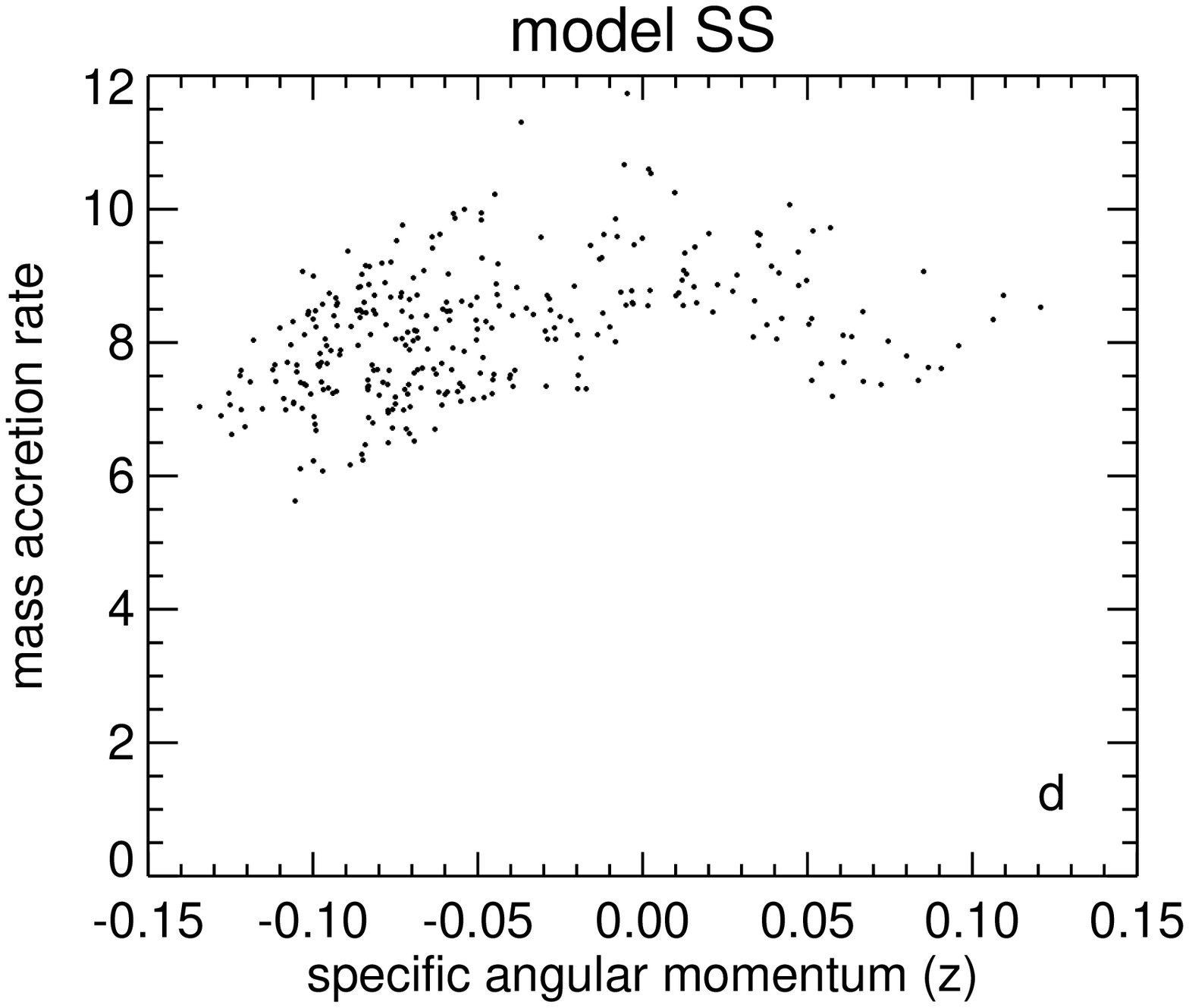} 
 \end{tabular}
\caption[]{\label{fig:correl}
The mass accretion rate is plotted versus the specific angular 
momentum of the accreted matter for models~JM, JS, IS and~SS.
Each dot displays the two quantities at one moment in time.
The times for which the dots are plotted are $t\!\ga\!1$, $t\!\ga\!2$,
$t\!\ga\!6$, and $t\!\ga\!3$, respectively.
}
%\picplace{20.cm}
\end{figure*}

In Fig.~\ref{fig:correl} the rate at which mass is accreted is plotted
versus the specific angular momentum.
Analogously to the differing correlations of the entropy, the four
models~JM, JS, IS, and~SS display different behaviours.
The correlation in model~SS is clearest: the mass accretion rate is
highest when the specific angular momentum is around zero and the rate
decreases for more positive and more negative values of the specific
angular momentum. 
A similar, but less clear, trend can be discerned for model~JM
(Fig.~\ref{fig:correl}).
Obviously, when the flow does not rotate around the accretor it falls
down the potential to the surface of the accretor and can thus be
absorbed more efficiently.

The maximum rate at which mass is accreted also decreases 
with increasing magnitude of specific angular momentum (independent of
sign) in models~JS and~IS (Figs.~\ref{fig:correl}b and~c). 
However, the smallest mass accretion rates are scattered fairly
uniformly along all momenta.

\subsection{Comparison with previous works\label{sec:works}}

Two of the previously published {\it two}-dimensional simulations
mentioned in the introductory chapter Sect.~\ref{sec:intro}
investigate a velocity gradient (as was done here); 
Taam \& Fryxell~(1989) and Anzer et al.~(1987).
The most important parameters of the first work are: the radius of the
accretor is $R\!=\!0.037\,R_{\rm A}$, the adiabatic index is
$\gamma\!=4/3$, the velocity gradient is $\varepsilon\!=\!0.005$
or~$\varepsilon\!=\!0.0625$, with a Mach number of ${\cal M}=4$ or
${\cal M}=12$.
Thus a model most similar to these conditions is model~SS
(cf.~Table~\ref{tab:models}).
When comparing the top two panels of Fig.~\ref{fig:valueS} with the
equivalent Figs.~10 and~11 in Taam \& Fryxell~(1989) one notices one
main difference: model~SS does not show the ``flaring events''
described by Taam \& Fryxell~(1989).
These flaring events are due to the collapse of otherwise fairly
stable disks around the accretor.
In quasi-regular intervals the disk changes its direction of rotation
and during these inversions the mass of the disk is accreted.
This yields short episodes of very high accretion rates, termed 
flaring events by Taam \& Fryxell~(1989).
In model~SS the disk is much less stable than in Sequence~2 of 
Taam \& Fryxell~(1989), consequently the buildup and collapse of the
disk is much more eratic, so no flaring events of the same magnitude
as in Taam \& Fryxell~(1989) is seen in model~SS.
That the disk is so stable in the simulation by 
Taam \& Fryxell~(1989) is due to the fact that their calculation is
two-dimensional, contrary to the three-dimensional models presented here.
Once a disk is formed in a two-dimensional calculation hardly any
matter can be accreted in radial direction.
In three-dimensions, however, matter can still be accreted via the
polar caps even if is disk is present in the equatorial plane.
If this matter is focussed from above and below the disk, it can 
also act to disrupt the integrity of the disk, shortening the lifetime
of the disk in three-dimensional calculations.

The most important parameters of the Anzer et al.~(1987) work are: 
the radius of the accretor is $R\!=\!0.14\,R_{\rm A}$, the adiabatic
index is $\gamma\!=1.5$, the velocity gradient is $\varepsilon\!=\!0.3$
and the Mach number is ${\cal M}=3$.
Models~IM or~SS from the present work most closely resemble these
parameters.
In Sect.~4 of Anzer et al.~(1987) they report finding a ratio of 0.22
between the numerically obtained (SPH simulation with 7500 particles)
specific angular momentum and the analytically estimated value.
This value is within a factor of two of the results shown in
Fig.~\ref{fig:angacc}.

Of the five previously published {\it three}-dimensional simulations
mentioned in the introduction, we will concentrate our comparison to
the results of Ishii et al.~(1993). 
In the other cases the numerics is very coarse or questionable,
e.g.~few zones or particles, local time stepping not appropriate to
the problem, etc.
The parameters which were used in the ``3D velocity inhomogeneous
case'' of Ishii et al.~(1993) are the following.
The radius of the accretor is $R\!=\!0.125\,R_{\rm A}$, the adiabatic
index is $\gamma\!=1$ (isothermal), the velocity gradient is
$\varepsilon\!=\!0.1$ and their Mach number is ${\cal M}=3$.
No model described in the present paper (cf.~Table~\ref{tab:models})
has such a low adiabatic index, but the other parameters are most
closely covered by model~IM or~KM.
The value that Ishii et al.~(1993) obtain numerically
(24\% is -0.11 devided by -0.45, these values are taken from
Sect.~3.5 in Ishii et al.~1993) is represented
by a star ($*$) in Fig.~\ref{fig:angacc}.
Their value (24\%) must be used with caution, since their model has
an adiabatic index of $\gamma=1$ and their numerical resolution
of the accretor is only two zones.
Extrapolating from the models presented in this paper, I would expect
a higher value, since the average specific angular momentum of
model~SS (smaller $\gamma$) is larger than of model~IS 
(larger $\gamma$), and the models with larger accretors (``M''-models)
are larger still.

\section{Conclusions\label{sec:conc}}

For the first time a comprehensive numerical {\it three}-dimensional
study is presented of wind-accretion with a velocity gradient using a
high resolution hydrodynamic code. 
I vary the following parameters: Mach number of the relative flow
(Mach~3 and~10), strength of the velocity gradient perpendicular to
this flow (3\% and 20\% over one accretion radius), radius 
of the accretor (0.02, 0.1 and 1 accretion radius), and adiabatic
index (5/3 and 4/3).
The results are compared among the models with differing parameters,
to some previously published simulations, and also to the
analytic estimates of the specific angular momentum of the matter that
is accreted (Eq.~(\ref{eq:specmomang}), which assumes that all angular
momentum in the accretion cylinder is actually accreted).

\begin{enumerate}
\item
All models with a small enough accretor (with a size less or equal than
0.1~accretion radii) exhibit active unstable phases, very
similar to the models without gradients.
The accretion rates of mass, linear and angular momentum fluctuate
with time, although not as strongly as published previously for 2D
models (e.g.~Fryxell \& Taam 1988).
Similarly to the 2D simulations, transient disks form around the
accretor that alternate their direction of rotation with time.
\item
Depending on the model parameters, the average specific angular
momentum accreted is roughly between 7\% and 70\% of the analytical
estimate.
For the models with small velocity gradients (3\%) the accreted specific
angular momentum is roughly a factor of~10 smaller than the value of a
vortex with Kepler velocity around the surface of the accretor.
This factor is roughly~3 for models with a large gradient of~20\%.
\item
The mass accretion rates of all models with velocity gradients are
equal, to within the fluctuation amplitudes, to the rates of the
models without gradients (published previously).
\item
The fluctuations of the mass accretion rate in the models with small
gradients (3\%) are also similar to the values of the models without
gradients, while the models with large gradients (20\%) exhibit larger
fluctuations.
So large gradients either amplify existing instability mechanisms or 
generate new ones.
\item
Marginal correlations are found, connecting the mass accretion rate,
the specific angular momentum, and the specific entropy during the
temporal evolution.
The mass accretion rate is maximal when the specific angular momentum
is zero, while the specific entropy tends to be smaller when the disks
are prograde (i.e.~when the specific angular momentum is negative, in
our units).
\end{enumerate}

Movies in mpeg format of the dynamical evolution of some  models are
available in the WWW at
{\tt http:\nix//www.mpa-garching.mpg.de\nix/\lower0.7ex\hbox{$\!$\~~$\!$}mor\nix/bhla.html}

\begin{acknowledgements}
I would like to thank Dr.~F.~Meyer, Dr.~U.~Anzer, Dr.~T.~Foglizzo,
and K.~Schenker
%and R.~Stehle 
for carefullay reading 
and correcting the the manuscript and suggesting improvements.
The calculations were done at the Rechenzentrum Garching.
\end{acknowledgements}


\begin{thebibliography}{}

\bibitem[]{anz} Anzer U., B\"orner G., Monaghan J.J., 1987, A\&A~176, 235

\bibitem[]{anz} Anzer U., B\"orner G., 1995, A\&A~299, 62

\bibitem[]{bal} Balsara D., Livio M., O'Dea C.P., 1994, ApJ 437, 83


\bibitem[]{ber} Berger M.J., Colella P., 1989, JCP~82, 64

\bibitem[]{bin} Binney J., Tremaine S., 1987,
      ``Galactic Dynamics'', Princeton University Press, New Jersey

\bibitem[]{bof} Boffin H.M.J., 1991, IAU~Symp.~151.

\bibitem[]{bon} Bondi H., 1952, MNRAS~112, 195

\bibitem[]{boh} Bondi H., Hoyle F., 1944, MNRAS~104, 273

\bibitem[]{cha} Chandrasekhar S., 1943, ApJ~97, 255

\bibitem[]{che} Chevalier R.A., 1996, ApJ~459, 322

\bibitem[]{col} Colella P., Woodward P.R., 1984, JCP~54, 174

\bibitem[]{dav} Davies R.E., Pringle J., 1980, MNRAS~191, 599

\bibitem[]{dod} Dodd K.N., McCrea W.H., 1952, MNRAS~112, 205

\bibitem[]{dra} Drazin P.G., Reid W.H., 1981, in {\it Hydrodynamic
                Stability}, Cambridge University Press, Cambridge

\bibitem[]{fry} Fryxell, B.A., Taam, R.E., 1988, ApJ, 335, 862.

\bibitem[]{ho1} Ho C., Taam R.E., Fryxell B.A., Matsuda T., Koide H.,
     Shima E., 1989, MNRAS~238, 1447

\bibitem[]{ho3} Hoyle F., Lyttleton R.A., 1939,
{\it Proc.\ Cam.\ Phil.\ Soc.}~35, 405

\bibitem[]{h4a} Hoyle F., Lyttleton R.A., 1940a,
{\it Proc.\ Cam.\ Phil.\ Soc.}~36, 323

\bibitem[]{h4b} Hoyle F., Lyttleton R.A., 1940b,
{\it Proc.\ Cam.\ Phil.\ Soc.}~36, 325

\bibitem[]{h4c} Hoyle F., Lyttleton R.A., 1940c,
{\it Proc.\ Cam.\ Phil.\ Soc.}~36, 424

\bibitem[]{hun} Hunt R., 1979, MNRAS~188, 83

\bibitem[]{ill} Illarionov A.F., Sunyaev R.A., 1975, A\&A~39, 185

\bibitem[]{ish} Ishii T., Matsuda T., Shima E., Livio M., Anzer U., 
                B\"orner G., 1993, ApJ~404, 706

\bibitem[]{liv} Livio M., Soker N., deKool M., Savonije G.J., 
                1986, MNRAS~222, 235

\bibitem[]{liv} Livio M., Soker N., Matsuda T., Anzer U., 
                1991, MNRAS~253, 633

\bibitem[]{mir} Mirabel I.F., Paul J., Cordier B., Morris M., 
                Wink J., 1991, A\&A~251, L43

\bibitem[]{mat} Matsuda T., Ishii T., Sekino N., Sawada K.,
                Shima E., Livio M., Anzer U., 1992, MNRAS~255, 183

\bibitem[]{pet} Petrich L.I., Shapiro S.L., Stark R.F., 
                Teukolsky S.A., 1989, ApJ~336, 313

\bibitem[]{r94a} Ruffert M., 1994a, ApJ~427, 342

\bibitem[]{r94b} Ruffert M., 1994b, A\&AS~106, 505

\bibitem[]{r95} Ruffert M., 1995, A\&AS~113, 133

\bibitem[]{r96} Ruffert M., 1996, A\&A, in press

\bibitem[]{ruf:arn} Ruffert M., Arnett D., 1994, ApJ~427, 351

\bibitem[]{ruf:anz} Ruffert M., Anzer U., 1995, A\&A~295, 108

\bibitem[]{ruf:mel} Ruffert M., Melia F., 1994, A\&A~288, L29

\bibitem[]{saw} Sawada K., Matsuda T., Anzer U., B\"orner G., 
         Livio M., 1989, A\&A~221, 263.

\bibitem[]{sha} Shapiro S.L., Lightman A.P., 1976, ApJ~204, 555

\bibitem[]{shi} Shima E., Matsuda T., Takeda H., Sawada K., 
                1985, MNRAS~217, 367

\bibitem[]{taa} Taam R.E., Fryxell B.A., 1989, ApJ~339, 297

\bibitem[]{wan} Wang Y.-M., 1981, A\&A~102, 36

%\bibitem[]{woo} Wood M.A., Oswald T.D., 1992, ApJ~394, L53

\end{thebibliography}
\end{document}